\documentclass[twocolumn,english,prl]{revtex4-1}

\setcitestyle{super}
\usepackage{amssymb}
\usepackage{amsmath}
\usepackage{graphicx}
\usepackage{amsfonts}
\usepackage[usenames, dvipsnames]{xcolor}
\usepackage{bm}
\usepackage{bbm}
\usepackage{color}
\usepackage{hyperref}
\hypersetup{pdfnewwindow=true,
	colorlinks=true,
	linkcolor=blue,
	citecolor=blue,
	urlcolor=blue}
\usepackage{soul}
\usepackage[caption=false, justification=centerlast]{subfig}
\usepackage{psfrag}
\usepackage{textcomp}
\usepackage{braket}
\usepackage{babel}

\usepackage[T1]{fontenc}
\usepackage[latin9]{inputenc}
\usepackage{amsmath}
\usepackage{amssymb}
\usepackage{graphicx}
\usepackage{wasysym}

\makeatletter
\usepackage{braket}
\usepackage{dsfont}

\usepackage{lineno}

\begin{document}
	
\title{Adaptive quantum error mitigation using pulse-based inverse evolutions }
\author{Ivan Henao$^{1}$}

\author{Jader P. Santos$^{1}$}

\author{Raam Uzdin$^{1}$}
\email{raam@mail.huji.ac.il}

\affiliation{$^{1}$Fritz Haber Research Center for Molecular Dynamics,Institute
	of Chemistry, The Hebrew University of Jerusalem, Jerusalem 9190401,
	Israel}


\begin{abstract}
	{\normalfont  $^*$\url{raam@mail.huji.ac.il}}\\
	
Quantum Error Mitigation (QEM) enables the extraction of high-quality results from the presently-available noisy quantum computers. In this approach, the effect of the noise on observables of interest can be mitigated using multiple measurements without additional hardware overhead. Unfortunately, current QEM techniques are limited to weak noise or lack scalability. In this work, we introduce a QEM method termed `Adaptive KIK' that adapts to the noise level of the target device, and therefore, can handle moderate-to-strong noise. The implementation of the method is experimentally simple --- it does not involve any tomographic information or machine-learning stage, and the number of different quantum circuits to be implemented is independent of the size of the system. Furthermore, we have shown that it can be successfully integrated with randomized compiling for handling both incoherent as well as coherent noise. Our method handles spatially correlated and time-dependent noise which enables to run shots over the scale of days or more despite the fact that noise and calibrations change in time. Finally, we discuss and demonstrate why our results suggest that gate calibration protocols should be revised when using QEM. We demonstrate our findings in the IBM quantum computers and through numerical simulations. 
\end{abstract}

\maketitle
\setcounter{section}{0}
\setcounter{figure}{0}
\setcounter{equation}{0}

\onecolumngrid
\section*{Introduction}

Quantum computers have reached a point where they outperform even the most powerful classical computers in specific tasks \cite{arute2019quantum,zhong2020quantum,madsen2022quantum}. 
However, these quantum devices still face considerable noise levels that need to be managed for quantum algorithms to excel in practical applications. Quantum error correction (QEC) is a prominent solution, although its implementation, particularly in complex problems such as Shor's factoring algorithm, might demand thousands of physical qubits for each encoded logical qubit \cite{fowler2012surface,o2017quantum}. 

A different approach, quantum error mitigation (QEM), has garnered substantial attention recently \cite{temme2017error,li2017efficient,endo2018practical,strikis2021learning,czarnik2021error,koczor2021exponential,huggins2021virtual,giurgica2020digital,cai2021quantum,mari2021extending,lowe2021unified,nation2021scalable,bravyi2021mitigating,kim2023scalable,van2023probabilistic,ferracin2022efficiently,endo2021hybrid,cai2022quantum}.
Its viability has been demonstrated through experiments involving superconducting circuits \cite{kandala2019error,song2019quantum,google2020hartree,urbanek2021mitigating,ferracin2022efficiently,van2023probabilistic,kim2023evidence,kim2023scalable,shtanko2023uncovering}, trapped
ions \cite{zhang2020error}, and circuit QED \cite{sagastizabal2019experimental}. QEM protocols aim to estimate ideal expectation values from noisy measurements, without the resource-intensive requirements of QEC. This positions them as potential solutions for achieving quantum advantage in practical computational tasks \cite{kim2023scalable,kim2023evidence}. Some QEM strategies require moderate hardware overheads and can be seen as intermediary solutions between NISQ (Noisy Intermediate-Scale Quantum) computers and devices that fully exploit QEC \cite{koczor2021exponential,huggins2021virtual}. These strategies aim to virtually refine the pure final state, by utilizing extra qubits for error mitigation without actively correcting errors. The approach introduced here fits into the more common class of QEM techniques that maintain the qubit count of the original circuit.

The objective of QEM is to reduce errors in post processing, rather than fixing them in real time. For instance, the zero-noise extrapolation (ZNE) method \cite{temme2017error,li2017efficient} employs circuits that mimic the ideal target evolution but amplify noise by a controlled factor.  
The noiseless expectation values are estimated via
extrapolation to the zero-noise limit, after fitting a noise scaling ansatz to the measured data. While the construction of circuits that correctly scale the noise is straightforward  if the noise is time independent  \cite{temme2017error} or if it is decribed by a global depolarizing channel  \cite{giurgica2020digital}, it has been observed that circuits designed to amplify depolarizing noise fail to achieve the intended noise scaling, when applied to more realistic noise models \cite{kim2023scalable}. Our experimental findings also show related issues when applying such circuits to QEM in a real system. Another strategy is to simplify the actual noise appearing in multi-qubit gates such  as the CNOT, CZ, Toffoli and Fredkin gates, by using randomized compiling, which renders the noise to a Pauli channel \cite{wallman2016noise,PhysRevX.11.041039}. A sufficiently sparse Pauli channel facilitates accurate characterization and noise amplification for ZNE \cite{kim2023evidence}. Additionally, as in other QEM methods, the performance of ZNE can be enhanced by integrating it with other error mitigation techniques \cite{majumdar2023best}, such as readout error mitigation \cite{nation2021scalable}.

In comparison to ZNE,  Probabilistic Error Cancellation (PEC) is a QEM scheme that relies on an experimental characterization of the noise to effectively suppress the associated error channel \cite{temme2017error,endo2018practical,strikis2021learning,song2019quantum,van2023probabilistic}. To this end, PEC  uses a Monte Carlo sampling of noisy operations that on average cancel out the noise, thereby providing an unbiased estimation of the noise-free expectation value. However, this objective can only be accomplished when precise and complete tomographic details of the noise process are accessible. In practice, the success of bias suppression in PEC is limited by the scalability and accuracy of gate set tomography in realistic scenarios.   Additionally, since noise characteristics evolve over time, the learning process for PEC must be carried out efficiently within a timescale that is shorter than the timescale in which the noise parameters change. A more realistic approach aims for a partial characterization of the noise, using tools like local gate set tomography \cite{endo2018practical} or learning of a sparse noise model \cite{van2023probabilistic}. The latter strategy was also employed to assist the implementation of ZNE in the experiment of Ref. \cite{kim2023evidence}. Alternatively, it is possible to learn a noise model by taking advantage of circuits that are akin to the target circuit but admit an efficient classical simulation \cite{strikis2021learning,czarnik2021error,lowe2021unified,czarnik2022improving}. By concatenating the outcomes from the ideal (simulated) circuits with their experimental counterparts, the noise-free expectation value can be estimated through some form of data regression \cite{czarnik2021error,czarnik2022improving}. Similar learning-based schemes have also been integrated with PEC \cite{strikis2021learning} and ZNE \cite{lowe2021unified}.

In this work, we introduce the `Adaptive KIK' method (`KIK' for brevity) for handling time dependent and spatially correlated noise in QEM. This technique bears a certain (misleading) similarity to a ZNE variant known as circuit (or `global' \cite{majumdar2023best}) unitary folding \cite{giurgica2020digital}, where noise is augmented through identity operations that comprise products of the target evolution and its inverse. While both methods utilize folding to mitigate noise, they differ in the error mitigation mechanism and the way the measured data is processed. Instead of extrapolating to the zero-noise limit, we combine appropriately folded circuits to effectively construct the `inverse noise channel' and approximate the ideal unitary evolution.
As opposed to PEC, the implementation of the KIK method does not involve any tomographic information or noise learning subroutine. More precisely, the coefficients that weight the folded circuits are analytically optimized according to a single experimental parameter that probes the intensity of the noise. Another distinctive aspect of KIK mitigation is a specific inversion of the target circuit for the folding procedure. This constitutes a pivotal difference with respect to circuit folding and has practical consequences, as we show experimentally. The combination of a proper inverse and coefficients adapted to the noise strength  allows us to mitigate moderate-to-strong noise and significantly outperform circuit folding ZNE in experiments and simulations.
Although we show that the weak noise limit of our theory has a clear connection with Richardson ZNE using circuit folding \cite{giurgica2020digital}, the correct inversion of the target circuit is still crucial in this limit. 
     
Recently, important results on the fundamental limitations of QEM protocols have been obtained \cite{takagi2022fundamental,quek2022exponentially}. These studies address the degradation in the statistical precision of generic QEM schemes, as noise accumulates in circuits of increasing size. In this work, instead of analyzing the degradation of statistical precision, our focus is on the accuracy of error mitigation. 
We obtain upper bounds for the bias between the ideal expectation value of an arbitrary observable and the value estimated using the KIK method, as a function of the accumulated noise. Our bounds show exponential suppression of the bias with respect to the number of foldings when the noise is below a certain threshold. This is in contrast with ZNE schemes which, in general, do not provide accuracy guarantees.  
 
We test the KIK method on a ten-swap circuit and in a CNOT calibration process, using the IBM quantum computing platform. In the ten-swap experiment, we demonstrate the success of our approach for mitigating strong noise. In the calibration experiment, it is illustrated that a noise-induced bias in gate parameters leads to coherent errors. KIK-based calibration can efficiently mitigate these coherent
errors by reducing the bias in the calibration measurements. Furthermore, we find that circuit
folding (which uses the CNOT as its own inverse) produces erroneous and inconsistent results. Our experimental findings are enhanced by complementing the KIK method with randomized compiling and readout mitigation. We also simulate
the fidelity obtained with a noisy ten-step Trotterization \cite{trotter1959product}
of the transverse Ising model on five qubits. For unmitigated fidelities
as low as 0.85, we show that KIK error mitigation produces final
fidelities beyond 0.99. 

\section*{Results}
\subsection*{The KIK formula for time-dependent noise }

To derive our results, we adopt the Liouville-space formalism of Quantum Mechanics \cite{gyamfi2020fundamentals}
(see Supplementary Note 1), in which density matrices that
describe quantum states are written as vectors, and quantum operations
as matrices that act on these vectors. In the following, we will employ
calligraphic fonts to denote quantum operations. For example, the
unitary evolution associated with an ideal (noise-free) quantum circuit
and its noisy implementation will be written as $\mathcal{U}$ and
$\mathcal{K}$, respectively.

In the standard representation involving superoperators
	and density matrices, the noisy evolution is governed by the equation
	\begin{equation}
		\frac{d}{dt}\rho=-i[H(t),\rho]+\hat{L}(t)[\rho].\label{eq:1 mast equation in HS}
	\end{equation}
	The ideal evolution is generated by the time-dependent Hamiltonian
	$H(t)$. On the other hand, the effect of noise is characterized by
	the superoperator $\hat{L}(t)$. In the following, we will refer to
	this superoperator as the `dissipator'. The equivalent of Eq.
	(\ref{eq:1 mast equation in HS}) in Liouville space is the equation
	\begin{equation}
		\frac{d}{dt}|\rho\rangle=\left(-i\mathcal{H}(t)+\mathcal{L}(t)\right)|\rho\rangle,\label{eq:2 mast equation in LS}
	\end{equation}
	where $|\rho\rangle$ is the vectorized form of $\rho$. Moreover,
	$\mathcal{H}(t)$ and $\mathcal{L}(t)$ are square matrices that represent
	the Hamiltonian $H(t)$ and the dissipator, respectively. We refer
	the reader to Supplementary Note 2 for more details.

The dynamics (\ref{eq:2 mast equation in LS}) gives
	rise to the noisy target evolution, which we have denoted by $\mathcal{K}$.
	As shown in Supplementary Note 3, we can write the solution to Eq. (\ref{eq:2 mast equation in LS})
	as $\mathcal{K}=\mathcal{U}e^{\Omega(T)}$, where
$\Omega(T)=\sum_{n=1}^{\infty}\Omega_{n}(T)$ is
	the so called Magnus expansion \cite{blanes2009magnus}. The time $T$ is the total
	evolution time and $\Omega_{n}(T)$ is the $n$th order Magnus term
	corresponding to $T$. Here, we are specifically interested in the
	first Magnus term $\Omega_{1}(T)$, for reasons that will be clarified
	below. In our framework, $\Omega_{1}(T)$ characterizes the impact
	of noise and is given by
	 
	\begin{equation}
		\Omega_{1}(T)=\int_{0}^{T}dt\mathcal{U}^{\dagger}(t)\mathcal{L}(t)\mathcal{U}(t),\label{eq:3 Omega1}
	\end{equation}
	where $\mathcal{U}(t)$ is the noise-free evolution at time $t$.
	In particular, $\mathcal{U}:=\mathcal{U}(T)$ is the unitary associated
	with the noise-free target circuit. 

Our basic approximation is the
	truncation of the Magnus series to first order. This leads to 
	\begin{equation}
		\mathcal{K}\approx\mathcal{U}e^{\Omega_{1}(T)}.\label{eq:4 K}
	\end{equation}
	Next, we apply the same approximation to a suitable inverse evolution
	$\mathcal{K}_\textrm{I}$, such that $\mathcal{K}_\textrm{I}$ reproduces the unitary
	$\mathcal{U}^{\dagger}$ in the absence of noise. We construct $\mathcal{K}_\textrm{I}$
	through an inverse driving  $\mathcal{H}_\textrm{I}(t)$ defined by
	\begin{equation}
		\mathcal{H}_\textrm{I}(t)=-\mathcal{H}(T-t).\label{eq:5 HI}
	\end{equation}
	The driving $\mathcal{H}_\textrm{I}(t)$ undoes the action of $\mathcal{H}(t)$,
	and it produces $\mathcal{U}^{\dagger}$.
	By using $\mathcal{H}_\textrm{I}(t)$,
	we find in Supplementary Note 3 that, to first order in the Magnus expansion,
	the solution to the corresponding noisy dynamics satisfies 

\begin{equation}
	\mathcal{K}_\textrm{I}\approx e^{\Omega_{1}(T)}\mathcal{U}^{\dagger}.\label{eq:6 KI}
\end{equation}
Note that this approximation does not mean that we keep only the linear term $\Omega_{1}(T)$, since all the powers of $\Omega_{1}(T)$ are included in the exponential $e^{\Omega_{1}(T)}$. In Eqs. (\ref{eq:6 KI}) and (\ref{eq:7 U_KIK}), we use the symbol `$\approx$' to denote equality up to the first Magnus term.

The fact that $\Omega_{1}(T)$ is also present in
	the inverse evolution  $\mathcal{K}_\textrm{I}$ allows us to express the
	error channel as $e^{\Omega_{1}(T)}\approx\left(\mathcal{K}_\textrm{I}\mathcal{K}\right)^{\frac{1}{2}}$. While $\mathcal{H}_\textrm{I}(t)$ is not the only alternative for generating
		$\mathcal{U}^{\dagger}$,
	it guarantees the generation of a noise channel that is identical, within our appoximation, to the noise channel of $\mathcal{K}$.
	Thus, by working within the first-order truncation of the Magnus expansion,
	we can combine Eqs. (\ref{eq:4 K}) and (\ref{eq:6 KI}) to obtain

	\begin{align}
		\mathcal{U} & \approx\mathcal{K}e^{-\Omega_{1}(T)}\nonumber \\
		& \approx\mathcal{K}\left(\mathcal{K}_\textrm{I}\mathcal{K}\right)^{-\frac{1}{2}}.\label{eq:7 U_KIK}
	\end{align}
	The `KIK formula' in the second line of (\ref{eq:7 U_KIK}) is
	our main result. In the next section, we discuss the implementation
	of the KIK method through polynomial expansions of the operator $\left(\mathcal{K}_\textrm{I}\mathcal{K}\right)^{-\frac{1}{2}}$ appearing in this formula.
	
	We stress that until now the only assumption regarding the nature of the noise is that (see Supplementary Note 2) 
\begin{equation}	
	 \mathcal{L}_\textrm{I}(t)=\mathcal{L}(T-t),\label{eq:Inverse dissipator} 
 \end{equation}
	 where $\mathcal{L}_\textrm{I}(t)$ is the dissipator acting alongside $\mathcal{H}_\textrm{I}(t)$. This relationship follows from the form of the driving  (\ref{eq:5 HI}), and is schematically explained in Fig.  \textcolor{red}{1}. 
	 As detailed in Supplementary Note 2, Eq. \eqref{eq:Inverse dissipator} relies on the time locality of the noise. That is, on the assumption that the dissipators $\mathcal{L}(t)$ and $\mathcal{L}_\textrm{I}(t)$ are only determined by the current time  $t$ and not by the previous history of the evolution. Therefore, Eq. \eqref{eq:Inverse dissipator} may be violated or only hold approximately in the presence of pronounced non Markovian noise. 
 	
Due to the generality of $\mathcal{L}(t)$,  Eq. (\ref{eq:7 U_KIK}) is applicable to quantum
circuits $\mathcal{K}$ that feature time-dependent and spatially
correlated noise, as well as gate-dependent errors. In Supplementary Note 3, we also discuss the scenario where noise
parameters drift during the experiment, which occurs for example due
to temperature variations or laser instability. We show that the impact
of noise drifts can be practically eliminated in our method, if the execution order
of the circuits $\mathcal{K}\left(\mathcal{K}_\textrm{I}\mathcal{K}\right)^{m}$
in Eq. (\ref{eq:2 KIK polynomials}) is properly chosen. As a final remark, we note that the time independent Lindblad master equation \cite{breuer2002theory} is a special case of Eq. \eqref{eq:1 mast equation in HS}. Therefore, our formalism goes beyond QEM proposals based on such a master equation, like the one adopted in Ref. \cite{sun2021mitigating}.

\subsection*{QEM using the KIK formula}

Since $\mathcal{K}$$\left(\mathcal{K}_\textrm{I}\mathcal{K}\right)^{-\frac{1}{2}}$ is
not directly implementable in a quantum device, we utilize polynomial
expansions of  $\left(\mathcal{K}_\textrm{I}\mathcal{K}\right)^{-\frac{1}{2}}$
such that 
\begin{equation}
{\normalcolor {\color{red}{\normalcolor \mathcal{U}_\textrm{KIK}^{(M)}=}}}\sum_{m=0}^{M}a_{m}^{(M)}\mathcal{K}\left(\mathcal{K}_\textrm{I}\mathcal{K}\right)^{m}.\label{eq:2 KIK polynomials}
\end{equation}
The notation $\mathcal{U}_\textrm{KIK}^{(M)}$ represents an $M$th-order
approximation to $\mathcal{U}_\textrm{KIK}$, with real coefficients $\{a_{m}^{(M)}\}_{m=0}^{M}$.
In this way, we estimate the error-free expectation of an observable $A$ as 
\begin{equation}
	{\left\langle A\right\rangle _\textrm{KIK}^{(M)}=\sum_{m=0}^{M}a_{m}^{(M)}\left\langle A\right\rangle _{m}},\label{eq: KIK expect value}
\end{equation}
where $\left\langle A\right\rangle _{m}$ is the expectation value measured after executing the circuit $\mathcal{K}\left(\mathcal{K}_\textrm{I}\mathcal{K}\right)^{m}$ on the initial state $\rho$. Before discussing the evaluation of the coefficients $a_{m}^{(M)}$, used in Eq. \eqref{eq: KIK expect value}, it is instructive to clarify some similarities and differences between the KIK method and ZNE based on  circuit folding.

The application of the KIK formula is operationally
	analogous to the use of circuit folding for ZNE \cite{giurgica2020digital,majumdar2023best}. However,
	there are two crucial differences between these two techniques. Circuit folding is a variant of unitary folding, first introduced in Ref. \cite{giurgica2020digital} as a user-friendly strategy
	for noise amplification in ZNE . It operates by inserting
	quantum gates that are logically equivalent to the identity operation,
	which leave the noiseless circuit unmodified. In the case of `circuit
	folding', identities are generated by folding the target
	circuit with a corresponding inverse circuit. Hence, the noise is
	scaled through evolutions that have the structure $\mathcal{U}\left(\mathcal{U}^{\dagger}\mathcal{U}\right)^{m}$
	\cite{giurgica2020digital}. Notably, excluding the trivial case of a global depolarizing channel \cite{giurgica2020digital}, a rigorous description of how noise manifests when executing $\mathcal{U}\left(\mathcal{U}^{\dagger}\mathcal{U}\right)^{m}$was never presented, to the best of our knowledge. In this sense, circuit folding and other variants of unitary folding can be considered  as a heuristic approach to QEM.  Upon measuring the observable of interest on these circuits,
	the noiseless expectation value is estimated by combining the results
	corresponding to different values of $m$, with weights that depend
	on the noise scaling ansatz. 

The similarity with respect to the KIK method comes
	from the fact that the circuits $\mathcal{K}\left(\mathcal{K}_\textrm{I}\mathcal{K}\right)^{m}$
	in Eq. (\ref{eq:2 KIK polynomials}) are noisy implementations of $\mathcal{U}\left(\mathcal{U}^{\dagger}\mathcal{U}\right)^{m}$.
	However, a key difference is that in our case $\mathcal{U}^{\dagger}$
	is performed using the driving (\ref{eq:5 HI}). Hereafter, we shall refer to this implementation as the `pulse inverse'. Conversely, unitary folding (and particularly
	circuit folding) relies on a circuit-based inversion, where gates
	that are their own inverses are executed in their original form. This is true for both foldings of single gates (or circuit layers) and for circuit foldings. A paradigmatic
	example would be the CNOT gate. In contrast, the driving (\ref{eq:5 HI}) reverses
	the pulse schedule for each gate in the target circuit, including
	CNOTs and other gates that are their own inverses. This translates
	into a very distinct execution of $\mathcal{U}^{\dagger}$, as illustrated
	in Fig. \textcolor{red}{1}(a). Even if $\mathcal{U}$ is just a single CNOT, we show in the section `Experimental results' that properly folded circuits correspond to products between the CNOT and its pulse inverse, while circuit folding (i.e. products of the CNOT with itself) leads to erroneous results. Regarding the implemetation of our method on cloud-based platforms, we are currently writing an open source Qiskit module that generates pulse-inverse circuits automatically, using only gate-level control. Consequently, users will not need to master pulse-level control to utilize our QEM technique.

\begin{figure}
	\centering{}\includegraphics[scale=0.45]{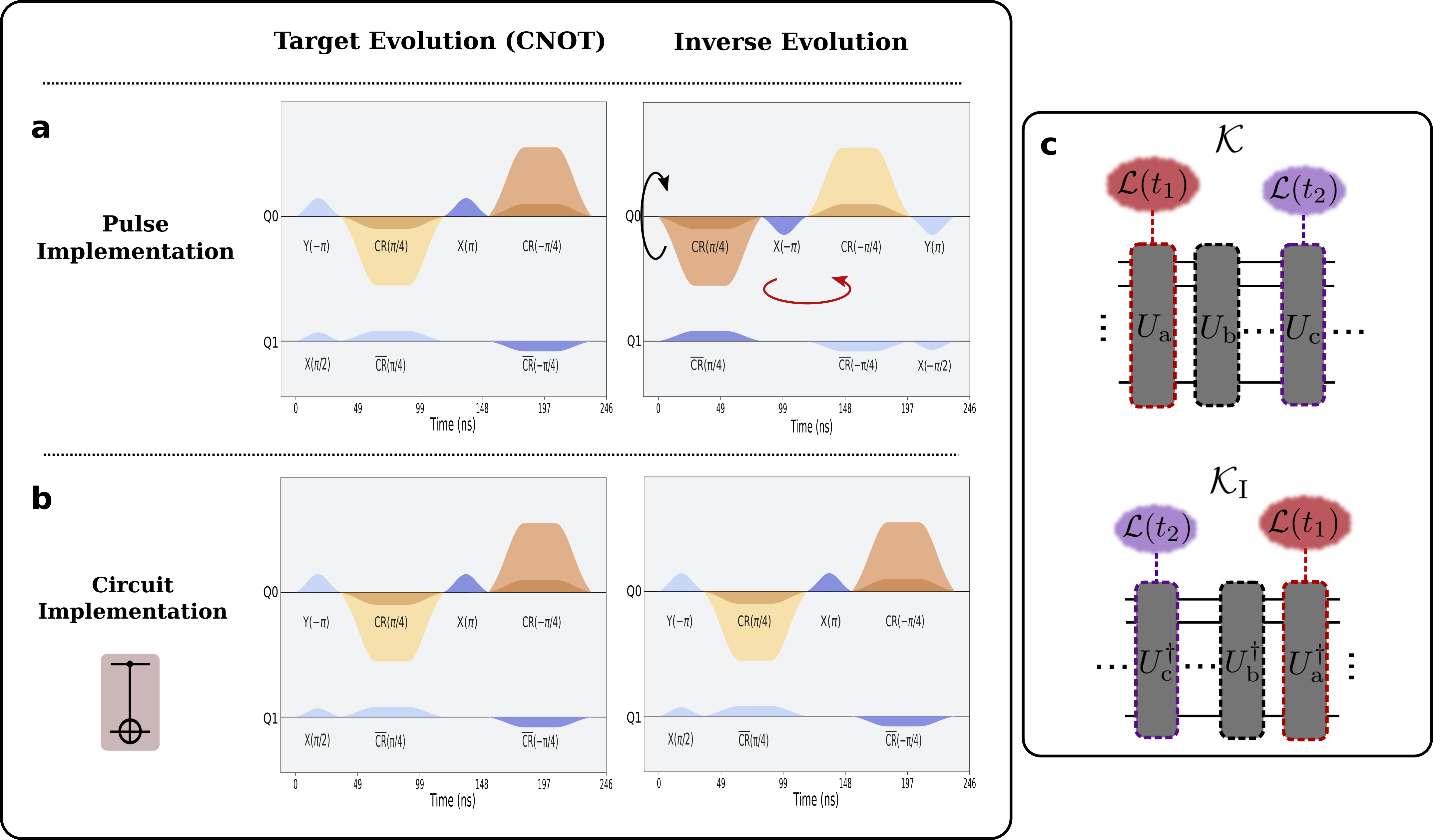}\caption{Illustration of the pulse inverse used in the KIK method. (a) Quantum gates are executed via classical control signals, or `pulses'. The left panel shows a pulse schedule used for a CNOT gate in the IBM quantum computing platform. The pulse schedule in the right panel performs the inverse of the CNOT through the inverse driving $\mathcal{H}_\textrm{I}(t)$. It is constructed from the original pulse schedule $\mathcal{H}(t)$, by inverting the amplitudes of the pulses (black curved arrow) and their time ordering (red curved arrow). (b) Instead of the pulse inverse, circuit folding and other variants of unitary folding \cite{giurgica2020digital,majumdar2023best} use the CNOT as its own inverse. Therefore, the pulse schedule for the inverse evolution is not modified. (c) Noisy implementations of $\mathcal{K}$ and $\mathcal{K}_\textrm{I}$. We assume that during the executions of $\mathcal{K}$ and $\mathcal{K}_\textrm{I}$ temporal variations of the noise due to external factors (e.g. temperature variations) are negligible. Thus, any time dependence in $\mathcal{L}(t)$ is induced by the time dependence of $\mathcal{H}(t)$. (Top) This leads to gate dependent noise depicted by different border colors in the gates $U_\textrm{a}$, $U_\textrm{b}$, and $U_\textrm{c}$. (Bottom) Since $\mathcal{H}_\textrm{I}(t)$ reverses the time ordering of $\mathcal{H}(t)$, the time ordering of $\mathcal{L}(t)$ is also reversed. However, the sign of $\mathcal{L}(t)$ does not change because otherwise the inverse evolution would undo the noise. }
	\end{figure}

Let us now discuss another major difference between our scheme and QEM protocols based on ZNE (including circuit folding). In the case of ZNE, the coefficients that weigh different noise amplification circuits are determined by the fitting of the noise scaling ansatz to experimental data. Rather than that, we ask how to choose these coefficients in such a way that $\mathcal{U}_\textrm{KIK}^{(M)}$ constitutes a good approximation to the KIK formula. This problem can be formulated in terms of the eigenvalues of the operators $\left(\mathcal{K}_\textrm{I}\mathcal{K}\right)^{-\frac{1}{2}}$ and $\sum_{m=0}^{M}a_{m}^{(M)}\left(\mathcal{K}_\textrm{I}\mathcal{K}\right)^{m}$. If $\lambda$ denotes a generic eigenvalue of $\mathcal{K}_\textrm{I}\mathcal{K}$, our goal is to find a polynomial $\sum_{m=0}^{M}a_{m}^{(M)}\lambda^{m}$ that is as close as possible to $\lambda^{-1/2}$. Depending on the noise strenght, we follow the two strategies presented in the following two sections. This will further clarify why our method cannot be not considered as a ZNE variant. 
	
  \subsection*{QEM in the weak noise regime }
  
	In the limit of weak noise, the circuit  $\mathcal{K}_\textrm{I}\mathcal{K}$  resembles the identity operation and therefore in this case it is reasonable to approximate the function $\lambda^{-\frac{1}{2}}$ by a truncated Taylor series around $\lambda$=1. The resulting Taylor polynomial leads to the Taylor mitigation coefficients  $a_{m}^{(M)}$=$a_{\textrm{Tay},m}^{(M)}$, derived in Supplementary Note 4. Explicitly,
	\begin{equation}
		a_{\textrm{Tay},m}^{(M)}=(-1)^{m}\frac{(2M+1)!!}{2^{M}[(2m+1)m!(M-m)!]}.\label{eq:Taylor coefficients}
	\end{equation}
    In the same supplementary note we show that $a_{\textrm{Tay},m}^{(M)}$ coincide with the coefficients
	obtained from Richardson ZNE, by assuming that noise scales linearly
	with respect to $m$. Nevertheless, it is worth  stressing that a distinctive characteristic of our approach is the pulse-based inverse $\mathcal{K}_\textrm{I}$. As proven in Supplementary Note 4, for gates that satisfy $\mathcal{U}^{2}=\mathcal{I}$, using the  circuit-based inverse $\mathcal{K}_\textrm{I}=\mathcal{K}$ introduces an additional error term that afflicts $\mathcal{U}_\textrm{KIK}^{(M)}$ (cf. Eq. (\ref{eq:2 KIK polynomials})) for any mitigation order $M$. Thus, ignoring the pulse inverse hinders QEM performance in paradigmatic gates such as the CNOT, swap, or Toffoli gate.

As a final remark, we note that circuit folding does not explicitly distinguish between
noise amplification using powers of $\mathcal{K}_\textrm{I}\mathcal{K}$
or $\mathcal{K}\mathcal{K}_\textrm{I}$, as both choices reproduce the identity
operation in the absence of noise. However, we show in Supplementary Note 3 that a correct application of the KIK formula involves powers
of $\mathcal{K}_\textrm{I}\mathcal{K}$.

\subsection*{QEM in the strong noise regime }

In this section, we present a strategy to adapt the coefficients $a_{m}^{(M)}$ to the noise strength, for handling moderate or strong noise. To this end, we introduce the quantity

\begin{equation}
\varepsilon_{\textrm{L2}}^{(M)}:=\int_{g(\mu)}^{1}\left(\sum_{m=0}^{M}a_{m}^{(M)}\lambda^{m}-\lambda^{-\frac{1}{2}}\right)^{2}d\lambda,\label{eq:3 L2 estimator}
\end{equation}
where $\mu=\textrm{Tr}\left(\rho'\rho\right)$, $\rho$ is the initial
state, and $\rho'$ is the state obtained by evolving $\rho$ with
the KIK cycle $\mathcal{K}_\textrm{I}\mathcal{K}$. 

Let us elaborate on the physical meaning of $\varepsilon_{\textrm{L2}}^{(M)}$. For a pure state $\rho$, $\mu$ is the survival probability under
the evolution $\mathcal{K}_\textrm{I}\mathcal{K}$. Note that, in this case, $\mu=1$ if  $\mathcal{K}_\textrm{I}\mathcal{K}=\mathcal{I}$. The lower integration limit $g(\mu)$ in Eq. \eqref{eq:3 L2 estimator}  is a monotonically increasing function of $\mu$, such that $0\leq g(\mu)\leq1$ for $0\leq\mu\leq1$ and $g(\mu)=1$ if $\mu=1$. Therefore, $g(\mu)$ serves as a proxy for the intensity of the noise affecting the circuit $\mathcal{K}_\textrm{I}\mathcal{K}$. More precisely, $g(\mu)$ represents an approximation to the smallest eigenvalue of $\mathcal{K}_\textrm{I}\mathcal{K}$, which equals 1 in the noiseless case. As the noise becomes stronger, both the smallest eigenvalue of $\mathcal{K}_\textrm{I}\mathcal{K}$ and $g(\mu)$ get closer to 0, which implies that the interval [$g(\mu)$,1] is representative of the region where all the eigenvalues of $\mathcal{K}_\textrm{I}\mathcal{K}$ lie. Now, letting $\lambda$  denote a general eigenvalue of this operator, the eigenvalues of $\left(\mathcal{K}_\textrm{I}\mathcal{K}\right)^{-\frac{1}{2}}$ and $\sum_{m=0}^{M}a_{m}^{(M)}\left(\mathcal{K}_\textrm{I}\mathcal{K}\right)^{m}$ can be written as $\lambda^{-\frac{1}{2}}$ and $\sum_{m=0}^{M}a_{m}^{(M)}\lambda^{m}$, respectively. Since the integrand of Eq. \eqref{eq:3 L2 estimator} quantifies the deviation between these quantities, $\varepsilon_{\textrm{L2}}^{(M)}$ represents the total error when using Eq. (\ref{eq:2 KIK polynomials}) to approximate the KIK formula \eqref{eq:7 U_KIK}. 

Figures \textcolor{red}{2}(a)  and \textcolor{red}{2}(b) illustrate the circuits involved in our adaptive approach to error mitigation. The experimental data comprise the expectation values measured on the noisy circuits $\mathcal{K}(\mathcal{K}_\textrm{I}\mathcal{K})^{m}$, shown in Fig. \textcolor{red}{2}(a), and the survival probility $\mu$ (Fig. \textcolor{red}{2}(b)). In the limit weak noise limit, the circuit of Fig. \textcolor{red}{2}(b) is not necessary and the $a_{m}^{(M)}$ become the Taylor coefficients given in Eq. \eqref{eq:Taylor coefficients} (which can also be obtained by setting $g(\mu)=1$ in the adapted coefficients).  

We point out that the L2 norm used to express $\varepsilon_{\textrm{L2}}^{(M)}$ in Eq. \eqref{eq:3 L2 estimator} is not the only possibility to quantify this error. However, it allows us to greatly simplify the derivation of  $a_{m}^{(M)}$. The adaptive aspect of our method is based on the minimization of the error $\varepsilon_{\textrm{L2}}^{(M)}$ with respect to these coefficients, under the
condition that $\mathcal{U}_\textrm{KIK}^{(M)}$ constitutes a trace-preserving
map. In this way, we obtain the `adapted' mitigation coefficients $a_{m}^{(M)}=a_{\textrm{Adap},m}^{(M)}$, which depend on $g(\mu)$ by virtue of Eq. (\ref{eq:3 L2 estimator}) (for brevity, this dependence is not explicit in the notation for the adapted coefficients but it is expressed through the subscript `$\textrm{Adap}$'). In particular, we obtain in Supplementary Note 4 the expressions  
\begin{align}
	a_{\textrm{Adap},0}^{(1)} & =1+\frac{1}{(1+\sqrt{g})^{3}}+\frac{3}{2(1+\sqrt{g})^{2}},\label{eq:44 optimal coeff a_0^(1)}\\
	a_{Adap,1}^{(1)} & =-\frac{5+3\sqrt{g}}{2(1+\sqrt{g})^{3}},\label{eq:45 optimal coeff a_1^(1)}
\end{align}
for $M=1,$ and  
\begin{align}
	a_{\textrm{Adap},0}^{(2)} & =1+\frac{16}{3(1+\sqrt{g})^{5}}-\frac{14}{3(1+\sqrt{g})^{4}}+\frac{4}{(1+\sqrt{g})^{2}},\label{eq:46 optimal coeff a_0^(2)}\\
	a_{\textrm{Adap},1}^{(2)} & =-4\frac{10+8\sqrt{g}+9g+3g^{\frac{3}{2}}}{3(1+\sqrt{g})^{5}},\label{eq:47 optimal coeff a_1^(2)}\\
	a_{\textrm{Adap},2}^{(2)} & =2\frac{13+5\sqrt{g}}{3(1+\sqrt{g})^{5}},\label{eq:48 optimal coeff a_2^(2)}
\end{align}
for $M=2$. The coefficients corresponding to $M=3$ are also derived in the same supplementary note. 

  \begin{figure}
	\centering{}\includegraphics[scale=0.095]{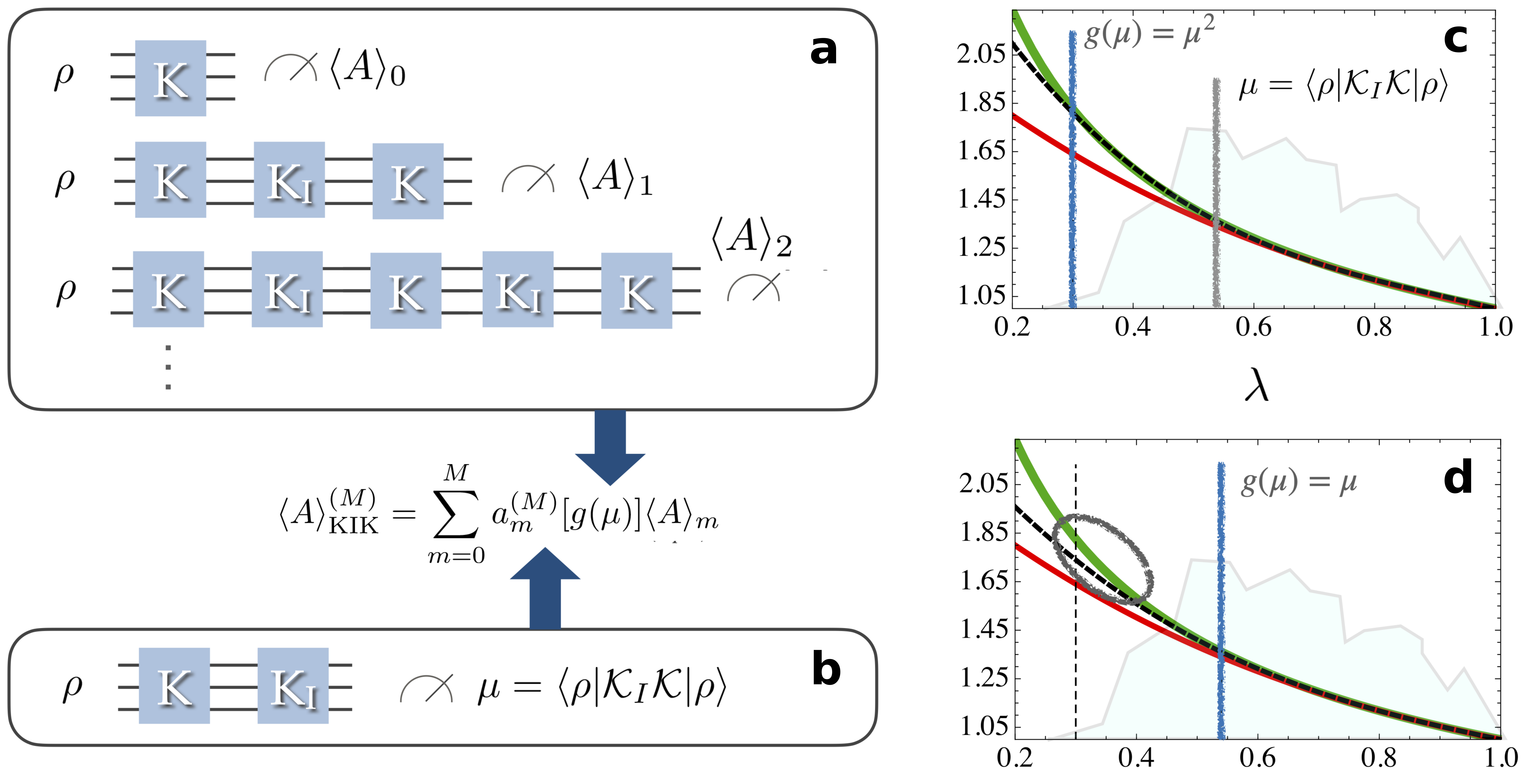}\caption{Adaptive KIK error mitigation. The estimate $\left\langle A\right\rangle _\textrm{KIK}^{(M)}$ of a noiseless expectation value involves the execution of the circuits shown in (a) and (b). In particular, the survival probability $\mu$ is used to evaluate the coefficients $a_{\textrm{Adap},m}^{(M)}[g(\mu)]$, for adaptive error mitigation (see main text for details). The green curve in Figs. (c) and (d) is the plot of $\lambda^{-1/2}$ and it contains the eigenvalues of the operation that effectively suppresess the error channel ( $\left(\mathcal{K}_\textrm{I}\mathcal{K}\right)^{-\frac{1}{2}}$ in Eq. \eqref{eq:7 U_KIK} ). The black dashed curves represent the polynomial approximations $\sum_{m=0}^{M}a_{m}^{(M)}\lambda^{m}$ that appear in the integrand of \eqref{eq:3 L2 estimator}, for third-order mitigation ($M=3$). The better these approximations, the more accurate the corresponding error mitigation. This accuracy is related to the argument $g(\mu)$ in the optimal coefficients $a_{m}^{(3)}=a_{\textrm{Adap},m}^{(3)}[g(\mu)]$, which are obtained by minimizing \eqref{eq:3 L2 estimator} over the interval  [$g(\mu)$,1]. Figures (c) and (d) correspond to $g(\mu)=\mu^{2}$ and $g(\mu)=\mu$, respectively.   In (c), $\lambda^{-1/2}$ is very well approximated by $\sum_{m=0}^{3}a_{\textrm{Adap},m}^{(3)}[\mu^{2}]\lambda^{m}$  in the interval where the eigenvalues of $\left(\mathcal{K}_{I}\mathcal{K}\right)^{-\frac{1}{2}}$are distributed (jagged line in the background). In (d), the interval [$\mu$,1] is too small to cover the full eigenvalue distribution and thus $\sum_{m=0}^{3}a_{\textrm{Adap},m}^{(3)}[\mu]\lambda^{m}$ starts to deviate significantly from the green curve, as shown by the gray ellipse. The red curve corresponds to the Taylor polynomial $\sum_{m=0}^{3}a_{\textrm{Adap},m}^{(3)}[1]\lambda^{m}$ and is the less effective approximation, as seen in both (c) and (d).}
\end{figure}

According to our previous remarks, we can recover the limit of weak noise by setting $g(\mu)=1$. As expected, in this limit  Eqs. \eqref{eq:44 optimal coeff a_0^(1)}-\eqref{eq:48 optimal coeff a_2^(2)} coincide with the coefficients    $a_{\textrm{Tay},m}^{(M)}$ in Eq. \eqref{eq:Taylor coefficients} (and similarly for $M=3$, see Supplementary Note 4). 
	
	An important question is how the choice of $g(\mu)$ affects the quality of our adaptive KIK scheme.  
	We consider functions $\{g(\mu)\}=\{1,\mu,\mu^{2}\}$ in the ten-swap experiment presented below, and $\{g(\mu)\}=\{1,\mu,\mu^{2},\mu^{2.5}\}$ for a simulation of the transverse Ising model on  five qubits, in Supplementary Note 5. In both cases, we observe that
	$g(\mu)=1$ is outperformed by the functions that explicitly depend on  $\mu$. This shows that the adaptive KIK method consistently produces better results, and demonstrates the usefulness of probing the noise strength through the survival probability $\mu$. For $M$ sufficiently large, the adaptive scheme and the  Taylor scheme produce similar results. Yet, the adaptive scheme enables to achieve substantially higher accuracies using lower mitigation orders. 
	This is of key importance in practical applications, as  low-order mitigation involves less circuits with lower depth (cf. Eq. (\ref{eq:2 KIK polynomials})) and is therefore more robust to noise drifts. In addition, the approximation of keeping only the first Magnus term becomes less accurates as $M$ increases. 
		
	The function $g(\mu)=\mu^{2}$ yields the best  error mitigation performance, both in the ten-swap experiment and in the simulation presented in Supplementary Note 5. To understand why this happens, it is instructive to consider Figs. \textcolor{red}{2}(c) and  \textcolor{red}{2}(d). These figures show  plots of $\lambda^{-1/2}$ (green solid curves), which denotes a generic eigenvalue of the noise inversion operation $\left(\mathcal{K}_\textrm{I}\mathcal{K}\right)^{-\frac{1}{2}}$, and the polynomial approximations involved in third-order error mitigation (cf. Eq. \eqref{eq:3 L2 estimator}). The polynomials with coefficients $a_{\textrm{Tay},m}^{(3)}$ (Taylor mitigation) and coefficients $a_{\textrm{Adap},m}^{(3)}$ (adaptive mitigation) correspond to the red solid and black dashed curves, respectively. The jagged line in the background depicts a possible distribution of the eigenvalues of $\left(\mathcal{K}_\textrm{I}\mathcal{K}\right)^{-\frac{1}{2}}$ (the height for a given value of $\lambda$ represents the density of eigenvalues close to that value). In Fig.  \textcolor{red}{2}(c), the adapted coefficients are evaluated at $g(\mu)=\mu^{2}$, and the interval [$\mu^{2}$,1] approximately covers the full region where the eigenvalues of $\left(\mathcal{K}_\textrm{I}\mathcal{K}\right)^{-\frac{1}{2}}$ are contained. Thus, the associated polynomial constitutes a very good approximation to the curve $\lambda^{-1/2}$, as seen in Fig.  \textcolor{red}{2}(c). In contrast, the black curve in Fig. \textcolor{red}{2}(d) corresponds to coefficients $a_{\textrm{Adap},m}^{(3)}$ evaluated at $g(\mu)=\mu$, which leads to a poor approximation outside the interval [$\mu$,1] (area enclosed by the gray ellipse). This behavior sheds light on the advantage provided by $g(\mu)=\mu^{2} $ in our experiments and simulations.  
	    Note also that all the polynomials converge as $\lambda$ tends to 1 but the Taylor polynomial (red curve) substantially separates from $\lambda^{-1/2}$ for small $\lambda$.

It is important to remark that Eq. \eqref{eq:3 L2 estimator} represents a measure of the distance between the polynomial \eqref{eq:2 KIK polynomials} and the KIK formula \eqref{eq:7 U_KIK}, in terms of the L2 norm. In this expression, we assume that the eigenvalues $\lambda$ of $\mathcal{K}_\textrm{I}\mathcal{K}$ are uniformly distributed across the integration interval. This is a conservative approach, given that no information besides $\mu$ is available, and in this sense it is also agnostic to the specific noise structure of $\mathcal{K}_\textrm{I}\mathcal{K}$. However, the evaluation of the  distance $\varepsilon_{\textrm{L2}}^{(M)}$ could benefit from additional knowledge about the eigenvalue distribution, which can be incorporated through a weight function $w(\lambda)\neq1$ in the integrand of Eq. \eqref{eq:3 L2 estimator}.  

We leave the study of experimental criteria for choosing $g(\mu)$ and the potential improvements that this possibility entails for the KIK method for future work. For example, by considering higher order moments such as $\mu_{2}:=\langle\rho|(\mathcal{K}_\textrm{I}\mathcal{K})^{2}|\rho\rangle$  it is possible to devise more systematic choices of $g(\mu)$, e.g. $g(\mu)=\mu-\sqrt{\mu_{2}-\mu^2}$. Yet, in the studied examples we observed no significant advantage over the simple heuristic choice $g(\mu)=\mu^2$. As for other modifications and improvements, one could also explore the use of norms other than the L2 norm employed in Eq. (12). Furthermore, the approximating polynomial can be determined in a non integral manner. For example, by using Lagrange polynomials or a two-point Taylor expansion \cite{lopez2002two}.
        
Finally, we remark that, apart from the circuits $\mathcal{K}\left(\mathcal{K}_\textrm{I}\mathcal{K}\right)^{m}$,
used for the error mitigation itself, the estimation of $\mu$ only
involves the additional circuit $\mathcal{K}_\textrm{I}\mathcal{K}$.
Therefore, our adaptive strategy is not based on any tomographic procedure or noise learning stage. Since $\mu$ is a survival probability, its variance is given by $\mu(1-\mu)$ and has the maximum value 0.25, irrespective of the size of the system. This allows for a scalable evaluation of the coefficients for adaptive KIK mitigation. Once these coefficients are determined, the next step is the estimation of the noise-free expectation value using  Eq. \eqref{eq: KIK expect value}. In the section `Fundamental limits and measurement cost of KIK error mitigation', we will present the corresponding measurement cost, for $1\leq M\leq3$, and discuss why and in what sense the KIK  method is scalable.

\subsection*{Experimental results}

In the experiments described below, the KIK mitigation of noise on
the target evolution $\mathcal{K}$ is complemented by an independent
mitigation of readout errors and a simple protocol for mitigating the coherent preparation error of the intial state $\rho=|00\rangle\langle00|$ \cite{landa2022experimental}.
The results of the section `Quantum error mitigation in a ten-swap circuit' also include the application
of randomized compiling \cite{wallman2016noise} to the evolutions $\mathcal{K}$ and
$\mathcal{K}_\textrm{I}$, where circuits logically equivalent to the corresponding ideal
evolutions are randomly implemented. This is useful
for turning coherent errors into incoherent noise, which can be addressed
by our method. Details concerning these experimental methods can be
found in Supplementary Note 6.\\

\textbf{KIK-based gate calibration for mitigating coherent errors.} A usual approach to handle coherent errors in QEM is to first transform
them into incoherent errors via randomized compiling \cite{wallman2016noise},
and then apply QEM. In this section, we discuss the application of
the KIK formula to directly mitigate the coherent errors caused by
a faulty calibration of 
a CNOT gate. 

The calibration process involves measurements and  adjustments of gate parameters for achieving the results that these measurements would produce in the absence of noise. Since noise affects measured expectation values, the resulting bias leads to incorrect adjustments, i.e. miscalibration.
This `noise-induced coherent error' effect may be small in each gate but it builds up to a subtantial error in sufficiently deep circuits.
Our idea is to complement the KIK
error mitigation for a whole circuit, with a KIK-based
calibration of the individual gates.

\begin{figure}
	\centering{}\includegraphics[scale=0.45]{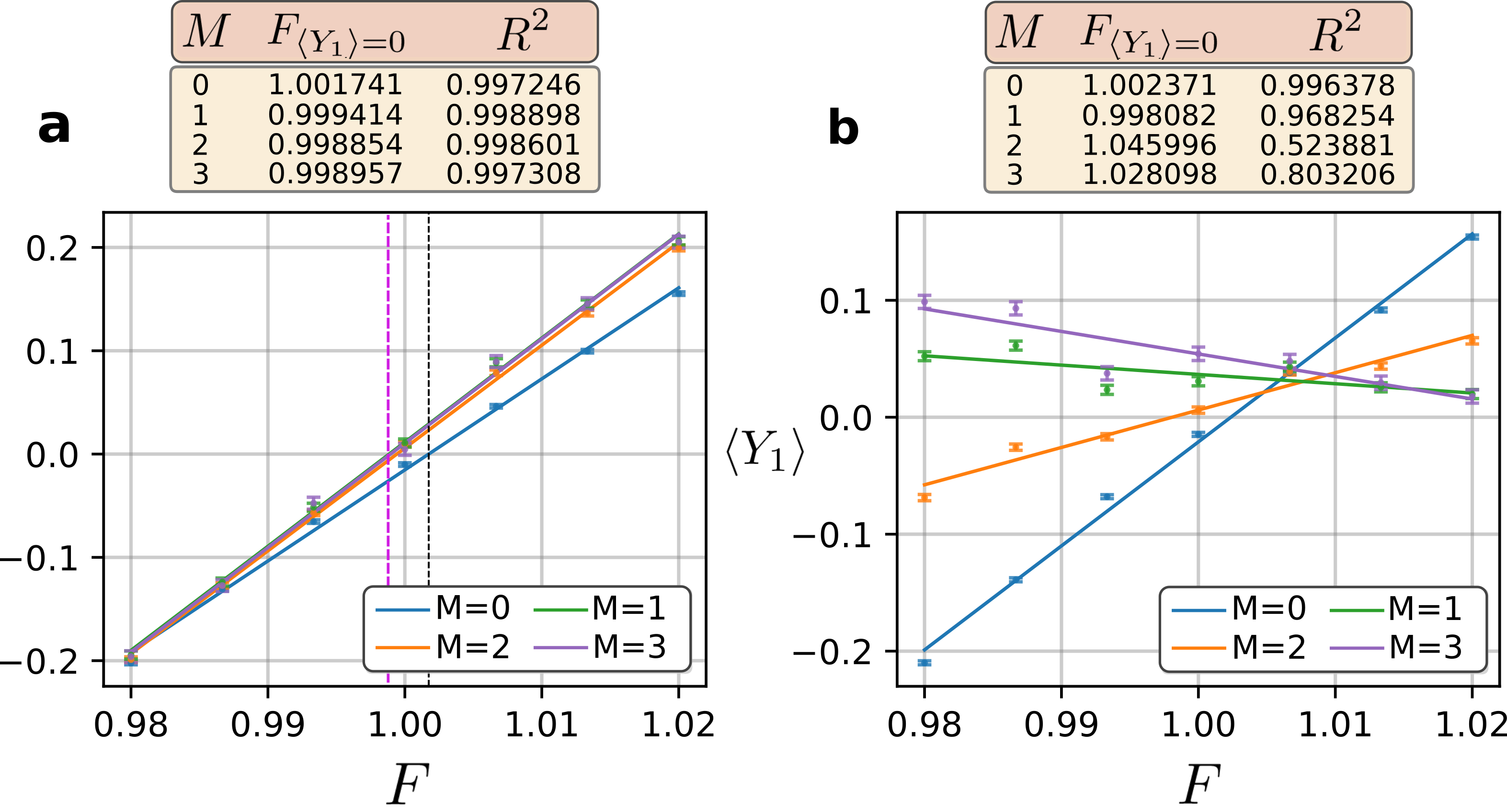}\caption{Calibration curve of the pulse amplitude of a CNOT gate in the IBM processor Jakarta, using the KIK method. In (a) and (b) $\mathcal{K}_\textrm{I}$ is given by the pulse inverse and the circuit inverse $\mathcal{K}_\textrm{I}=\mathcal{K}$, respectively. The initial state is $\rho=|\psi\rangle\langle\psi|$, with $|\psi\rangle=\frac{1}{\sqrt{2}}\left(|0\rangle+|1\rangle\right)\otimes|0\rangle$. The default amplitude is increased by the factors $F$ shown in the $x$ axis of the figure, and for each factor we apply Eq. (\ref{eq:2 KIK polynomials}) to evaluate the expectation value $\left\langle Y_{1}\right\rangle $, where $Y_{1}$ is the $y$-Pauli matrix acting on the target qubit. The factor $F_{\left\langle Y_{1}\right\rangle =0}$ corresponds to the ideal expectation value $\left\langle Y_{1}\right\rangle =0$ and yields the calibrated amplitude. The factors $F_{\left\langle Y_{1}\right\rangle =0}$ associated with the magenta and black dashed lines are different, which indicates a shift in the amplitude obtained without KIK calibration. In Fig. \textcolor{red}{3}(b), we see that the convergence achieved for increasing $M$ in Fig. \textcolor{red}{3}(a) is spoiled by the use of the circuit inverse.}
\end{figure}

Figure \textcolor{red}{3} shows the results of our calibration test of a CNOT in the IBM processor Jakarta. We apply the gate on the initial
state $\rho=\frac{1}{\sqrt{2}}\left(|0\rangle+|1\rangle\right)\otimes|0\rangle$,
and measure the expectation value of the Pauli matrix $Y$ acting on the target qubit (i.e. the qubit prepared in the state $|0\rangle$), denoted by $Y_{1}$. We repeat this procedure for different amplitudes of the cross resonance pulse \cite{alexander2020qiskit}, which constitutes the two-qubit interaction in the IBM CNOT implementation. 
Experimental details can be found in Supplementary Note 6. Each data point of
Fig. \textcolor{red}{3} is obtained by applying Taylor mitigation
(i.e. by applying Eq. \eqref{eq: KIK expect value} with the coefficients (\ref{eq:Taylor coefficients})), for $0\leq M\leq3$, and linear regression
(least squares) is used to determine the line that best fits the
experimental data. We also verify that in this case error mitigation with
the adapted coefficients $a_{\textrm{Adap},m}^{(M)}$ does not yield a noticeable
advantage. This indicates that noise is sufficiently weak, which is
further supported by the quick convergence of the lines corresponding
to $M\geq1$ in Fig. \textcolor{red}{3}(a). 

Keeping in mind that the calibrated amplitude must reproduce the ideal
expectation value ${\color{blue}{\normalcolor \left\langle Y_{1}\right\rangle =0}}$,
we can see from Fig. \textcolor{red}{3}(a) that the predicted amplitude
without QEM ($M=0$) and with QEM are different. Since the CNOT is
subjected to stochastic noise, without QEM the measured expectation
values will be shifted and the corresponding linear regression results
in a calibrated amplitude that is also shifted with respect to the
correct value. This is illustrated by the separation between the black and magenta dashed lines in Fig. \textcolor{red}{3}(a). The magenta line represents the calibrated amplitude using KIK error mitigation, while the black one is the amplitude obtained without  noise mitigation. Calibration based on the black line leads to a noise-induced coherent error. It is important to stress that
the benefit of this calibration procedure would manifest when combined with
QEM of the target circuit in which the CNOTs participate.
The reason is that the calibrated field is consistent with gates of
reduced (stochastic) noise (due to the use of QEM in the calibration
process), and therefore it is not useful if the target circuit is
implemented without QEM. 

In Fig. \textcolor{red}{3} we also observe that a proper implementation
of KIK QEM requires the pulse-based inverse $\mathcal{K}_\textrm{I}$ (Fig.
\textcolor{red}{3}(a)),  performed through the driving \eqref{eq:5 HI},
while the use of another CNOT for $\mathcal{K}_\textrm{I}$ (Fig. \textcolor{red}{3}(b))
does not show the expected convergence as the mitigation order $M$
increases. Note also that although a CNOT is its own inverse in the
noiseless scenario, it leads to a coefficient of determination $R^{2}$
whose values show a poor linear fit. This further illustrates the importance of using the pulse inverse instead of the circuit inverse, characteristic of ZNE based on global folding. We point out that odd powers of the CNOT gate are a common choice for the application of local folding ZNE \cite{majumdar2023best,PhysRevA.102.012426,PhysRevA.105.042406}, where the goal is to amplify the noise on local sectors of the circuit rather than globally. As such, we believe that in practice this procedure would display  inconsistencies similar to those observed in our CNOT experiment. More generally, we show in Supplementary Note 4 that foldings of any self-inverse gate with itself produce a residual error that is not present when the pulse inverse is applied.\\

\textbf{Quantum error mitigation in a ten-swap circuit.} In Fig. \textcolor{red}{4}(a), we show the results of QEM for a circuit
$\mathcal{K}$ given by a sequence of 10 swap gates. The experiments
were executed in the IBM quantum processor Quito. The schematic of
$\mathcal{K}$ is illustrated in Fig. \textcolor{red}{4}(b). 

\begin{figure}
	\centering{}\includegraphics[scale=0.45]{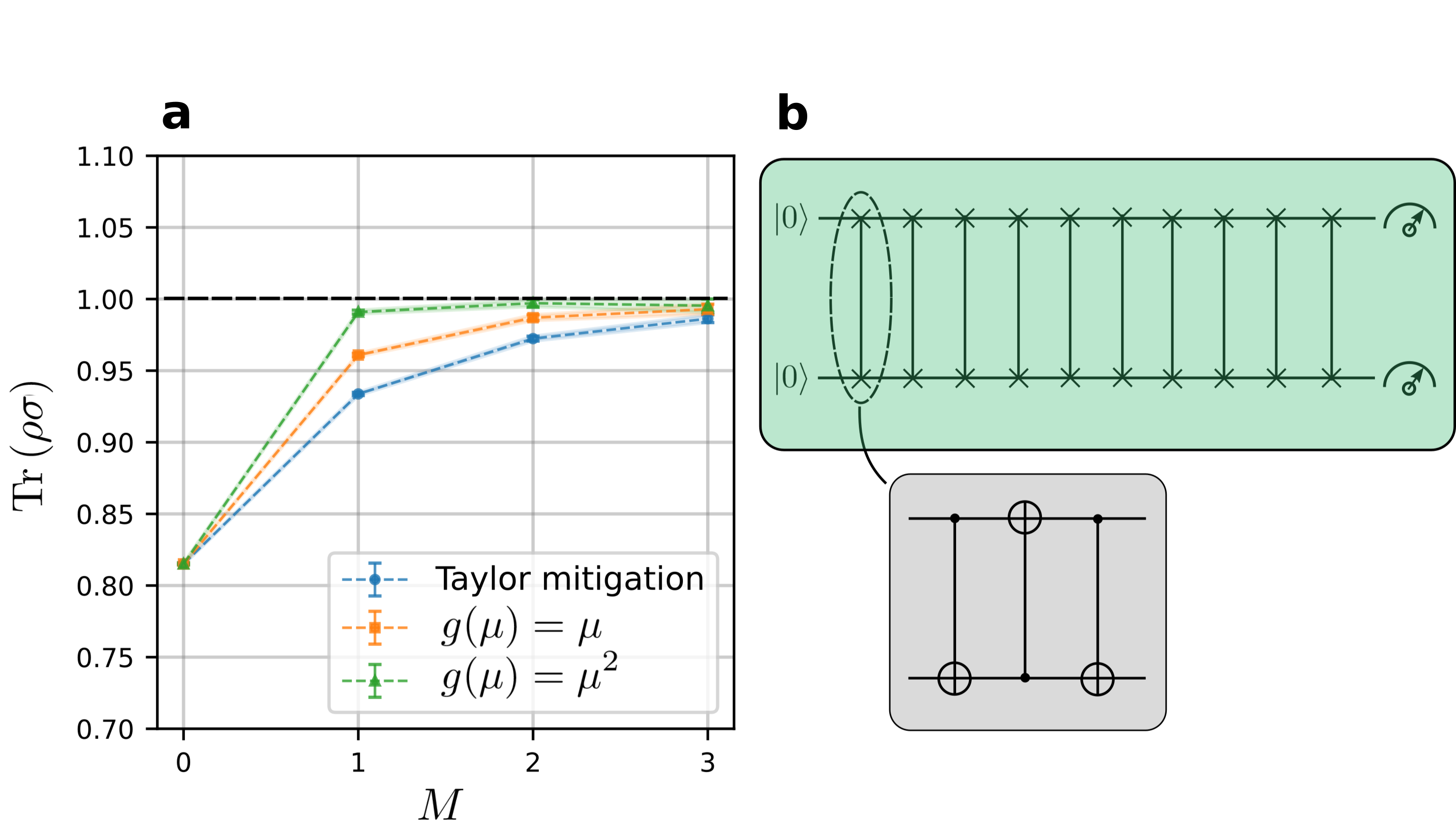}\caption{Experimental QEM in the IBM processor Quito. (a) Error-mitigated survival probability for the circuit of Fig. \textcolor{red}{4}(b), as a function of the mitigation order. The ideal survival probability is 1 (dashed black line). Green and orange curves show QEM adapted to the noise intensity, and the blue curve stands for mitigation assuming weak noise (Taylor mitigation). The thickness of the lines stands for the experimental error bars. We see that Taylor mitigation is outperformed by adapted mitigation. (b) The circuit used in the experiments. Each swap is implemented as a sequence of three CNOTs. }
\end{figure}

We mitigate errors in the survival probability $\textrm{Tr}\left(\rho\sigma\right)$,
where $\sigma$ is the noisy final state that results from applying
$\mathcal{K}$ to $\rho$. To perform QEM, we consider the truncated
expansion (\ref{eq:2 KIK polynomials}) with mitigation orders $1\leq M\leq3$.
The blue curve in Fig. \textcolor{red}{4}(a) corresponds to Taylor
mitigation $a_{m}^{(M)}=a_{\textrm{Tay},m}^{(M)}$. Coefficients $a_{m}^{(M)}=a_{\textrm{Adap},m}^{(M)}$
that are adapted with functions $g(\mu)=\mu$ and $g(\mu)=\mu^{2}$
in Eq. (\ref{eq:3 L2 estimator}) give rise to the orange and green
curves, respectively. Furthermore, for $\mathcal{K}_\textrm{I}$ we perform
the pulse inverse according to the pulse schedule described by Eq. \eqref{eq:5 HI}. 

In Fig. \textcolor{red}{4}(a) we observe that the adapted coefficients
$a_{\textrm{Adap},m}^{(M)}$ outperform Taylor mitigation. This shows that,
beyond the limit of weak noise, QEM can be substantially improved
by adapting it to the noise intensity. Within our Magnus truncation approximation, we observe that the ideal survival probability is almost fully recovered. The small residual bias is of order $10^{-3}$ and can be associated with small experimental imperfections (e.g. small errors in the detector calibration), or with the higher-order Magnus terms discarded in our framework. 
In Supplementary Note 7, we also provide a numerical example where neglecting higher-order Magnus
terms leads to an eventual saturation of the QEM accuracy. However, in this example we find that fourth-order QEM ($M=4$) yields a relative error as low as $10^{-4}$, which further illustrates the accuracy achieved by the KIK formula. 

Due to experimental limitations, it was not possible to implement
the ten-swap circuit using CNOTs calibrated through the KIK method.
Specifically, we could not guarantee that calibration circuits and
error mitigation circuits would run sequentially, and without the
interference of intrinsic (noncontrollable) calibrations of  the processor. Moreover, this demonstration requires that all the parameters
of the gate are calibrated using the KIK method, and not just the
cross resonance amplitude. However, we numerically verify in Supplementary Note 6 that coherent errors vanish for a gate calibrated using KIK QEM,
to the point that randomized compiling is no longer needed.

\subsection*{Fundamental limits and measurement cost of KIK error mitigation}

\textbf{Fundamental limits of KIK error mitigation.} The performance of QEM protocols is often analyzed using two figures
	of merit. One of them is the bias between the noisy expectation value
	of an observable and its ideal counterpart, and the other is the statistical
	precision of the error-mitigated expectation value. The bias defines
	the QEM accuracy and is evaluated in the limit of infinite
	measurements. However, any experiment has a limited precision
	because it always involves a finite number of samples. In QEM protocols,
	the estimation of ideal expectation values is usually accompanied
	by an increment of statistical uncertainty, which can be exponential
	in worst-case scenarios \cite{takagi2022fundamental,quek2022exponentially}. This results in a sampling overhead
	for achieving a given precision, as compared to the number of samples
	required without using QEM.
	
	In Supplementary Note 8, we derive the accuracy bounds

	\begin{align}
		\varepsilon_\textrm{KIK}^{(M)} & \leq\sqrt{\textrm{Tr}\left(A^{2}\right)-\frac{\left[\textrm{Tr}\left(A\right)\right]^{2}}{\textrm{Tr}\left(I\right)}}\left|1-\sum_{m=0}^{M}{\color{red}{\normalcolor a_{\textrm{Adap},m}^{(M)}}}(\mu)e^{-2(m+1/2)\intop_{0}^{T}\left\Vert \mathcal{L}(t)\right\Vert dt}\right|, \textrm{for }M=1,2,3,\label{eq:bound1}\\
		& \leq\sqrt{\textrm{Tr}\left(A^{2}\right)-\frac{\left[\textrm{Tr}\left(A\right)\right]^{2}}{\textrm{Tr}\left(I\right)}}\left|1-\sum_{m=0}^{M}{\color{red}{\normalcolor a_{\textrm{Adap},m}^{(M)}}}(1)e^{-2(m+1/2)\intop_{0}^{T}\left\Vert \mathcal{L}(t)\right\Vert dt}\right|, \textrm{\textrm{for }}M=1,2,3,\label{eq:bound2}\\
		& \leq\frac{(2M+1)!!}{2^{M+1}(M+1)!}\sqrt{\textrm{Tr}\left(A^{2}\right)-\frac{\left[\textrm{Tr}\left(A\right)\right]^{2}}{\textrm{Tr}\left(I\right)}}\left(e^{2\intop_{0}^{T}\left\Vert \mathcal{L}(t)\right\Vert dt}-1\right)^{M+1}.\label{eq:bound3}
	\end{align}
	These are upper bounds
	on the bias $\varepsilon_\textrm{KIK}^{(M)}$, for an arbitrary observable
	$A$ and an arbitrary initial state. We also note that the only approximation in Eqs. \eqref{eq:bound1}-\eqref{eq:bound3} and any of our derivations is the truncation of the Magnus expansion to its dominant term. Importantly, this does not exclude errors of moderate or strong magnitude associated with such a term. On the other hand,
	discarding Magnus terms beyond first order naturally leads to a saturation
	of accuracy. Such a saturation manifests in a residual bias that cannot
	be reduced by indefinitely increasing the mitigation order. Therefore,
	for the tighter bounds \eqref{eq:bound1} and \eqref{eq:bound2} we restrict ourselves to the mitigation
	orders used in our experiments and simulations, given by $1\leq M\leq3$.

	On the other hand, the loosest bound \eqref{eq:bound3} provides a clearer picture
	of how the bias associated with the first Magnus term is suppressed
	by increasing $M$. The quantity $\int_{0}^{T}\left\Vert \mathcal{L}(t)\right\Vert dt$
	is the integral of the spectral norm of the dissipator $\left\Vert \mathcal{L}(t)\right\Vert $,
	over the total evolution time $(0,T)$. This parameter serves as a
	quantifier of the noise accumulated  during the execution of the target
	evolution $\mathcal{K}$.   Since $\frac{(2M+1)!!}{2^{M+1}(M+1)!}\leq\frac{3}{8}$,
	Eq. \eqref{eq:bound3} implies that $\varepsilon_{KIK}^{(M)}$ is exponentially suppressed
	if the accumulated noise is such that 
	\begin{equation}
		e^{2\int_{0}^{T}\left\Vert \mathcal{L}(t)\right\Vert dt}<2.\label{eq:low accumulated noise}
	\end{equation}
	In the case of noise acting locally on individual gates, $\mathcal{L}(t)$ is given by a sum of local dissipators and one can show that  $\left\Vert \mathcal{L}(t)\right\Vert $ is upper bounded by the summation of all the gate errors in the circuit.
	
	We remark that, in the NISQ era, errors escalate in quantum algorithms
	due to the lack of QEC. Thus, NISQ computers can perform useful computations
	only if the accumulated noise $\int_{0}^{T}\left\Vert \mathcal{L}(t)\right\Vert dt$
	is below a certain value. Our notion of scalabitility is that under the contraint of moderate acumulated noise the KIK method is scale independent. In particular, when $\int_{0}^{T}\left\Vert \mathcal{L}(t)\right\Vert dt$ is sufficiently small to satisfy Eq.
	(\ref{eq:low accumulated noise}), the exponential error mitigation
	referred above is applicable to circuits of any size and topology. While achieving a low accumulated noise in big circuits is technologically challenging, if this condition is met the KIK method and the resources that it requires are agnostic to the size of the circuit. Moreover, it is worth noting that Eq. \eqref{eq:low accumulated noise} represents a sufficient condition for scalable error mitigation. The possibility of extending this scalability to values of $\int_{0}^{T}\left\Vert \mathcal{L}(t)\right\Vert dt$ that violate Eq. \eqref{eq:low accumulated noise} depends on the tightness of the accuracy bounds \eqref{eq:bound1}-\eqref{eq:bound3}, and constitutes an open problem.
	
	Equations \eqref{eq:bound1}-\eqref{eq:bound3}  are applicable to both adaptive mitigation and Taylor
	mitigation. In contrast, the tightest bound \eqref{eq:bound1} is exclusive of adaptive mitigation. The coefficients $a_{\textrm{Adap},m}^{(M)}$ in this
	bound are evaluated at $g(\mu)=\mu$. Importantly, \eqref{eq:bound1} is upper bounded by \eqref{eq:bound2} and \eqref{eq:bound3} for any $0\leq\mu\leq1$, as proven in Supplementary Note 8. According to our experiments and simulations,
	we believe that even tighter bounds can be obtained for $g(\mu)=\mu^{2}$
	or other choices of $g(\mu)$. This topic is left for future investigation. 
	
	Lastly, we stress that the condition (\ref{eq:low accumulated noise}) does not imply that the KIK method is restricted to error mitigation for weak noise. This is related to the reiterated  fact that Eqs. (\ref{eq:bound1})-(\ref{eq:bound3}) and particularly (\ref{eq:bound3}) probably overestimate the actual bias between the error-mitigated expectation value and its ideal counterpart. More importantly, we have shown experimentally and numerically the substantial advantage achieved by the adaptive KIK strategy, as compared to QEM under the assumption of weak noise. This further indicates that the regime of validity of our method likely goes beyond the prediction of Eq. (\ref{eq:bound3}).\\
	
	\textbf{Measurement cost of KIK error mitigation.} For the sampling overhead, we adopt the variance as the measure of
	statistical precision. Let $\textrm{Var}_{0}\left(A\right)$ denote
	the variance in the estimation of the expectation value $\left\langle A\right\rangle $,
	without using error mitigation, and $\textrm{Var}_{M}\left(A\right)$
	the variance associated with KIK mitigation of order $M\geq1$. The
	sampling overhead is defined as the increment in the number of samples
	needed to achieve the same precision as in the unmitigated case. Suppose
	that $N$ measurements constitute the shot budget for KIK mitigation.
	For a given value of $M$, the sampling overhead is evaluated by minimizing
	$\textrm{Var}_{M}\left(A\right)$ over the distribution of measurements
	between the different circuits $\mathcal{K}\left(\mathcal{K}_\textrm{I}\mathcal{K}\right)^{m}$.
	If $N_{m}$ measurements are allocated to $\mathcal{K}\left(\mathcal{K}_\textrm{I}\mathcal{K}\right)^{m}$,
	then 
	\begin{equation}
		\textrm{Var}_{M}\left(A\right)=\sum_{m=0}^{M}\left(a_{m}^{(M)}(g)\right)^{2}\frac{\textrm{var}_{m}\left(A\right)}{N_{m}},\label{eq:variance for N shots},
	\end{equation}
	where $\textrm{var}_{m}\left(A\right)$ denotes the variance that results from measuring $A$ on the circuit $\mathcal{K}\left(\mathcal{K}_{I}\mathcal{K}\right)^{m}$. 
	
	Taking into account the constraint $\sum_{m=0}^{M}N_{m}=N$, the minimization
	of Eq. \eqref{eq:variance for N shots} with respect to $\{N_{m}\}_{m}$ yields $N_{m}=\left|a_{m}^{(M)}\right|N$.
	Of course, these values have to approximated to the closest integer
	in practice. Now, we assume that $\textrm{var}_{m}\left(A\right)=\textrm{var}_{n}\left(A\right)$
	for all $0\leq m,n\leq M$. Since, for reasons previously discussed, we are interested in low mitigation orders $1\leq M\leq3$, $\left(\mathcal{K}_\textrm{I}\mathcal{K}\right)^{m}$ does not deviate too much from the identity operation and therefore the assumption stated above is reasonable. In this way, replacing $N_{m}=\left|a_{m}^{(M)}\right|N$
	into Eq. \eqref{eq:variance for N shots} yields 
	
	\begin{equation}
		\textrm{Var}_{M}\left(A\right)=\sum_{m=0}^{M}\left|a_{m}^{(M)}(g)\right|\frac{\textrm{var}_{0}\left(A\right)}{N}.\label{eq:variance for N shots minimized}
	\end{equation}

	The quantity $\frac{\textrm{var}_{0}\left(A\right)}{N}$ is the variance
	obtained without using error mitigation. Accordingly, 
	
		\begin{figure}
		\centering{}\includegraphics[scale=0.5]{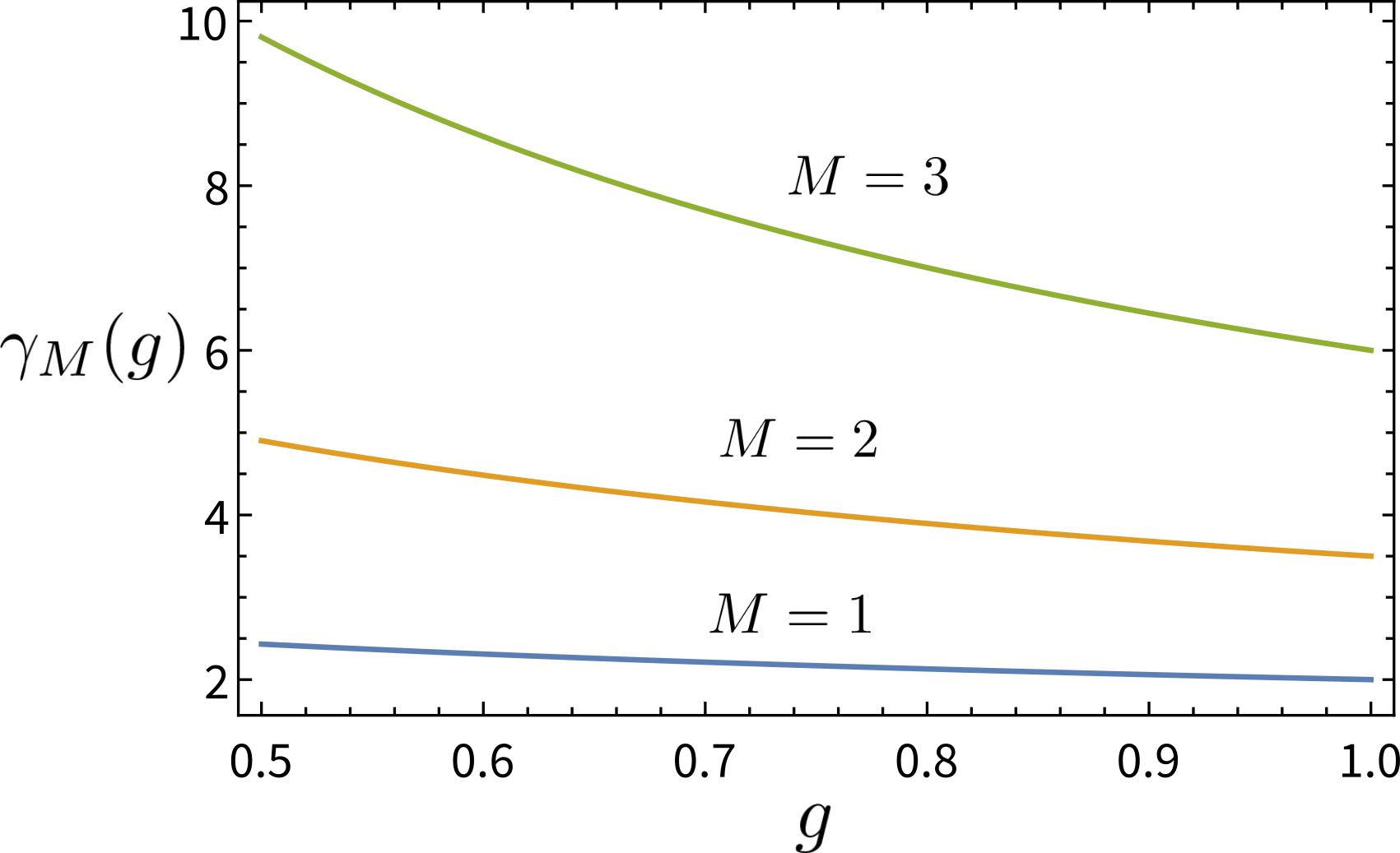}\caption{Variance overhead $\gamma_{M}(g)$ of KIK error mitigation, for mitigation orders $1\leq M\leq3$. The graph shows $\gamma_{M}(g)$ in terms of the function $g=g(\mu)$, used to evaluate the adaptive coefficients $a_{m}^{(M)}(g)$ (cf. Eq. \eqref{eq:sampling overhead}). The overheads for the Taylor coefficients (low noise limit) correspond to the values at $g=1$.}
	\end{figure}
	
	\begin{equation}
		\gamma_{M}(g)=\sum_{m=0}^{M}\left|a_{m}^{(M)}(g)\right|\label{eq:sampling overhead}
	\end{equation}
	represents the sampling overhead. In Fig.  \textcolor{red}{5}, we
	show the sampling overheads for $1\leq M\leq3$, as a function of
	$g=g(\mu)$. As expected, larger noise strengths (corresponding to
	smaller values of $g$) lead to  larger values of $\gamma_{M}(g)$. However, as shown in Fig. \textcolor{red}{5}, these sampling overheads are quite moderate and do not represent an obstacle for scalable error mitigation. In addition, we show in Supplementary Note 3 that our method is robust to noise drifts and miscalibrations that may result from larger sampling overheads, e.g. when higher mitigation orders ($M\geq4$) are considered.

\section*{Discussion}

Quantum error mitigation (QEM) is becoming a standard practice in
NISQ experiments. However, QEM methods that are free from intrinsic
scalability issues lack a physically rigorous formulation, or are
unable to cope with significant levels of noise. The KIK method allows
for scalable QEM whenever the  noise accumulated in the target circuit is not too high, as implied 
by our upper bounds on the QEM accuracy (cf. Eqs. \eqref{eq:bound1}-\eqref{eq:bound3}).
This QEM technique is based on a master equation analysis that incorporates
time-dependent and spatially correlated noise, and does not require
that the noise is trace-preserving. As such, based on elementary simulations
we observe that it can also mitigate leakage noise,
which can take place in superconducting circuits. In the limit of weak noise, the KIK method reproduces some features of zero noise extrapolation using circuit unitary
folding, and outperforms it. This is achieved thanks to the use of
pulse-based inverses for the implementation of QEM circuits, and the
adaptation of QEM parameters to the noise intensity for handling moderate
and strong noise. 

The shot overhead of our method depends only on the noise level and not on the size of the target circuit. For moderate noise, the sampling overhead for mitigation order three or lower is smaller than ten. While the KIK method can be adapted to the strength of the noise, this only requires measuring a single experimental parameter whose sampling cost is negligible and independent of the size of the system. Usually, the performace of QEM techniques may be compromised in experiments involving a large number of samples. When considering long runs, the system needs to be recalibrated multiple times, and noise parameters can undergo significant drifts. This poses challenges in the context of noise learning for QEM protocols that rely on this approach. We show in Supplementary Note 3 that our approach is resilient to drifts in the noise and calibration parameters (the latter holds if randomized compiling is applied). This enables it to be applied in calculations over runtimes of days or even weeks, including pauses for calibrations,  maintenance, or execution of supporting jobs.   On a similar basis, it is possible to   parallelize the error mitigation task, by averaging over data collected from different quantum processors or  platforms with spatially differentiated noise profiles (see Supplementary Note 3). 

We have demonstrated our findings using the IBM quantum processors
Quito and Jakarta. In  Quito, we implemented KIK error mitigation in a circuit
composed of 10 swap gates (30 sequential CNOTs). Despite the substantial
noise in this setup, the tiny bias between the error-mitigated expectation value and the ideal result demonstrates that, at least in this experiment, our theoretical approximations are quite consistent with the actual noise in the system. Using the processor Jakarta, we also showed that even the
calibration of a basic building block of quantum computing, such as
the CNOT gate, can be affected by unmitigated noise. As a consequence,
calibrated gate parameters feature erroneous values leading to coherent
errors. These errors can be avoided by incorporating the KIK method
in the calibration process. The integration of randomized compiling
into our technique also enables the mitigation of coherent errors in the CNOT gates.
This is possible because randomized compiling transforms coherent
errors into incoherent noise, which can be  addressed by the KIK method. 

Despite these successful demonstrations, we believe that there is
room for improvement by exploring some of the possibilities mentioned
in the section `QEM in the strong noise regime'. We also hope that the performance shown
here can be exploited for new demonstrations of quantum algorithms
on NISQ devices, with the potential of achieving quantum advantage
in applications of interest. 

\section*{Data availability}

Code employed in the execution of the experiments as well as raw experimental data and data underlying figures is hosted at http://dx.doi.org/10.5281/zenodo.7652322. All additional data are provided in the supplementary information. 

\section*{acknowledgments}

	We acknowledge the use of IBM Quantum services for this work. The views expressed are those of the authors, and do not reflect the official policy or position of IBM or the IBM Quantum team. Raam Uzdin is grateful for support from the Israel Science Foundation (Grant No. 2556/20). 

\section*{Author contributions}

Raam Uzdin conceived the method, set the theoretical framework, including most of the analytical derivations, and performed numerical simulations. Jader P. Santos designed and executed the experiments, and performed numerical simulations. Ivan Henao derived some theoretical results, in particular the performance bounds.  All the authors were involved in the analysis of theoretical and experimental results, and in the writing and presentation of the paper.

\section*{Competing interests}

The authors declare no competing financial or non-financial interests.

\vskip 6 mm

\bibliographystyle{naturemag}
\bibliography{Refs.bib}



\maketitle
\setcounter{section}{0}
\setcounter{figure}{0}
\setcounter{equation}{0}

\onecolumngrid


\providecommand{\tabularnewline}{\\}

\makeatother

\newcommand{\fakesubsection}[1]{\addcontentsline{toc}{subsection}{\hspace{2cm} #1}}

\global\long\def\thesection{S-\Roman{section}}
\setcounter{section}{0}
\global\long\def\thefigure{\arabic{figure}}
\setcounter{figure}{0}
\global\long\def\thetable{\arabic{table}}
\setcounter{table}{0}
\global\long\def\theequation{\arabic{equation}}
\setcounter{equation}{0}



	
	\title{Adaptive quantum error mitigation using pulse-based inverse evolutions }
	\author{Ivan Henao$^{1}$}
	\email{ivhenao@gmail.com}
	
	\author{Jader P. Santos$^{1}$}
	\email{jader.pereira.santos@gmail.com }
	
	\author{Raam Uzdin$^{1}$}
	\email{raam@mail.huji.ac.il}
	
	\affiliation{$^{1}$Fritz Haber Research Center for Molecular Dynamics,Institute
		of Chemistry, The Hebrew University of Jerusalem, Jerusalem 9190401,
		Israel}
	
	\maketitle
	\setcounter{section}{0}
	\setcounter{figure}{0}
	\setcounter{equation}{0}
	
	\onecolumngrid

	\onecolumngrid
	
	
	\providecommand{\tabularnewline}{\\}
	
	\makeatother
	
	\renewcommand{\figurename}{Supplementary Figure}
	\renewcommand{\tablename}{Supplementary Table}
	
	\global\long\def\thesection{S-\Roman{section}}
	\setcounter{section}{0}
	\global\long\def\thefigure{\arabic{figure}}
	\setcounter{figure}{0}
	\global\long\def\thetable{\arabic{table}}
	\setcounter{table}{0}
	\global\long\def\theequation{\arabic{equation}}
	\setcounter{equation}{0}
	
	\tableofcontents
	
	\section*{Supplementary Note 1: Quantum mechanics in Liouville space}
	
	In the standard description of Quantum Mechanics, a system of dimension
	$d$ is represented by a density matrix $\rho$ of dimension $d\times d$.
	Moreover, a CPTP (completely positive and trace preserving)\textcolor{red}{{}
	}quantum operation can be expressed as 
	\begin{equation}
		\rho'=\sum_{i}K_{i}\rho K_{i}^{\dagger},\label{eq:S1 Kraus representation}
	\end{equation}
	where $\rho'$ is a density matrix and $\{K_{i}\}$ are Kraus operators
	that satisfy the completeness relation $\sum_{i}K_{i}^{\dagger}K_{i}=I$,
	and $I$ is the $d\times d$ identity matrix. Observables correspond
	to hermitian operators $A$, and the associated expectation value
	for a system in a state $\rho$ reads 
	\begin{equation}
		\left\langle A\right\rangle =\textrm{Tr}\left(A\rho\right).\label{eq:S2 expect value in Hilbert space}
	\end{equation}

	The Liouville space formalism is an alternative formulation that is
	particularly useful to simplify notation and handle quantum operations.
	In this framework, a density matrix is replaced by a vector $|\rho\rangle$
	of dimension $d^{2}$ and a quantum operation is a matrix of dimension
	$d^{2}\times d^{2}$. Using the calligraphic notation $\mathcal{O}$
	for a quantum operation, the analogous of Eq. (\ref{eq:S1 Kraus representation})
	in Liouville space is given by 
	\begin{equation}
		|\rho'\rangle=\mathcal{\mathcal{O}}|\rho\rangle.\label{eq:S3 quantum operation in L space}
	\end{equation}
	Here, we adopt the approach of Ref. \cite{gyamfi2020fundamentals},
	where $|\rho\rangle$ is the column vector whose first $d$ components
	correspond to the first row of $\rho$, the next $d$ components correspond
	to the second row of $\rho$, and so forth. More formally, the vector
	representation of a $d\times d$ generic matrix $B$ (not necessarily
	a density matrix) is given by $|B\rangle=\left(B_{11},B_{12},...,B_{1d},B_{21},B_{22},...,B_{2d},...,B_{d1},B_{d2},...,B_{dd}\right)^{\textrm{T}}$,
	where $B_{ij}$ is the $ij$ entry of $B$. With this convention,
	in Liouville space the quantum operation (\ref{eq:S1 Kraus representation})
	takes the form \cite{gyamfi2020fundamentals}
	\begin{equation}
		|\rho'\rangle=\sum_{i}K_{i}\otimes K_{i}^{\ast}|\rho\rangle,\label{eq:S4 Kraus in L space}
	\end{equation}
	where $K_{i}^{\ast}$ is the element-wise complex conjugate of $K_{i}$.
	For example, a unitary operation $\rho'=U\rho U^{\dagger}$ is written
	as $|\rho'\rangle=\mathcal{U}|\rho\rangle$, where $\mathcal{U}=U\otimes U^{\ast}$. 
	
	Equation (\ref{eq:S4 Kraus in L space}) is obtained by following
	the rule to vectorize a product of three matrices $B$, $C$ and $D$.
	Denoting the associated vector as $|BCD\rangle$, this rule states
	that \cite{gyamfi2020fundamentals} 
	\begin{equation}
		|BCD\rangle=B\otimes D^{\textrm{T}}|C\rangle,\label{eq:S5 triple-prod identity}
	\end{equation}
	where the superscript $\textrm{T}$ denotes transposition. Setting
	$B=K_{i}$, $C=\rho$, and $D=K_{i}^{\dagger}$, Eq. (\ref{eq:S4 Kraus in L space})
	follows by applying (\ref{eq:S5 triple-prod identity}) to (\ref{eq:S1 Kraus representation})
	and using the linearity property of the vectorization. 
	
	Finally, to express the expectation value (\ref{eq:S2 expect value in Hilbert space})
	in Liouville space one writes $A$ as a row vector $\langle A|$ defined
	by $\langle A|=\left(A_{11}^{\ast},A_{12}^{\ast},...,A_{1d}^{\ast},...,A_{d1}^{\ast},A_{d2}^{\ast},...,A_{dd}^{\ast}\right)$.
	In this way, the hermiticity of $A$ leads to 
	\begin{align}
		\langle A|\rho\rangle & =\sum_{i,j}A_{ij}^{*}\rho_{ij}\nonumber \\
		& =\sum_{i,j}A_{ji}\rho_{ij}\nonumber \\
		& =\langle A\rangle.\label{eq:S6 expect value in L space}
	\end{align}

	\section*{Supplementary Note 2: Dynamical description of noise for the target evolution and its inverse}
	
	In this section, we set the framework for the derivation of the KIK
	formula (Eq. (1) in the main text). For the sake of clarity and completness, we will discuss again some topics addressed in the main text and rewrite a few equations that were already introduced. We consider a continuous-time
	description of the system evolution, modeled by the master equation
	\begin{equation}
		\frac{d}{dt}\rho=-i[H(t),\rho]+\hat{L}(t)[\rho].\label{eq:S7 master eq in Hilber space}
	\end{equation}
	Here, $H(t)$ is a time-dependent Hamiltonian and $\hat{L}(t)$ is
	a time-dependent dissipator that accounts for the non-unitary contribution
	to the dynamics, which is induced by external noise. The hat symbol
	in $\hat{L}(t)$ is used to emphasize that it represents a superoperator
	in Hilbert space.
	
	To specify the form of $\hat{L}(t)$ one could invoke a microscopic
	description of the dynamics, where the system is coupled to some external
	environment and the total system obeys the Schrodinger equation. The
	time-independent case is extensively studied in \cite{breuer2002theory},
	and various time-dependent Markovian master equations have been derived
	\cite{dann2018time}. The dissipator is often given in
	the Lindblad form, which represents the most general dissipator for
	a Markovian and CPTP evolution. For our purposes, this is not necessary.
	For example, $\hat{L}(t)$ could incorporate leakage noise, which
	does not preserve probability and thus is not trace preserving. 
	
	Now, let us rewrite Eq. (\ref{eq:S7 master eq in Hilber space}) in
	Liouville space. Using the linearity of the vectorization operation,
	we have that 
	\begin{equation}
		\frac{d}{dt}|\rho\rangle=-i|[H(t),\rho]\rangle+|\hat{L}(t)[\rho]\rangle,\label{eq:S8 mast eq in L space 1}
	\end{equation}
	where $|[H(t),\rho]\rangle$ and $|\hat{L}(t)[\rho]\rangle$ are the
	vectors corresponding to $[H(t),\rho]$ and $\hat{L}(t)[\rho]$, respectively.
	By applying the rule (\ref{eq:S5 triple-prod identity}) to the commutator
	$[H(t),\rho]=H(t)\rho-\rho H(t)$, we obtain 
	\begin{align}
		|[H(t),\rho]\rangle & =|H(t)\rho\rangle-|\rho H(t)\rangle\nonumber \\
		& =\left(H(t)\otimes I-I\otimes H(t)^{\textrm{T}}\right)|\rho\rangle\nonumber \\
		& :=\mathcal{H}(t)|\rho\rangle,\label{eq:S9 L space commutator =00005BH,rho=00005D}
	\end{align}
	where $D$ is identified with the identity $I$ for the product $H(t)\rho$,
	and with $H(t)$ for $\rho H(t)$. In both cases, $C$ is associated
	with $\rho$. For a general dissipator $\hat{L}(t)$ we can simply
	write $|\hat{L}(t)[\rho]\rangle=\mathcal{L}(t)|\rho\rangle$, because
	$\rho$ can always be associated with $C$ in Eq. (\ref{eq:S5 triple-prod identity}). 
	
	In this way, the Liouville-space representation of (\ref{eq:S7 master eq in Hilber space})
	reads 
	\begin{equation}
		\frac{d}{dt}|\rho\rangle=\left(-i\mathcal{H}(t)+\mathcal{L}(t)\right)|\rho\rangle.\label{eq:S10 mast eq in L space 2}
	\end{equation}
	We note that both $\mathcal{H}(t)$ and $\mathcal{L}(t)$ are linear
	operators that correspond to matrices of dimension $d^{2}\times d^{2}$. While we do not impose any physical constraint on $\mathcal{L}(t)$, in what follows we introduce and physically justify a relationship between $\mathcal{L}(t)$ and the dissipator that affects the pulse inverse evolution. This relationship is crucial for the derivation of the KIK formula in Supplementary Note 3. \\
	\\
	\fakesubsection{Noise for pulse-based inverse evolution}
	\textbf{Noise for pulse-based inverse evolution.} Suppose that applying the driving $\mathcal{H}(t)$ in Eq. (\ref{eq:S10 mast eq in L space 2})
	during a total time $T$ leads to an ideal unitary evolution $\mathcal{U}=\mathcal{T}e^{-\int_{0}^{T}i\mathcal{H}(t)dt}$,
	where $\mathcal{T}$ is the time-ordering operator. Even in the circuit
	model of quantum computing, where unitary operations are composed
	of discrete quantum gates, each elementary gate is itself generated
	by a pulse schedule that can be represented as a time-dependent Hamiltonian.
	Hence, any quantum circuit is ultimately generated by some pulse schedule
	$\mathcal{H}(t)$. 
	
	Different pulse schedules can result in the same ideal evolution $\mathcal{U}$,
	and naturally the same is true for its inverse $\mathcal{U}^{\dagger}$.
	However, in the presence of noise this equivalence does not hold in
	general. The derivation of the KIK formula relies on relating the
	pulse schedule for $\mathcal{U}^{\dagger}$ with the pulse schedule
	for $\mathcal{U}$ in a specific manner. Denoting the driving that
	generates $\mathcal{U}^{\dagger}$ by $\mathcal{H}_{\textrm{I}}(t)$, this
	relationship reads
	
	\begin{equation}
		\mathcal{H}_{\textrm{I}}(t)=-\mathcal{H}(T-t).\label{eq:S11 inverse driving}
	\end{equation}
	
	In combination with Eq. (\ref{eq:S11 inverse driving}), the other
	ingredient for obtaining the KIK formula has to do with how noise
	comes into play for a given driving $\mathcal{H}(t)$. On the one
	hand, Eq. (\ref{eq:S10 mast eq in L space 2}) describes noise that
	acts locally in time, i.e. that $\mathcal{L}(t)$ only depends on
	the current instant $t$ and not on the previous history of the evolution.
	On the other hand, this time dependence can have two origins. One
	of them is the time-dependence of $\mathcal{H}(t)$ itself, and the
	other are intrinsic fluctuations of the noise that are related e.g.
	to changes in the environment or miscalibrations that occur during
	the execution of an experiment. The second possibility is discussed
	in detail in Supplementary Note 3. As for
	the influence of the driving $\mathcal{H}(t)$ on the noise, we assume
	that $\mathcal{L}(t)$ does not depend on the sign of $\mathcal{H}(t)$,
	but only on its amplitude. This reflects the fact that noise cannot
	be undone when running the reverse schedule described in Eq. (\ref{eq:S11 inverse driving}).
	Taking into account (\ref{eq:S11 inverse driving}), the dissipator
	$\mathcal{L}_{\textrm{I}}(t)$ for the ``pulse inverse'' $\mathcal{H}_{\textrm{I}}(t)$
	should then satisfy 
	\begin{equation}
		\mathcal{L}_{\textrm{I}}(t)=\mathcal{L}(T-t).\label{eq:S12 inverse dissipator}
	\end{equation}
	
	\begin{figure}
		\centering\includegraphics[scale=0.5]{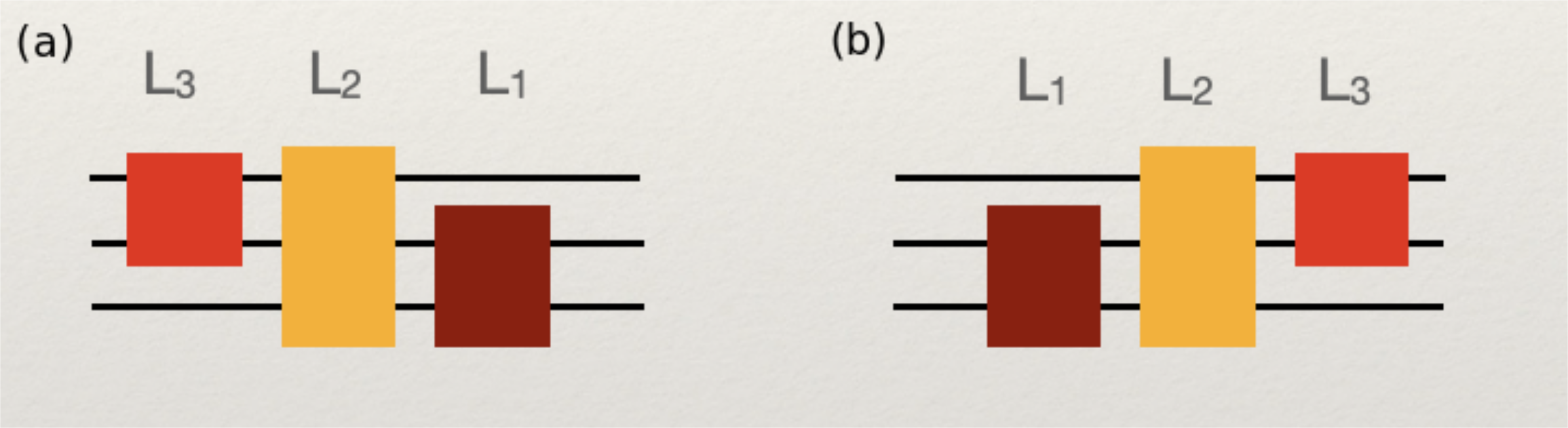}\caption{Time dependence of dissipators $\mathcal{L}(t)$ and $\mathcal{L}_{\textrm{I}}(t)$.
			(a) Illustrative noisy quantum circuit, composed of three gates that
			act on three qubits. The total dissipator $\mathcal{L}(t)$ is given
			by a sequence of dissipators $\mathcal{L}_{i}$, $1\protect\leq i\protect\leq3$,
			which affect each gate. (b) Upon reversing the drive pulse schedule
			(cf. Eq. (\ref{eq:S11 inverse driving})), the order of the gates
			and the corresponding dissipators is also reversed. Hence, if the
			sequence of dissipators in (a) is $\mathcal{L}_{1}\mathcal{L}_{2}\mathcal{L}_{3}$,
			in (b) it is given by $\mathcal{L}_{3}\mathcal{L}_{2}\mathcal{L}_{1}$.}
		\label{fig:supplementary}
	\end{figure}
	
	Equation (\ref{eq:S12 inverse dissipator}) implies that if different
	gates are affected by different noise mechanisms (e.g. very fast gates
	may be prone to leakage noise due to non-adiabatic couplings to the
	levels outside the computational basis), the order in which these
	noise mechanisms operate is reversed when applying the pulse inverse
	(\ref{eq:S11 inverse driving}), as shown in Supplementary Figure 1.
	This is a consequence of the time locality of both $\mathcal{L}(t)$
	and $\mathcal{L}_{\textrm{I}}(t)$, and the reversed time schedule that $\mathcal{H}_{\textrm{I}}(t)$
	imprints on the corresponding inverse gates. To summarize, our sole
	assumptions on the noise are: 
	\begin{enumerate}
		\item The linearity and the time locality of the dissipator $\mathcal{L}(t)$.
		\item The relationship between the dissipator $\mathcal{L}_{\textrm{I}}(t)$, which
		affects the pulse inverse $\mathcal{H}_{\textrm{I}}(t)$ (cf. Eq. (\ref{eq:S11 inverse driving}))
		at time $t$, and $\mathcal{L}(t)$. This relationship is encapsulated
		by Eq. (\ref{eq:S12 inverse dissipator}). 
	\end{enumerate}
	We remark that $\mathcal{L}(t)$ is not restricted to have a Lindblad
	form or to give rise to a trace preserving map. For example, it can
	incorporate leakage noise, which does not conserve the total probability
	and therefore is not trace preserving.
	
	Following Eq. (\ref{eq:S10 mast eq in L space 2}), the noisy evolutions
	$\mathcal{K}$ and $\mathcal{K}_{\textrm{I}}$ that appear in the KIK formula
	are thus given by 
	\begin{align}
		\mathcal{K} & =\mathcal{T}e^{\int_{0}^{T}(-i\mathcal{H}(t)+\mathcal{L}(t))dt},\label{eq:S13 K}\\
		\mathcal{K}_{\textrm{I}} & =\mathcal{T}e^{\int_{0}^{T}(-i\mathcal{H}_{\textrm{I}}(t)+\mathcal{L}_{\textrm{I}}(t))dt}.\label{eq:S14 KI}
	\end{align}
	It is important to remark that no restrictions are imposed on the
	pulse $\mathcal{H}(t)$, so long as it reproduces the noise-free evolution
	$\mathcal{U}$. On the other hand, we will see in Supplementary Note 3
	that Eqs. (\ref{eq:S11 inverse driving}) and (\ref{eq:S12 inverse dissipator})
	allow us to approximate the noise channel for $\mathcal{K}$ as $\left(\mathcal{K}_{\textrm{I}}\mathcal{K}\right)^{\frac{1}{2}}$,
	which is why the form of the inverse driving in (\ref{eq:S11 inverse driving})
	is important for our main finding. 
	
	We also note that the level of control required for implementing $\mathcal{H}_{\textrm{I}}(t)$
	is very similar to that used for $\mathcal{H}(t)$. In essence, we
	only need to time-reverse the pulse schedule corresponding to $\mathcal{H}(t)$
	and flip its sign. In this work, we use the pulse-gate capabilities
	of the IBM processors to implement $\mathcal{H}_{\textrm{I}}(t)$. No stretching
	of the pulses or any modification of their shape is involved. Therefore,
	the powers of $\mathcal{K}_{\textrm{I}}\mathcal{K}$ that enter the implementation
	of the KIK method are basically as easy to execute $\mathcal{K}$
	itself. 
	
	\section*{Supplementary Note 3: Derivation of the KIK formula}
	
	In this section, we derive the KIK formula 
	\begin{equation}
		\mathcal{U}_{\textrm{KIK}}=\mathcal{K}\left(\mathcal{K}_{\textrm{I}}\mathcal{K}\right)^{-\frac{1}{2}},\label{eq:S15 KIK formula}
	\end{equation}
	where $\mathcal{U}_{\textrm{KIK}}$ is a first-order Magnus approximation to
	the ideal evolution $\mathcal{U}$ that we will clarify in what follows.
	Hereafter, we will refer to $\mathcal{K}$ and $\mathcal{K}_{\textrm{I}}$
	as target evolution and inverse evolution, respectively. In particular,
	$\mathcal{K}$ is the noisy evolution over which we intend to perform
	error mitigation. For now, we assume that the noise-free unitary is
	given by $\mathcal{U}=\mathcal{T}e^{-\int_{0}^{T}i\mathcal{H}(t)dt}$,
	meaning that the pulse schedule $\mathcal{H}(t)$ is perfectly calibrated.
	Hence, the KIK formula is useful to mitigate errors caused by the
	dissipator $\mathcal{L}(t)$. On the other hand, we will see in Supplementary Note 6 that randomized compiling \cite{wallman2016noise}
	complements and enhances the error mitigation achieved with the KIK
	method. Accordingly, integrating randomized compiling into our QEM
	technique also allows for the mitigation of coherent errors, related
	to miscalibrations of $\mathcal{H}(t)$. 
	
	To arrive at Eq. (\ref{eq:S15 KIK formula}) we shall proceed as follows.
	We consider that the driving $\mathcal{H}(t)$ acts in the time interval
	$(0,T)$ and the driving $\mathcal{H}_{\textrm{I}}(t)$ is applied in the interval
	$(T,2T)$. Thus, the total evolution at time $t=2T$ is $\mathcal{K}_{\textrm{I
	}}\mathcal{K}$.
	For any time $t\in(0,2T)$ the dynamics is modeled according to 
	
	\begin{equation}
		\frac{d}{dt}|\rho\rangle=\left(-i\tilde{\mathcal{H}}(t)+\tilde{\mathcal{L}}(t)\right)|\rho\rangle,\label{eq:S16 master eq for t in (0,2T)}
	\end{equation}
	where 
	\begin{align}
		\tilde{\mathcal{H}}(t) & =\begin{cases}
			\begin{array}{c}
				\mathcal{H}(t)\textrm{ for }t\in(0,T)\\
				\mathcal{H}_{\textrm{I}}(t-T)\textrm{ for }t\in(T,2T),
		\end{array}\end{cases}\nonumber \\
		\tilde{\mathcal{L}}(t) & =\begin{cases}
			\begin{array}{c}
				\mathcal{L}(t)\textrm{ for }t\in(0,T)\\
				\mathcal{L}_{\textrm{I}}(t-T)\textrm{ for }t\in(T,2T).
		\end{array}\end{cases}\label{eq:S17 H(t) and L(t) for t in (0,2T)}
	\end{align}
	Note that the action of $\mathcal{H}_{\textrm{I}}$ and $\mathcal{L}_{\textrm{I
	}}$
	on $(T,2T)$ requires the time shift by $T$ as described in Eqs.
	(\ref{eq:S17 H(t) and L(t) for t in (0,2T)}). With this notation,
	the ideal evolution for $t\in(0,2T)$ is given by $\tilde{\mathcal{U}}(t)=\mathcal{T}e^{-i\int_{0}^{t}\mathcal{\tilde{\mathcal{H}}}(t')dt'}$,
	and therefore, $\tilde{\mathcal{U}}(T)=\mathcal{U}$, and $\tilde{\mathcal{U}}(2T)=\mathcal{I}$.
	Similarly, for the noisy evolution we have that $\tilde{\mathcal{K}}(t)=\mathcal{T}e^{\int_{0}^{t}\left(-i\mathcal{\tilde{\mathcal{H}}}(t')+\tilde{\mathcal{L}}(t')\right)dt'}$,
	$\tilde{\mathcal{K}}(T)=\mathcal{K}$, and $\tilde{\mathcal{K}}(2T)=\mathcal{K}_{I}\mathcal{K}$. 
	
	By expressing Eq. (\ref{eq:S16 master eq for t in (0,2T)}) in the
	interaction picture we can write the evolution operator in the form
	$\mathcal{K}=\mathcal{U}\mathcal{N}$, where the noise channel $\mathcal{N}$
	is the solution to Eq. (\ref{eq:S16 master eq for t in (0,2T)}) in
	interaction picture, at time $t=T$. Next, using the Magnus expansion
	\cite{blanes2009magnus}, we will find that $\mathcal{N}$
	can be approximated by $\left(\mathcal{K}_{\textrm{I}}\mathcal{K}\right)^{\frac{1}{2}}$.
	These are the main ingredients for the derivation of the KIK formula
	(\ref{eq:S15 KIK formula}).
	
	To define the transformed states and operators in interaction picture,
	we use the noiseless evolution $\tilde{\mathcal{U}}(t)$. Denoting
	interaction picture vectors and matrices with the subscript ``int'',
	we have that 
	\begin{align}
		|\rho_{\textrm{int}}(t)\rangle & =\tilde{\mathcal{U}}^{\dagger}(t)|\rho(t)\rangle,\label{eq:S18 rho in int picture}\\
		\tilde{\mathcal{L}}_{\textrm{int}}(t) & =\tilde{\mathcal{U}}^{\dagger}(t)\tilde{\mathcal{L}}(t)\tilde{\mathcal{U}}(t),\label{eq:S19 L in int picture}\\
		\frac{d}{dt}|\rho_{\textrm{int}}(t)\rangle & =\tilde{\mathcal{L}}_{\textrm{int}}(t)|\rho_{\textrm{int}}(t)\rangle.\label{eq:S20 Master Eq in int picture}
	\end{align}
	The solution $|\rho_{\textrm{int}}(t)\rangle=\tilde{\mathcal{K}}_{\textrm{int}}(t)|\rho_{\textrm{int}}(0)\rangle$
	to Eq. (\ref{eq:S20 Master Eq in int picture}) is related to the
	original (Schrodinger-picture) solution by $\tilde{\mathcal{K}}_{\textrm{int}}(t)=\tilde{\mathcal{U}}^{\dagger}(t)\tilde{\mathcal{K}}(t)$.
	Therefore, 
	\begin{align}
		\tilde{\mathcal{K}}(t) & =\tilde{\mathcal{U}}(t)\tilde{\mathcal{K}}_{\textrm{int}}(t)\nonumber \\
		& =\tilde{\mathcal{U}}(t)e^{\Omega(t)},\label{eq:S21 K for t in (0,2T)}
	\end{align}
	where we express $\tilde{\mathcal{K}}_{\textrm{int}}(t)$ in terms
	of the Magnus expansion $\Omega(t)=\sum_{n=1}^{\infty}\Omega_{n}(t)$
	\cite{blanes2009magnus}. The first-order Magnus term $\Omega_{1}(t)$
	is central to our analysis, and is given by 
	\begin{equation}
		\Omega_{1}(t)=\int_{0}^{t}\tilde{\mathcal{L}}_{\textrm{int}}(t')dt'.\label{eq:S22 Omega1}
	\end{equation}
	Regarding higher order terms $\Omega_{n\geq2}(t)$, we only mention
	that they contain nested commutators that obey time ordering. For
	example, $\Omega_{2}(t)=\frac{1}{2}\int_{0}^{t}dt'\int_{0}^{t'}dt''[\mathcal{L}_{\textrm{int}}(t'),\mathcal{L}_{\textrm{int}}(t'')]$. 
	
	Setting $t=T$ and $t=2T$ in Eq. (\ref{eq:S21 K for t in (0,2T)})
	leads us to the exact solutions $\mathcal{K}=\mathcal{U}e^{\Omega(T)}$ and $\mathcal{K}_{\textrm{I}}\mathcal{K}=e^{\Omega(2T)}$. If we keep only the first Magnus term in the corresponding Magnus expansions,
	\begin{align}
		\mathcal{K} & \approx\mathcal{U}e^{\Omega_{1}(T)},\label{eq:S23 K with Magnus expansion}\\
		\mathcal{K}_{\textrm{I}}\mathcal{K} & \approx e^{\Omega_{1}(2T)}.\label{eq:S24 KIK with Magnus expansion}
	\end{align}
	From these expressions, our final step in the derivation of (\ref{eq:S15 KIK formula})
	is to show that 
	\begin{equation}
		\Omega_{1}(2T)=2\Omega_{1}(T)\Leftrightarrow\int_{0}^{T}\tilde{\mathcal{L}}_{\textrm{int}}(t')dt'=\int_{T}^{2T}\tilde{\mathcal{L}}_{\textrm{int}}(t')dt'.\label{eq:S25 Omega1(T)=00003D(1/2)Omega1(2T)}
	\end{equation}
	Taking into account Eqs. \eqref{eq:S23 K with Magnus expansion} and \eqref{eq:S24 KIK with Magnus expansion}, Eq. (\ref{eq:S25 Omega1(T)=00003D(1/2)Omega1(2T)})
	implies that the noise channel $\mathcal{N}=e^{\Omega(T)}$ for the evolution $\mathcal{K}$ 
	can be approximated by 
	\begin{equation}
		\mathcal{N}\approx\mathcal{N}_{\textrm{KIK}}:=\left(\mathcal{K}_{\textrm{I}}\mathcal{K}\right)^{\frac{1}{2}}.\label{eq:S26 KIK formula for noise chanel}
	\end{equation}
	In this way, the KIK formula is obtained by multiplying $\mathcal{K}\approx\mathcal{U}\mathcal{N}_{\textrm{KIK}}$
	by the inverse $\mathcal{N}_{\textrm{KIK}}^{-1}=\left(\mathcal{K}_{\textrm{I}}\mathcal{K}\right)^{-\frac{1}{2}}$.\textcolor{green}{{} }
	
	Before proving Eq. \eqref{eq:S25 Omega1(T)=00003D(1/2)Omega1(2T)}, it is instructive to write also the inverse evolution $\mathcal{K}_{\textrm{I}}$ in the first Magnus approximation. Using Eqs. \eqref{eq:S23 K with Magnus expansion}-\eqref{eq:S25 Omega1(T)=00003D(1/2)Omega1(2T)}, we have that $\mathcal{K}_{\textrm{I}}\mathcal{K}\approx\mathcal{K}_{\textrm{I}}\mathcal{U}e^{\Omega_{1}(T)}\approx e^{2\Omega_{1}(T)}$. Therefore, we can multiply the expression  $\mathcal{K}_{\textrm{I}}\mathcal{U}e^{\Omega_{1}(T)}\approx e^{2\Omega_{1}(T)}$ from the right hand side by $e^{-\Omega_{1}(T)}\mathcal{U}^{\dagger}$, to obtain
	
	\begin{equation}
		\mathcal{K}_{\textrm{I}}\approx e^{\Omega_{1}(T)}\mathcal{U}^{\dagger}.\label{eq:S26.1 KIK formula for KI}
	\end{equation}
	
	Let us now prove Eq. (\ref{eq:S25 Omega1(T)=00003D(1/2)Omega1(2T)}).
	First, we note that $\tilde{\mathcal{U}}(t)=\mathcal{U}^{\dagger}(t-T)\mathcal{U}(T)=\mathcal{U}(2T-t)$
	for $t\in(T,2T)$. In addition, for the same time interval Eqs. (\ref{eq:S12 inverse dissipator})
	and (\ref{eq:S17 H(t) and L(t) for t in (0,2T)}) lead to $\tilde{\mathcal{L}}(t)=\mathcal{L}(2T-t)$.
	Therefore,
	\begin{align}
		\int_{T}^{2T}\tilde{\mathcal{L}}_{\textrm{int}}(t)dt & =\int_{T}^{2T}\tilde{\mathcal{U}}^{\dagger}(t)\tilde{\mathcal{L}}(t)\tilde{\mathcal{U}}(t)dt\nonumber \\
		& =\int_{T}^{2T}\mathcal{U}^{\dagger}(2T-t)\mathcal{L}(2T-t)\mathcal{U}(2T-t)dt\nonumber \\
		& =\int_{0}^{T}\mathcal{U}^{\dagger}(t')\mathcal{L}(t')\mathcal{U}(t')dt'\nonumber \\
		& =\int_{0}^{T}\tilde{\mathcal{L}}_{\textrm{int}}(t)dt,\label{eq:S27  proof of equality of integrals in (0,T) and (T,2T)}
	\end{align}
	where the last line follows by performing the change of variable $t'=2T-t$. 
	
	To conclude this section, we stress that Eq. (\ref{eq:S12 inverse dissipator})
	is key for the proof of Eq. (\ref{eq:S25 Omega1(T)=00003D(1/2)Omega1(2T)}).
	In turn, within our characterization of noise it is specifically the
	inverse driving (\ref{eq:S11 inverse driving}) which provides the
	form taken by the dissipator (\ref{eq:S12 inverse dissipator}). This
	shows the crucial role of using the pulse schedule (\ref{eq:S11 inverse driving})
	for the inverse evolution, rather than a different alternative that
	generates the ideal unitary $\mathcal{U}^{\dagger}$ in the absence
	of noise. \\
	\\
	\fakesubsection{Relation between $\mathcal{K}_{\textrm{I}}\mathcal{K}$ and $\mathcal{K}\mathcal{K}_{\textrm{I}}$
		in the KIK formula}
	\textbf{Relation between $\mathcal{K}_{\textrm{I}}\mathcal{K}$ and $\mathcal{K}\mathcal{K}_{\textrm{I}}$
		in the KIK formula.} In the following, we show that 
	\begin{equation}
		\mathcal{K}\mathcal{K}_{\textrm{I}}=\mathcal{U}\mathcal{K}_{\textrm{I}}\mathcal{K}\mathcal{U}^{\dagger}.\label{eq:S28 KKI and KIK}
	\end{equation}
	Clearly, this implies that we cannot substitute $\mathcal{K}_{\textrm{I}}\mathcal{K}$
	by $\mathcal{K}\mathcal{K}_{\textrm{I}}$ in the KIK formula or in the corresponding
	expansions. In particular, the coincidence with Richardson ZNE applying
	circuit unitary folding, discussed in Supplementary Note 4,
	is sound whenever noise amplification is performed using the correct
	ordering $\mathcal{K}_{\textrm{I}}\mathcal{K}$. This is different from the
	heuristic approach taken in Ref. \cite{giurgica2020digital},
	where $\mathcal{K}\mathcal{K}_{\textrm{I}}$ could be an equally valid choice
	because it also reproduces the identity operation in the absence of
	noise. 
	
	More specifically, we show that the relation (\ref{eq:S28 KKI and KIK})
	holds under the same approximation that leads to Eq. (\ref{eq:S15 KIK formula}).
	Namely, when the Magnus expansion used to express the evolution $\mathcal{K}\mathcal{K}_{\textrm{I}}$
	is also truncated to the first Magnus term. Following our noise model,
	this evolution is the solution to the equation 
	\begin{equation}
		\frac{d}{dt}|\rho\rangle=\left(-i\bar{\mathcal{H}}(t)+\bar{\mathcal{L}}(t)\right)|\rho\rangle,\label{eq:S29 master eq for KKI}
	\end{equation}
	where
	\begin{align}
		\bar{\mathcal{H}}(t) & =\begin{cases}
			\begin{array}{c}
				\mathcal{H}_{\textrm{I}}(t)\textrm{ for }t\in(0,T)\\
				\mathcal{H}(t-T)\textrm{ for }t\in(T,2T),
		\end{array}\end{cases}\nonumber \\
		\bar{\mathcal{L}}(t) & =\begin{cases}
			\begin{array}{c}
				\mathcal{L}_{\textrm{I}}(t)\textrm{ for }t\in(0,T)\\
				\mathcal{L}(t-T)\textrm{ for }t\in(T,2T).
		\end{array}\end{cases}\label{eq:S30 H(t) and L(t) for t in (0,2T) for KKI}
	\end{align}
	
	In interaction picture, the first Magnus term for the solution of
	(\ref{eq:S29 master eq for KKI}) at time $t=2T$ reads $\int_{0}^{2T}\bar{\mathcal{L}}_{\textrm{int}}(t)dt$,
	where $\bar{\mathcal{L}}_{\textrm{int}}(t)=\mathcal{\bar{\mathcal{U}}}^{\dagger}(t)\bar{\mathcal{L}}(t)\mathcal{\bar{\mathcal{U}}}(t)$.
	Taking this into account, we will show that 
	\begin{equation}
		\int_{0}^{2T}\bar{\mathcal{L}}_{\textrm{int}}(t)dt=\mathcal{U}(T)\int_{0}^{2T}\tilde{\mathcal{L}}_{\textrm{int}}(t)dt\mathcal{U}^{\dagger}(T).\label{eq:S31first Magnus terms for KIK and KKI}
	\end{equation}
	Accordingly, up to first order in the Magnus expansion we have that
	$\mathcal{K}\mathcal{K}_{\textrm{I}}\approx e^{\int_{0}^{2T}\bar{\mathcal{L}}_{\textrm{int}}(t)dt}=\mathcal{U}(T)e^{\int_{0}^{2T}\tilde{\mathcal{L}}_{\textrm{int}}(t)dt}\mathcal{U}^{\dagger}(T)\approx\mathcal{U}(T)\mathcal{K}_{\textrm{I}}\mathcal{K}\mathcal{U}^{\dagger}(T)$,
	which is tantamount to Eq. (\ref{eq:S28 KKI and KIK}). 
	
	To prove Eq. (\ref{eq:S31first Magnus terms for KIK and KKI}), we
	derive the following two alternative forms of $\int_{0}^{T}\bar{\mathcal{L}}_{\textrm{int}}(t)dt$:
	
	\begin{align}
		\int_{0}^{T}\bar{\mathcal{L}}_{\textrm{int}}(t)dt & =\mathcal{U}(T)\int_{0}^{T}\tilde{\mathcal{L}}_{\textrm{int}}(t)dt\mathcal{U}^{\dagger}(T),\label{eq:31.1 first Magnus term of KI}\\
		\int_{0}^{T}\bar{\mathcal{L}}_{\textrm{int}}(t)dt & =\int_{T}^{2T}\bar{\mathcal{L}}_{\textrm{int}}(t)dt.\label{eq:S32 firt Magnus term of KI}
	\end{align}
	Noting that $\bar{\mathcal{U}}(t)=\mathcal{U}(t)$ for $t\in(0,T)$,
	we obtain 
	\begin{align}
		\int_{0}^{T}\bar{\mathcal{L}}_{\textrm{int}}(t)dt & =\int_{0}^{T}\mathcal{\bar{\mathcal{U}}}^{\dagger}(t)\mathcal{L}_{\textrm{I}}(t)\mathcal{\bar{\mathcal{U}}}(t)dt\nonumber \\
		& =\int_{0}^{T}\mathcal{U}(t)\mathcal{L}(T-t)\mathcal{U}^{\dagger}(t)dt\nonumber \\
		& =\int_{0}^{T}\mathcal{U}(T-t')\mathcal{L}(t')\mathcal{U}^{\dagger}(T-t')dt'\nonumber \\
		& =\mathcal{U}(T)\left(\int_{0}^{T}\mathcal{U}^{\dagger}(t')\mathcal{L}(t')\mathcal{U}(t')dt'\right)\mathcal{U}^{\dagger}(T),\label{eq:S33 first Magnus term of KKI 1}
	\end{align}
	where the change of variable $t'=T-t$ is performed in the third line,
	and in the last line we use the relation $\mathcal{U}(T)=\mathcal{U}(T-t')\mathcal{U}(t')$.
	This proves Eq. (\ref{eq:31.1 first Magnus term of KI}). 
	
	For the time interval $t\in(T,2T)$, the evolution $\mathcal{\bar{\mathcal{U}}}(t)$
	reads $\mathcal{\bar{\mathcal{U}}}(t)=\mathcal{U}(t-T)\mathcal{U}^{\dagger}(T)$.
	Therefore, 
	\begin{align}
		\int_{T}^{2T}\bar{\mathcal{L}}_{\textrm{int}}(t)dt & =\int_{T}^{2T}\bar{\mathcal{U}}^{\dagger}(t)\mathcal{L}(t-T)\mathcal{\bar{\mathcal{U}}}(t)dt\nonumber \\
		& =\int_{T}^{2T}\mathcal{U}(T)\mathcal{U}^{\dagger}(t-T)\mathcal{L}(t-T)\nonumber \\
		& \quad\quad\quad\mathcal{U}(t-T)\mathcal{U}^{\dagger}(T)dt\nonumber \\
		& =\mathcal{U}(T)\left(\int_{0}^{T}\mathcal{U}^{\dagger}(t')\mathcal{L}(t')\mathcal{U}(t')dt'\right)\mathcal{U}^{\dagger}(T),\label{eq:S34 first Magnus term of K}
	\end{align}
	which (in combination with (\ref{eq:S33 first Magnus term of KKI 1}))
	proves Eq. (\ref{eq:S32 firt Magnus term of KI}). Equation (\ref{eq:S31first Magnus terms for KIK and KKI})
	follows straightforwardly by combining Eqs. (\ref{eq:31.1 first Magnus term of KI})
	and (\ref{eq:S32 firt Magnus term of KI}).
	
	We note that the simulations studied in Ref. \cite{giurgica2020digital}
	are based on the assumption of a global depolarizing channel that
	is identical for $\mathcal{K}$ and $\mathcal{K}_{\textrm{I}}$. Because global
	depolarizing noise commutes with any unitary $\mathcal{U}$, in this
	case the total noise channel for both $\mathcal{K}_{\textrm{I}}\mathcal{K}$
	and $\mathcal{K}\mathcal{K}_{\textrm{I}}$ is simply another depolarizing channel
	with an increased error rate. Hence, for this simple model both the
	KIK formula and circuit unitary folding can be applied using either
	$\mathcal{K}_{\textrm{I}}\mathcal{K}$ or $\mathcal{K}\mathcal{K}_{\textrm{I}}$. However,
	as we have shown here, in a more realistic scenario the
	proper time ordering corresponding to $\mathcal{K}_{\textrm{I}}\mathcal{K}$
	is crucial for a correct application of QEM. \\ 
	\\
	\fakesubsection{Robustness of KIK error mitigation to noise drifts and spatially varying noise profiles}
	\textbf{Robustness of KIK error mitigation to noise drifts and spatially varying noise profiles.} Until now, we have approached the time dependence of $\tilde{\mathcal{L}}(t)$
	(cf. Eq. (\ref{eq:S17 H(t) and L(t) for t in (0,2T)})) as being a
	consequence of the time dependence associated with the pulse schedules,
	$\tilde{\mathcal{H}}(t)$. In this framework, any implementation of
	$\mathcal{K}$ or $\mathcal{K}_{\textrm{I}}$ would be affected by the same
	dissipators $\mathcal{L}(t)$ and $\mathcal{L}_{\textrm{I}}(t-T)$. However,
	it is more realistic to include the possibility of noise sources that
	also change in time. For example, a varying temperature or external
	electromagnetic field can be such that the dissipator $\mathcal{L}(t)$
	acting during a given implementation of $\mathcal{K}$ differs from
	the dissipator $\mathcal{L}'(t)$, associated with an execution of
	the same evolution at a later time. In the present section we discuss
	a technique to collect the QEM data that minimizes the effect of noise
	drifts. As we shall see, this is possible by distributing the circuits
	for QEM into suitable sets, and separately applying the KIK formula
	(\ref{eq:S15 KIK formula}) to each of these sets.

	As discussed in the main text (see also Supplementary Note 4),
	performing QEM with the KIK method involves executing circuits of
	the form $\mathcal{K}\left(\mathcal{K}_{\textrm{I}}\mathcal{K}\right)^{m}$,
	for $0\leq m\leq M$, where $M$ is the mitigation order. Therefore,
	the time for running $\mathcal{K}\left(\mathcal{K}_{\textrm{I}}\mathcal{K}\right)^{m}$
	is $(2m+1)T$, where $T$ is the evolution time of $\mathcal{K}$
	or $\mathcal{K}_{\textrm{I}}$. In the computation of expectation values, it
	is necessary to implement each $\mathcal{K}\left(\mathcal{K}_{\textrm{I}}\mathcal{K}\right)^{m}$
	a certain number of times $N_{m}$. Following standard terminology
	in quantum computing, a single execution of a circuit, including the
	preparation of the initial state $|\rho\rangle$ and the measurement
	of the final state, is dubbed a ``shot''. Hence, $N_{m}$ shots
	are used for each $\mathcal{K}\left(\mathcal{K}_{\textrm{I}}\mathcal{K}\right)^{m}$,
	and the ``shot budget'' to collect all the QEM data characteristic
	of the KIK method reads $N=\sum_{m=0}^{M}N_{m}$ (note that this excludes
	the shots invested in the estimation of the survival probability $\mu=\langle\rho|\mathcal{K}_{\textrm{I}}\mathcal{K}|\rho\rangle$,
	in the case of adaptive mitigation). Assuming for now that the time
	for preparing $|\rho\rangle$ and the time for measuring the corresponding
	final states are negligible with respect to $T$, performing $N$
	shots takes a total time 
	\begin{equation}
		t_{N}=\sum_{m=0}^{M}N_{m}(2m+1)T.\label{eq:S34.4  total time for KIK method}
	\end{equation}
	
	For our analysis, it is useful to extend the time domain of $\tilde{\mathcal{L}}(t)$,
	to account for the behavior of the noise under repetitions of the
	evolutions $\mathcal{K}$ and $\mathcal{K}_{\textrm{I}}$. In this way, the
	time for an arbitrary repetition of $\mathcal{K}$ or $\mathcal{K}_{\textrm{I}}$
	can be expressed as $t+2kT$, with $0\leq t\leq2T$ and $k$ a positive
	integer, and stationary noise is characterized by the condition $\tilde{\mathcal{L}}(t+2kT)=\tilde{\mathcal{L}}(t)$,
	where $\tilde{\mathcal{L}}(t)$ is the dissipator in Eq. (\ref{eq:S17 H(t) and L(t) for t in (0,2T)}).
	Conversely, noise drifts take place within the total time interval
	$(0,t_{N})$ if $\tilde{\mathcal{L}}(t+2kT)\neq\tilde{\mathcal{L}}(t)$
	for some $k$. 
	
	\begin{figure}
		\begin{centering}
			\includegraphics[scale=0.7]{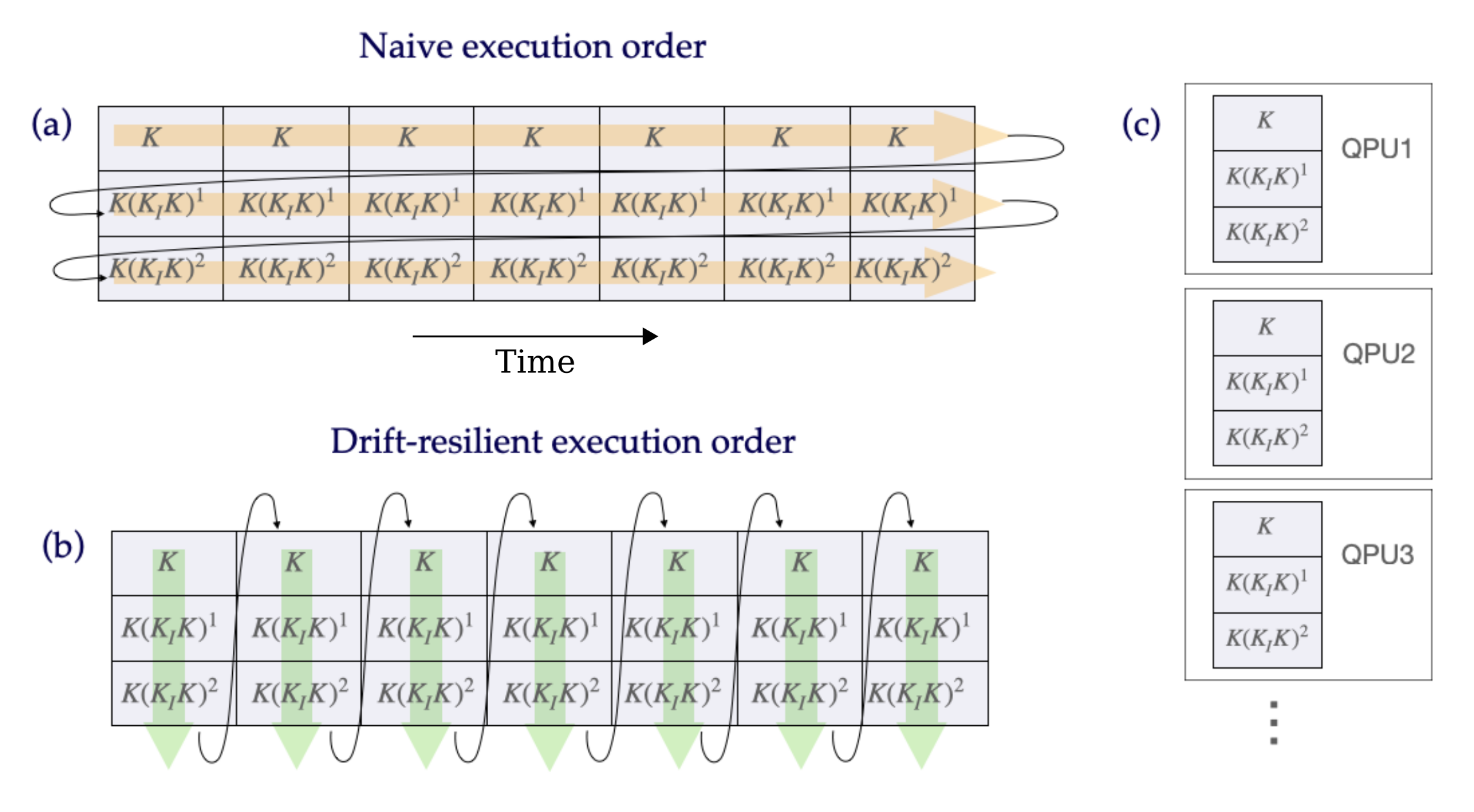}\caption{Two strategies to collect data for second-order error mitigation,
				using the KIK method. The arrows indicate the order of implementation
				of different circuits. In (a), the implementations (shots) are separated
				into repetitions of each circuit $\mathcal{K}\left(\mathcal{K}_{\textrm{I}}\mathcal{K}\right)^{m}$,
				for $0\protect\leq m\protect\leq2$ (horizontal groups). In (b), the
				shots are divided into $S$ sets (vertical groups) that include implementations
				of all the circuits $\mathcal{K}\left(\mathcal{K}_{\textrm{I}}\mathcal{K}\right)^{m}$. (a) and (b) depict circuit distributions with respect to time, for addressing noise drifts. The  temporal sequence of circuit sets shown in (b) is rearranged as a ``spatial distribution'', where different sets are implemented in different Quantum Processing Units (QPUs). This is useful for a parallel implementation of the KIK method.}
			\par\end{centering}
	\end{figure}

	Let us now suppose that noise unavoidably drifts in the interval $(0,t_{N})$.
	In this scenario, the consistency of the evolutions $\mathcal{K}$
	or $\mathcal{K}_{\textrm{I}}$ in different shots can break down and prevent
	a correct implementation of the KIK formula (\ref{eq:S15 KIK formula}).
	However, we can avoid or at least alleviate this effect through a proper distribution of the shot budget $N$. Consider Supplementary Figure 2, where two strategies for implementing the circuits $\{\mathcal{K}\left(\mathcal{K}_{\textrm{I}}\mathcal{K}\right)^{m}\}_{m=0}^{2}$
	(second-order mitigation) are illustrated. In the case of Supplementary Figure 2(a), all the shots corresponding to a given $N_{m}$ are sequentially
	implemented, i.e., $N_{0}$ shots are first performed, followed by
	$N_{1}$ shots, and finally by $N_{2}$ shots. On the other hand,
	the strategy of Supplementary Figure 2(b) relies on dividing the
	$N$ shots into $S$ sets $\{n_{0},n_{1},n_{2}\}$ of $N_{S}=n_{0}+n_{1}+n_{2}$
	shots each, where $n_{m}$ shots are employed for the circuit $\mathcal{K}\left(\mathcal{K}_{\textrm{I}}\mathcal{K}\right)^{m}$.
	Therefore, all the mitigation circuits $\mathcal{K}\left(\mathcal{K}_{\textrm{I}}\mathcal{K}\right)^{m}$
	appear in each set. Without loss of generality for our argumentation,
	we can focus on the simple case where $n_{m}=N_{S}/3$ for $0\leq m\leq2$,
	i.e. when the shots of each set are equally distributed into the different
	circuits $\mathcal{K}\left(\mathcal{K}_{\textrm{I}}\mathcal{K}\right)^{m}$. 
	
	Since each set $\{n_{0},n_{1},n_{2}\}$ contains data produced by
	all the circuits $\mathcal{K}\left(\mathcal{K}_{\textrm{I}}\mathcal{K}\right)^{m}$,
	the KIK formula can be individually applied to these data sets. Let
	\begin{equation}
		\mathcal{U}_{\textrm{KIK},s}=\mathcal{K}_{s}\left(\mathcal{K}_{\textrm{I};s}\mathcal{K}_{s}\right)^{-\frac{1}{2}}\label{eq:S34.5 KIK formula for sth set}
	\end{equation}
	denote the KIK formula corresponding to evolutions $\mathcal{K}_{s}$
	and $\mathcal{K}_{\textrm{I},s}$ that are executed in shots of the $s$th
	set $\{n_{0},n_{1},n_{2}\}^{(s)}$. If there were no noise drifts,
	$\mathcal{K}_{s}=\mathcal{K}_{1}=\mathcal{K}$ and $\mathcal{K}_{\textrm{I};s}=\mathcal{K}_{\textrm{I};1}=\mathcal{K}_{\textrm{I}}$
	for $1\leq s\leq S$. Therefore, the two strategies depicted in Supplementary Figure 2 would lead to the same result. Nevertheless, the
	non-stationary character of the noise may cause that $\mathcal{K}_{s}$
	or $\mathcal{K}_{\textrm{I};s}$ change significantly when moving between different
	sets $\{n_{0},n_{1},n_{2}\}^{(s)}$, or even within a fixed set. The
	second possibility is less likely though, if the time $\sum_{m=0}^{M}n_{m}(2m+1)T$
	invested in implementing each set $\{n_{0},n_{1},n_{2}\}^{(s)}$ is
	smaller than the characteristic time for noise drifts to be significant.
	In other words, if the time scale over which noise drifts occur is
	sufficiently large to allow a consistent execution of $\mathcal{U}_{\textrm{KIK},s}$.
	Assuming that $N_{S}$ is sufficiently small (equivalently, $S$ sufficiently
	large) for this to happen, for any $1\leq s\leq S$ we can implement
	the formula (\ref{eq:S34.5 KIK formula for sth set}) without worrying
	about variations in the evolutions $\mathcal{K}_{s}$ or $\mathcal{K}_{\textrm{I};s}$.
	In this way, for the shot budget $N$ we utilize the ``average KIK
	formula'' 
	\begin{figure}
		
		\begin{centering}
			\includegraphics[scale=0.5]{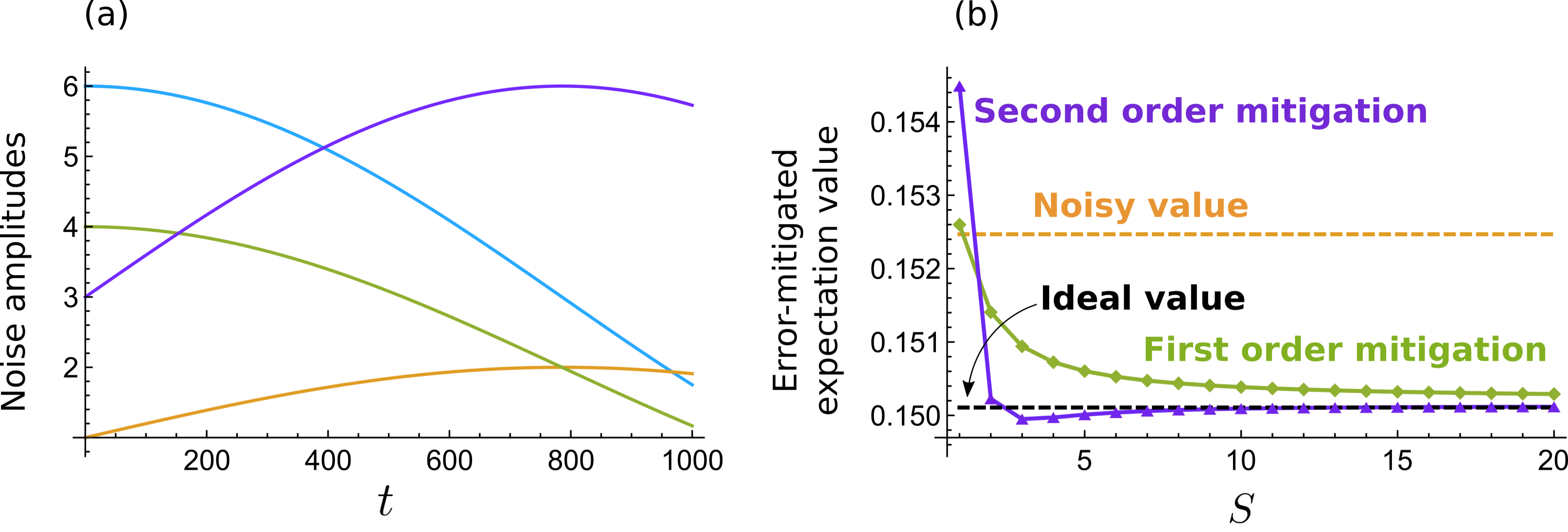}\caption{KIK method applied to a system with noise drifts. (a) The four time-dependent
				amplitudes $f_{k}(t)$ that appear in the dissipator (\ref{eq:S34.7 fluctuating dissipator}),
				and characterize the drift in noise parameters. (b) First order mitigation
				and second order mitigation using the average formula (\ref{eq:S 34.6 average KIK formula}),
				as a function of the number of sets $S$ (see Supplementary Figure 2
				and main text for details). }
			\par\end{centering}
	\end{figure}
	
	\begin{equation}
		\mathcal{U}_{\textrm{KIK},\textrm{av}}^{(M)}=\frac{1}{S}\sum_{s=1}^{S}\mathcal{U}_{\textrm{KIK},s}^{(M)}.\label{eq:S 34.6 average KIK formula}
	\end{equation}

	A pertinent question when executing a computation is whether it can be parallelized and save time. The parallelization can be carried out using subsets of qubits in the same quantum processing unit 
	(QPU), in a different QPU on the same chip, or on a different platform in a remote location. In the context of QEM, a recent proposal is to integrate this approach with virtual distillation \cite{koczor2021exponential,huggins2021virtual} techniques \cite{jnane2022multicore}.
	
	The KIK method can easily be parallelized by assigning different sets to different QPUs as shown in  Supplementary Figure 2(c). In this case, the index $s$ in Eq.  \eqref{eq:S34.5 KIK formula for sth set} would label different QPUs, rather than different instants of time, and the corresponding error-mitigated expectation value would also be computed using \eqref{eq:S 34.6 average KIK formula}.
	The justification is the same as in the case of noise drifts discussed above.
	
	Let us now  illustrate the application of this methodology to KIK mitigation under noise drifts. In Supplementary Figure 3, we consider
	two qubits subjected to a dissipator 
	\begin{equation}
		\tilde{\mathcal{L}}(t)=\xi\sum_{k=1}^{4}f_{k}(t)\mathcal{L}_{k},\label{eq:S34.7 fluctuating dissipator}
	\end{equation}
	where $\xi=0.05$,
	\begin{align}
		\mathcal{L}_{k} & =A_{k}\otimes A_{k}^{\ast}-\frac{1}{2}A_{k}^{\dagger}A_{k}\otimes I-\frac{1}{2}I\otimes\left(A_{k}^{\dagger}A_{k}\right)^{\textrm{T}},\label{eq:S34.8 Lk Lindbladian}\\
		A_{1} & =\left(\begin{array}{cc}
			0 & 1\\
			0 & 0
		\end{array}\right)\otimes\left(\begin{array}{cc}
			1 & 0\\
			0 & 1
		\end{array}\right),\label{eq:S34.8.1 A1}\\
		A_{2} & =\left(\begin{array}{cc}
			1 & 0\\
			0 & 0
		\end{array}\right)\otimes\left(\begin{array}{cc}
			1 & 0\\
			0 & 1
		\end{array}\right),\label{eq:S34.8.2 A2}\\
		A_{3} & =\left(\begin{array}{cc}
			1 & 0\\
			0 & 1
		\end{array}\right)\otimes\left(\begin{array}{cc}
			0 & 1\\
			0 & 0
		\end{array}\right),\label{eq:S34.8.3 A3}\\
		A_{4} & =\left(\begin{array}{cc}
			1 & 0\\
			0 & 1
		\end{array}\right)\otimes\left(\begin{array}{cc}
			1 & 0\\
			0 & 0
		\end{array}\right),\label{eq:S34.8.4 A4}
	\end{align}
	and 
	\begin{align}
		f_{1}(t) & =3\left(1+\textrm{cos}(2t/T)\right),\label{eq:S34.8.5 f1}\\
		f_{2}(t) & =1+\textrm{sin}(2t/T),\label{eq:S34.8.6 f2}\\
		f_{3}(t) & =2\left(1+\textrm{cos}(2t/T)\right),\label{eq:S34.8.7 f3}\\
		f_{4}(t) & =3\left(1+\textrm{sin}(2t/T)\right).\label{eq:S34.8.8 f4}
	\end{align}
	The noise amplitudes $f_{k}(t)$ are plotted in Supplementary Figure 3(a), for $1\leq t\leq1000$. Moreover, we assume a time-independent
	Hamiltonian (cf. Eq. (\ref{eq:S9 L space commutator =00005BH,rho=00005D}))
	\begin{align}
		\mathcal{H} & =H\otimes I-I\otimes H^{\textrm{T}},\label{eq:S34.8.9 H Liouville}\\
		H & =3X\otimes X+I\otimes X,\label{eq:S34.8.10 H Hilbert}
	\end{align}
	where $I=\left(\begin{array}{cc}
		1 & 0\\
		0 & 1
	\end{array}\right)$ and $X=\left(\begin{array}{cc}
		0 & 1\\
		1 & 0
	\end{array}\right)$.
	
	$\tilde{\mathcal{L}}(t)$ is the Liouville-space representation of
	the Hilbert space dissipator that satisfies $\hat{L}(t)[\rho]=\xi\sum_{k=1}^{4}f_{k}(t)\hat{L}_{k}$,
	which follows by applying the vectorization rule (\ref{eq:S5 triple-prod identity})
	to each term $\hat{L}_{k}[\rho]=A_{k}\rho A_{k}^{\dagger}-\frac{1}{2}A_{k}^{\dagger}A_{k}\rho-\frac{1}{2}\rho A_{k}^{\dagger}A_{k}$.
	Since the associated master equation 
	\begin{equation}
		\frac{d}{dt}|\rho\rangle=\left(-i\mathcal{H}+\tilde{\mathcal{L}}(t)\right)|\rho\rangle\label{eq:S34.8.11 master equation}
	\end{equation}
	has GKSL (Gorini--Kossakowski--Sudarshan--Lindblad) form, it is
	guaranteed that its integration results in a completely positive and
	trace-preserving evolution. We also stress that $\tilde{\mathcal{L}}(t)$
	is now defined in the total time interval $(0,t_{N})$, to account
	for the many repetitions of the pulses $\mathcal{H}(t)$ and $\mathcal{H}_{\textrm{I}}(t-T)$
	that come with the $N$ shots. 
	
	We numerically simulate QEM using the average formula (\ref{eq:S 34.6 average KIK formula}),
	by setting 
	
	\begin{align}
		N & =1000,\label{eq:S34.8.12 N shots}\\
		N_{S} & =N/S,\label{eq:34.8.13 NS shots}\\
		n_{1} & =n_{2}=n_{3}=N_{S}/3,\label{eq:S34.8.13 n_m shots}
	\end{align}
	and $1\leq S\leq20$. An execution of $\mathcal{K}$ or $\mathcal{K}_{\textrm{I}}$
	is performed in a unit of time $T=1$, for which we assume that noise is essentially time independent. In other words, a more precise
	description of the noise occurring during the $N$ shots is given
	by the discrete dissipator
	
	\begin{equation}
		\tilde{\mathcal{L}}_{\textrm{disc}}(t)=\tilde{\mathcal{L}}(n),\textrm{ if }n\leq t\leq n+1,\label{eq:S34.9 discrete fluctuating dissipator}
	\end{equation}
	with $\tilde{\mathcal{L}}(t)$ satisfying Eq. (\ref{eq:S34.7 fluctuating dissipator})
	and $0\leq n\leq N-1$.
	
	Supplementary Figure 3(b) shows the error-mitigated expectation
	value $\left\langle \rho\right\rangle _{\textrm{mit}}=\left\langle \rho\left|\mathcal{U}_{\textrm{KIK},\textrm{av}}^{(M)}\right|\rho\right\rangle $,
	which quantifies the overlap with the initial state $\rho=I/2\otimes|0\rangle\langle0|$.
	We apply Taylor error mitigation, using the coefficients $a_{\textrm{Tay},m}^{(M)}$
	given in Eq. (\ref{eq:S39 coefficients of the truncated Taylor expansion}).
	If $S=1$, the standard strategy represented in Supplementary Figure 2(a) is recovered. We observe in Supplementary Figure 3(b) that
	in this case $\left\langle \rho\right\rangle _{\textrm{mit}}$ deviates
	drastically from the noiseless expectation value $\left\langle \rho\right\rangle _{\textrm{id}}=\left\langle \rho\left|\mathcal{U}\right|\rho\right\rangle $.
	As $S$ increases, second-order mitigation quickly approaches $\left\langle \rho\right\rangle _{\textrm{id}}$
	and converges to it at $S\approx10$. We also stress that for $S\geq10$
	the quality of error mitigation is maintained, both for $M=1$ and
	$M=2$, which shows that averaging the KIK formula over more sets
	does not degrade the performance of the KIK method. The success of
	this strategy is explained because $\left\langle \rho\right\rangle _{\textrm{mit}}=\frac{1}{S}\sum_{s=1}^{S}\left\langle \rho\left|\mathcal{U}_{\textrm{KIK},s}^{(M)}\right|\rho\right\rangle $,
	and if noise is approximately stationary for each
	$\mathcal{U}_{\textrm{KIK},s}^{(M)}$ then the corresponding $\left\langle \rho\left|\mathcal{U}_{\textrm{KIK},s}^{(M)}\right|\rho\right\rangle $
	is not affected by the action of noise drifts. However, the number
	of shots $N_{S}$ associated with each set is not sufficiently large
	to achieve the accuracy shown in Supplementary Figure 3(b). This
	accuracy is achieved after averaging over the total number of sets.
	The convergence observed in Supplementary Figure 3(b) confirms that
	in this example increasing the number of sets enhances the QEM performance,
	and solves the noise drift problem.

	\section*{Supplementary Note 4: Implementations of the inverse of the noise channel $\mathcal{N}_{\textrm{KIK}}$}
	
	The KIK formula (\ref{eq:S15 KIK formula}) provides a compact approximation
	for the ideal target evolution \textit{$\mathcal{U}$}. However, performing
	QEM with this formula also requires being able to physically implement
	the inverse 
	\begin{equation}
		\mathcal{N}_{\textrm{KIK}}^{-1}=\left(\mathcal{K}_{\textrm{I}}\mathcal{K}\right)^{-\frac{1}{2}}.\label{eq:S35 Inverse of Noise channel}
	\end{equation}
	In this section, we compute various approximations to this inverse,
	given as polynomials of the KIK cycle $\mathcal{K}_{\textrm{I}}\mathcal{K}$. \\
	\\
	\fakesubsection{Weak noise limit and Taylor approximation}
	\textbf{Weak noise limit and Taylor approximation.} First, we consider the limit of weak noise. As mentioned in the main
	text, in this case any eigenvalue $\lambda$ of $\mathcal{K}_{\textrm{I}}\mathcal{K}$
	is close to 1, and we can obtain the eigenvalues of $\left(\mathcal{K}_{\textrm{I}}\mathcal{K}\right)^{-\frac{1}{2}}$
	by Taylor expanding $\lambda^{-\frac{1}{2}}$ around $\lambda=1$.
	Note that here we assume that noise is such that $\mathcal{K}_{\textrm{I}}\mathcal{K}$
	is still diagonalizable, and therefore the eigenvalues of $\left(\mathcal{K}_{\textrm{I}}\mathcal{K}\right)^{-\frac{1}{2}}$
	can be obtained as $\lambda^{-\frac{1}{2}}$. 
	
	A Taylor expansion of $\lambda^{-\frac{1}{2}}$ around $\lambda=1$
	is thus equivalent to expand $\left(\mathcal{K}_{\textrm{I}}\mathcal{K}\right)^{-\frac{1}{2}}$
	around the identity $\mathcal{I}$. Namely, 
	\begin{align}
		\left(\mathcal{K}_{\textrm{I}}\mathcal{K}\right)^{-\frac{1}{2}} & =\sum_{m=0}^{\infty}c_{m}\left(\mathcal{K}_{\textrm{I}}\mathcal{K}-\mathcal{I}\right)^{m}\nonumber \\
		& =\sum_{m=0}^{\infty}c_{m}\sum_{k=0}^{m}\frac{m!(-1)^{m-k}}{k!(m-k)!}\left(\mathcal{K}_{\textrm{I}}\mathcal{K}\right)^{k},\label{eq:S36 Taylor expansion for N^-1}
	\end{align}
	where 
	\begin{equation}
		c_{m}=\frac{1}{m!}\frac{d^{m}\lambda^{-\frac{1}{2}}}{d\lambda^{m}}\left|_{\lambda=1}\right.=\left(-1\right)^{m}\frac{(2m-1)!!}{m!2^{m}}.\label{eq:S37 coefficients in the Taylor expansion of N^-1}
	\end{equation}
	Since the series (\ref{eq:S36 Taylor expansion for N^-1}) involves
	infinite powers of $\mathcal{K}_{\textrm{I}}\mathcal{K}$, we must truncate
	it to some fixed order $M$ for the implementation of $\left(\mathcal{K}_{\textrm{I}}\mathcal{K}\right)^{-\frac{1}{2}}$.
	In this way, 
	\begin{align}
		\left(\mathcal{K}_{\textrm{I}}\mathcal{K}\right)^{-\frac{1}{2}} & \approx\sum_{m=0}^{M}c_{m}\sum_{k=0}^{m}\frac{m!(-1)^{m-k}}{k!(m-k)!}\left(\mathcal{K}_{\textrm{I}}\mathcal{K}\right)^{k}\nonumber \\
		& =\sum_{m=0}^{M}a_{\textrm{Tay},m}^{(M)}\left(\mathcal{K}_{\textrm{I}}\mathcal{K}\right)^{m},\label{eq:S38 truncated expansion of N^-1}
	\end{align}
	where 
	\begin{equation}
		a_{\textrm{Tay},m}^{(M)}=(-1)^{m}\frac{(2M+1)!!}{2^{M}[(2m+1)m!(M-m)!]}.\label{eq:S39 coefficients of the truncated Taylor expansion}
	\end{equation}
	\\
	\fakesubsection{Richardson ZNE using Circuit folding and linear scaling of noise}
	\textbf{Richardson ZNE using Circuit folding and linear scaling of noise.}
	In the following, we show that the expansion coefficients (\ref{eq:S39 coefficients of the truncated Taylor expansion})
	predict the result of QEM using Richardson ZNE and noise amplification
	through a method known as circuit folding \cite{giurgica2020digital},
	under the assumption of a linear scaling of the noise. To put this
	result into context, we start by presenting the basics of Richardson
	ZNE and circuit folding.
	
	The goal of ZNE is to infer the noise-free expectation value of an
	observable $A$, by measuring this observable at different levels
	of noise and then extrapolating to the zero-noise limit. Therefore,
	the application of ZNE requires assuming a certain functional dependence
	$\langle A\rangle(\lambda)$, between the expectation value $\langle A\rangle$
	and some noise parameter $\lambda$ over which the experimentalist
	should have control. By measuring expectation values $\langle A\rangle_{k}$
	corresponding to different levels of noise $\lambda_{k}$, an experimentalist
	can fit the data $[\lambda_{k},\langle A\rangle_{k}]$ to the model
	$\langle A\rangle(\lambda)$ and thereby estimate the noiseless expectation
	value as $\langle A\rangle(0)$. In the case of Richardson extrapolation,
	for $M+1$ data points $[\lambda_{k},\langle A\rangle_{k}]$, $\langle A\rangle(\lambda)$
	is taken as a polynomial in $\lambda$ of degree $M$.
	
	There exists a unique polynomial $P(\lambda)$ of degree $M$ that
	intersects all the points $[\lambda_{k},\langle A\rangle_{k}]$. This
	polynomial can be constructed as $P(\lambda)=\sum_{m=0}^{M}\langle A\rangle_{m}l_{m}(\lambda)$,
	where 
	
	\begin{equation}
		l_{m}(\lambda):=\prod_{0\leq k\leq M,k\neq m}\frac{\lambda-\lambda_{k}}{\lambda_{m}-\lambda_{k}}\label{eq:S39.1 Lagrange polynomial}
	\end{equation}
	is a Lagrange polynomial of degree $M$. Noting that $l_{m}(\lambda_{k})=\delta_{km},$
	it follows that $P(\lambda_{k})=\langle A\rangle_{k}$ for $0\leq k\leq M$.
	Therefore, the noise-free expectation value is estimated by
	\begin{align}
		\langle A\rangle(0) & =P(0)\nonumber \\
		& =\sum_{m=0}^{M}\langle A\rangle_{m}l_{m}(0)\nonumber \\
		& =\sum_{m=0}^{M}\langle A\rangle_{m}\prod_{0\leq k\leq M,k\neq m}\frac{\lambda_{k}}{\lambda_{k}-\lambda_{m}}.\label{eq:S39.2}
	\end{align}
	
	Equation (\ref{eq:S39.2}) gives $\langle A\rangle(0)$ in terms of
	$\langle A\rangle_{k}$ and the noise strengths $\lambda_{k}$. One
	of the first techniques proposed to artificially increase the value
	of $\lambda$ is pulse stretching \cite{temme2017error},
	which involves pulse control from the user. In addition, we point
	out that pulse stretching also assumes that the noise is time-independent.
	Unitary folding is an alternative that does not require this level
	of control. If $\mathcal{U}$ describes the target ideal evolution,
	unitary folding operates by adding quantum gates to $\mathcal{U}$
	that in the noise-free case are just identity operations. This can
	be done either by using ``circuit foldings'' $\mathcal{U}^{\dagger}\mathcal{U}$,
	or by inserting products between gates and their own inverses. Noting
	that the polynomial (\ref{eq:S38 truncated expansion of N^-1}) contains
	powers of the (noisy) implementation of ${\normalcolor \mathcal{U}^{\dagger}\mathcal{U}}$,
	we are specifically interested in the connection between this polynomial
	and the use of circuit folding for ZNE, rather than
	folding at the level of gates. In this context, the assumption of
	linear scaling of the noise means that each power $\left(\mathcal{K}_{\textrm{I}}\mathcal{K}\right)^{k}$
	increases the noise characteristic of $\mathcal{K}$ by a factor of
	$2k$, i.e. that the noise increases proportionally to the depth of
	the circuit $\left(\mathcal{K}_{\textrm{I}}\mathcal{K}\right)^{k}$. If $\lambda_{0}$
	corresponds to the natural noise in the target circuit $\mathcal{K}$,
	then, the folding $\mathcal{K}\left(\mathcal{K}_{\textrm{I}}\mathcal{K}\right)^{k}$
	results in $\lambda_{k}=(2k+1)\lambda_{0}$. By substituting this
	expression of $\lambda_{k}$ into $\prod_{0\leq k\leq M,k\neq m}\frac{\lambda_{k}}{\lambda_{k}-\lambda_{m}}$,
	we obtain: 
	\begin{align}
		\prod_{0\leq k\leq M,k\neq m}\frac{\lambda_{k}}{\lambda_{k}-\lambda_{m}} & =\prod_{0\leq k\leq M,k\neq m}\frac{2k+1}{2(k-m)}\nonumber \\
		& =\frac{\prod_{0\leq k\leq M,k\neq m}(2k+1)}{2^{M}\prod_{0\leq k\leq M,k\neq m}(k-m)}\nonumber \\
		& =\frac{(2M+1)!!/(2m+1)}{2^{M}[(-m)(1-m)...(-1)][(1)(2)...(M-m)]}\nonumber \\
		& =(-1)^{m}\frac{(2M+1)!!}{2^{M}[(2m+1)m!(M-m)!]},\label{eq:S39.3}
	\end{align}
	which coincides with the coefficient $a_{\textrm{Tay},m}^{(M)}$ {[}cf. Eq.
	(\ref{eq:S39 coefficients of the truncated Taylor expansion}){]}. 
	
	Taking into account that $\langle A\rangle_{m}=\langle A|\mathcal{K}\left(\mathcal{K}_{\textrm{I}}\mathcal{K}\right)^{m}|\rho\rangle$,
	it follows that the application of the KIK formula in the weak noise
	limit reproduces the prediction of Richardson ZNE, with circuit folding
	and linear scaling of noise {[}cf. Eqs. (\ref{eq:S39.2}) and (\ref{eq:S39.3}){]}.
	As a final remark, it is worth noting that in circuit folding the
	realization of $\mathcal{U}^{\dagger}$ using circuit level of control
	does not involve modifying gates that are their own inverse. A fundamental
	example in this respect is the CNOT gate, which is the basic unit
	of two-qubit interactions. In contrast, the pulse
	inverse $\mathcal{H}_{\textrm{I}}(t)$ used in our method reverses also the
	schedules of the CNOT gates, to keep consistency with the pulse-based
	inverse $\mathcal{K}_{\textrm{I}}$ appearing in the KIK formula (\ref{eq:S15 KIK formula}). 
	
	In what follows, we also show that using $\mathcal{K}=\mathcal{K}_{\textrm{I}}$
	in the case of target circuits that are their own inverse introduces
	an additional error term that is absent when $\mathcal{K}_{\textrm{I}}$ is
	the pulse inverse. This further demonstrates the importance of the
	correct implementation of $\mathcal{K}_{\textrm{I}}$ for KIK QEM.\\
	\\
	\fakesubsection{Error induced by the circuit inverse $\mathcal{K}_{\textrm{I}}=\mathcal{K}$
		in circuits that satisfy $\mathcal{U}^{2}=\mathcal{I}$}
	\textbf{Error induced by the circuit inverse $\mathcal{K}_{\textrm{I}}=\mathcal{K}$
		in circuits that satisfy $\mathcal{U}^{2}=\mathcal{I}$.}
	Previously, we showed that the coefficients $a_{\textrm{Tay},m}^{(M)}$ reproduce
	Richardson ZNE when noise is linearly scaled through circuit folding.
	However, even in this limit of weak noise the KIK method provides
	insights that elude a naive application of circuit folding. An example
	of this was already given in Supplementary Note 3, by showing
	that, even though $\mathcal{K}\mathcal{K}_{\textrm{I}}$ and$\mathcal{K}_{\textrm{I}}\mathcal{K}$
	are equivalent in the noiseless scenario, the product $\mathcal{K}_{\textrm{I}}\mathcal{K}$
	is the correct choice for using the KIK formula (\ref{eq:S15 KIK formula}).
	This is valid in particular for Eq. (\ref{eq:S38 truncated expansion of N^-1}).
	
	Here, we will see that ignoring the pulse inverse $\mathcal{K}_{\textrm{I}}$
	also has negative consequences for QEM using Eq. (\ref{eq:S38 truncated expansion of N^-1}),
	which for simplicity we call ``Taylor mitigation''. Specifically,
	we consider circuits such that 
	\begin{equation}
		\mathcal{U}^{2}=\mathcal{I},\label{eq:S39.4 U^2=00003DI}
	\end{equation}
	which suggests the application of (\ref{eq:S38 truncated expansion of N^-1})
	with $\mathcal{K}_{\textrm{I}}=\mathcal{K}$. In this way, the $M$th order
	approximation to $\mathcal{U}$ (cf. Eq. (2) in the main text) reads
	\begin{align}
		\mathcal{U}_{\textrm{KIK}}^{(M)} & =\sum_{m=0}^{M}a_{\textrm{Tay},m}^{(M)}\mathcal{K}\left(\mathcal{K}\mathcal{K}\right)^{m}\nonumber \\
		& =\sum_{m=0}^{M}a_{\textrm{Tay},m}^{(M)}\left(\mathcal{K}\right)^{2m+1}\nonumber \\
		& =\sum_{m=0}^{M}a_{\textrm{Tay},m}^{(M)}\left(\mathcal{U}e^{\Omega_{1}}\right)^{2m+1},\label{eq:S39.5}
	\end{align}
	where $\mathcal{K}$ has been replaced by Eq. (\ref{eq:S23 K with Magnus expansion})
	and $\Omega_{1}=\Omega_{1}(T)$. 
	
	Next, we approximate $\mathcal{U}e^{\Omega_{1}}$ by $\mathcal{U}e^{\Omega_{1}}\approx\mathcal{U}+\mathcal{U}\Omega_{1}$,
	and keep only terms that are linear in $\Omega_{1}$ in $\mathcal{U}_{\textrm{KIK}}^{(M)}$.
	Using $\mathcal{U}^{2m+1}=\mathcal{U}$, we have that 
	
	\begin{equation}
		\left(\mathcal{U}+\mathcal{U}\Omega_{1}\right)^{2m+1}\approx\mathcal{U}+\sum_{k=1}^{2m+1}\mathcal{U}^{k}\Omega_{1}\mathcal{U}^{2m+1-k},\label{eq:S39.6}
	\end{equation}
	for $m\geq1$. Therefore, 
	\begin{align}
		\mathcal{U}_{\textrm{KIK}}^{(M)} & \approx a_{\textrm{Tay},0}^{(M)}\left(\mathcal{U}+\mathcal{U}\Omega_{1}\right)+\sum_{m=1}^{M}a_{\textrm{Tay},m}^{(M)}\left(\mathcal{U}e^{\Omega_{1}}\right)^{2m+1}\nonumber \\
		& \approx a_{\textrm{Tay},0}^{(M)}\left(\mathcal{U}+\mathcal{U}\Omega_{1}\right)+\sum_{m=1}^{M}a_{\textrm{Tay},m}^{(M)}\left[\mathcal{U}+\sum_{k=1}^{2m+1}\mathcal{U}^{k}\Omega_{1}\mathcal{U}^{2m+1-k}\right]\nonumber \\
		& =\mathcal{U}+a_{\textrm{Tay},0}^{(M)}\mathcal{U}\Omega_{1}+\sum_{m=1}^{M}a_{\textrm{Tay},m}^{(M)}\sum_{k=1}^{2m+1}\mathcal{U}^{k}\Omega_{1}\mathcal{U}^{2m+1-k},\label{eq:S39.7}
	\end{align}
	where we have applied $\sum_{m=0}^{M}a_{\textrm{Tay},m}^{(M)}=1$. 
	
	Now we divide the sum $\sum_{k=1}^{2m+1}\mathcal{U}^{k}\Omega_{1}\mathcal{U}^{2m+1-k}$
	into two sums such that one of them contains only even powers $\mathcal{U}^{k}$,
	and the other contains only odd powers $\mathcal{U}^{k}$. Taking
	into account that $\mathcal{U}^{-1}=\mathcal{U}$ and that any even
	power of $\mathcal{U}$ yields $\mathcal{I}$, we obtain
	\begin{align}
		\sum_{k=1}^{2m+1}\mathcal{U}^{k}\Omega_{1}\mathcal{U}^{2m+1-k} & =\sum_{k=1}^{m+1}\mathcal{U}^{2k-1}\Omega_{1}\mathcal{U}^{2m+2-2k}+\sum_{k=1}^{m}\mathcal{U}^{2k}\Omega_{1}\mathcal{U}^{2m+1-2k}\nonumber \\
		& =\sum_{k=1}^{m+1}\mathcal{U}\Omega_{1}+\sum_{k=1}^{m}\Omega_{1}\mathcal{U}\nonumber \\
		& =(m+1)\mathcal{U}\Omega_{1}+m\Omega_{1}\mathcal{U}.\label{eq:S39.8}
	\end{align}
	Substituting Eq. (\ref{eq:S39.8}) into Eq. (\ref{eq:S39.7}), we
	find that 
	\begin{align}
		\mathcal{U}_{\textrm{KIK}}^{(M)} & \approx\mathcal{U}+a_{\textrm{Tay},0}^{(M)}\mathcal{U}\Omega_{1}+\sum_{m=1}^{M}a_{\textrm{Tay},m}^{(M)}\left[(m+1)\mathcal{U}\Omega_{1}+m\Omega_{1}\mathcal{U}\right]\nonumber \\
		& =\mathcal{U}+a_\textrm{{Tay},0}^{(M)}\mathcal{U}\Omega_{1}+\sum_{m=0}^{M}a_{\textrm{Tay},m}^{(M)}\left[(m+1)\mathcal{U}\Omega_{1}+m\Omega_{1}\mathcal{U}\right]-a_{\textrm{Tay},0}^{(M)}\mathcal{U}\Omega_{1}\nonumber \\
		& =\mathcal{U}+\sum_{m=0}^{M}a_{\textrm{Tay},m}^{(M)}\left[(m+1)\mathcal{U}\Omega_{1}+m\Omega_{1}\mathcal{U}\right],\label{eq:S39.9}
	\end{align}
	where in the second line we added and subtracted the term corresponding
	to $m=0$ in the sum. 
	
	Finally, we use again $\sum_{m=0}^{M}a_{\textrm{Tay},m}^{(M)}=1$, and $\sum_{m=0}^{M}a_{\textrm{Tay},m}^{(M)}m=-\frac{1}{2}$.
	In this way, for any $M\geq1$,
	\begin{equation}
		\mathcal{U}_{\textrm{KIK}}^{(M)}\approx\mathcal{U}+\frac{1}{2}[\mathcal{U},\Omega_{1}].\label{eq:S39.10}
	\end{equation}
	Because Taylor mitigation corresponds to weak noise, in the limit
	when $M$ tends to infinity $\mathcal{U}_{\textrm{KIK}}^{(M)}$ should converge
	to the evolution $\mathcal{U}_{\textrm{KIK}}$ in the KIK formula (\ref{eq:S15 KIK formula}).
	By construction, this is the case if $\mathcal{K}_{\textrm{I}}$ is given by
	the pulse inverse. However, we see from Eq. (\ref{eq:S39.10}) that
	when $\mathcal{K}_{\textrm{I}}=\mathcal{K}$ the term $\frac{1}{2}[\mathcal{U},\Omega_{1}]$
	remains in the approximation $\mathcal{U}_{\textrm{KIK}}^{(M)}$, irrespective
	of the mitigation order $M$. Note also that this term cannot be avoided
	in general by considering higher-order (nonlinear) contributions in
	$\Omega_{1}$. In this way, Eq. (\ref{eq:S39.10}) shows that using
	$\mathcal{K}_{\textrm{I}}=\mathcal{K}$ instead of the pulse inverse leads
	to an inconsistent application of QEM, which is afflicted by an additional
	error term $\frac{1}{2}[\mathcal{U},\Omega_{1}]$. 
	
	Finally, we remark that for global depolarizing noise the associated
	noise channel commutes with any unitary $\mathcal{U}$, and one can
	easily check that in this case the KIK formula (\ref{eq:S15 KIK formula})
	is exact when $\mathcal{K}_{\textrm{I}}=\mathcal{K}$. Thus, while the pulse-based
	inverse $\mathcal{K}_{\textrm{I}}$ becomes unnecessary for such a simplified
	noise model \cite{giurgica2020digital}, it is of paramount importance
	in practical applications under realistic noise. \\
	\\
	\fakesubsection{Coefficients for Adaptive QEM based on the KIK formula, for mitigation orders $M=1,2,3$}
	\textbf{Coefficients for Adaptive QEM based on the KIK formula, for mitigation orders $M=1,2,3$.} To obtain the coefficients $a_{\textrm{Adap},m}^{(M)}$, we minimize the quantity
	\begin{equation}
		\varepsilon_{\textrm{L2}}^{(M)}:=\int_{g(\mu)}^{1}\left(\sum_{m=0}^{M}a_{m}^{(M)}\lambda^{m}-\lambda^{-\frac{1}{2}}\right)^{2}d\lambda\label{eq:S40 L2-norm error for adapted coefficients}
	\end{equation}
	with respect the first $M$ coefficients $\left\{ a_{m}^{(M)}\right\} _{m=0}^{M-1}$.
	Therefore, we have to solve the equations 
	\begin{equation}
		\frac{\partial\varepsilon_{\textrm{L2}}^{(M)}}{\partial a_{m}^{(M)}}=0,\quad\textrm{for }0\leq m\leq M-1.\label{eq:S40.1 derivatives of e_L2^(M) w.r.t. a_m^(M)}
	\end{equation}
	The $M$th coefficient is obtained by imposing the constraint $a_{\textrm{Adap},M}^{(M)}=1-\sum_{m=0}^{M-1}a_{\textrm{Adap},m}^{(M)}$.
	This is equivalent to the normalization condition $\sum_{m=0}^{M}a_{\textrm{Adap},m}^{(M)}=1$,
	which ensures that the map ${\normalcolor {\color{red}{\normalcolor \mathcal{U}_{\textrm{KIK}}^{(M)}=}}}\sum_{m=0}^{M}a_{\textrm{Adap},m}^{(M)}[g(\mu)]\mathcal{K}\left(\mathcal{K}_{\textrm{I}}\mathcal{K}\right)^{m}$
	is trace-preserving if $\mathcal{K}$ and $\mathcal{K}_{\textrm{I}}$ are trace-preserving.
	For the sake of notational simplicity, in the following we will write
	$g(\mu)$ as $g$. 
	
	By explicitly evaluating $\varepsilon_{\textrm{L2}}^{(M)}$ with $M=1,2,3$,
	we find that 
	\begin{equation}
		\varepsilon_{\textrm{L2}}^{(1)}=(1-g)\left(a_{0}^{(1)}\right)^{2}+\frac{1}{3}(1-g^{3})\left(a_{1}^{(1)}\right)^{2}+(1-g^{2})a_{0}^{(1)}a_{1}^{(1)}-4(1-g^{\frac{1}{2}})a_{0}^{(1)}-\frac{4}{3}(1-g^{\frac{3}{2}})a_{1}^{(1)}-\textrm{ln}(g),\label{eq:S41 L2 norm error for M=00003D1}
	\end{equation}
	\begin{align}
		\varepsilon_{\textrm{L2}}^{(2)} & =(1-g)\left(a_{0}^{(2)}\right)^{2}+\frac{1}{3}(1-g^{3})\left(a_{1}^{(2)}\right)^{2}+\frac{1}{5}(1-g^{5})\left(a_{2}^{(2)}\right)^{2}+(1-g^{2})a_{0}^{(2)}a_{1}^{(2)}+\frac{2}{3}(1-g^{3})a_{0}^{(2)}a_{2}^{(2)}\nonumber \\
		& \quad+\frac{1}{2}(1-g^{4})a_{1}^{(2)}a_{2}^{(2)}-4(1-g^{\frac{1}{2}})a_{0}^{(2)}-\frac{4}{3}(1-g^{\frac{3}{2}})a_{1}^{(2)}-\frac{4}{5}(1-g^{\frac{5}{2}})a_{2}^{(2)}-\textrm{ln}(g),\label{eq:S42 L2 norm error for M=00003D2}
	\end{align}
	
	\begin{align}
		\varepsilon_{\textrm{L2}}^{(3)} & =(1-g)\left(a_{0}^{(3)}\right)^{2}+\frac{1}{3}(1-g^{3})\left(a_{1}^{(3)}\right)^{2}+\frac{1}{5}(1-g^{5})\left(a_{2}^{(3)}\right)^{2}+\frac{1}{7}(1-g^{7})\left(a_{3}^{(3)}\right)^{2}\nonumber \\
		& \quad+(1-g^{2})a_{0}^{(3)}a_{1}^{(3)}+\frac{2}{3}(1-g^{3})a_{0}^{(3)}a_{2}^{(3)}+\frac{1}{2}(1-g^{4})a_{0}^{(3)}a_{3}^{(3)}+\frac{1}{2}(1-g^{4})a_{1}^{(3)}a_{2}^{(3)}+\frac{2}{5}(1-g^{5})a_{1}^{(3)}a_{3}^{(3)}\nonumber \\
		& \quad+\frac{1}{3}(1-g^{6})a_{2}^{(3)}a_{3}^{(3)}-4(1-g^{\frac{1}{2}})a_{0}^{(3)}-\frac{4}{3}(1-g^{\frac{3}{2}})a_{1}^{(3)}-\frac{4}{5}(1-g^{\frac{5}{2}})a_{2}^{(3)}-\frac{4}{7}(1-g^{\frac{7}{2}})a_{3}^{(3)}-\textrm{ln}(g).\label{eq:S43 L2 norm error for M=00003D3}
	\end{align}
	By taking the partial derivatives (\ref{eq:S40.1 derivatives of e_L2^(M) w.r.t. a_m^(M)})
	and solving the resulting linear equations, the corresponding coefficients
	read 
	
	\begin{align}
		a_{\textrm{Adap},0}^{(1)} & =1+\frac{1}{(1+\sqrt{g})^{3}}+\frac{3}{2(1+\sqrt{g})^{2}},\label{eq:S44 optimal coeff a_0^(1)}\\
		a_{\textrm{Adap},1}^{(1)} & =-\frac{5+3\sqrt{g}}{2(1+\sqrt{g})^{3}},\label{eq:S45 optimal coeff a_1^(1)}
	\end{align}
	for $M=1,$ 
	\begin{align}
		a_{\textrm{Adap},0}^{(2)} & =1+\frac{16}{3(1+\sqrt{g})^{5}}-\frac{14}{3(1+\sqrt{g})^{4}}+\frac{4}{(1+\sqrt{g})^{2}},\label{eq:S46 optimal coeff a_0^(2)}\\
		a_{\textrm{Adap},1}^{(2)} & =-4\frac{10+8\sqrt{g}+9g+3g^{\frac{3}{2}}}{3(1+\sqrt{g})^{5}},\label{eq:S47 optimal coeff a_1^(2)}\\
		a_{\textrm{Adap},2}^{(2)} & =2\frac{13+5\sqrt{g}}{3(1+\sqrt{g})^{5}},\label{eq:S48 optimal coeff a_2^(2)}
	\end{align}
	for $M=2,$ and 
	\begin{align}
		a_{\textrm{Adap},0}^{(3)} & =\frac{31+97\sqrt{g}+276g+300g^{\frac{3}{2}}+270g^{2}+114g^{\frac{5}{2}}+28g^{3}+4g^{\frac{7}{2}}}{4(1+\sqrt{g})^{7}},\label{eq:S51.1 optimal coeff a_0^(3)-1}\\
		a_{\textrm{Adap},1}^{(3)} & =-5\frac{29+35\sqrt{g}+84g+44g^{\frac{3}{2}}+26g^{2}+6g^{\frac{5}{2}}}{4(1+\sqrt{g})^{7}},\label{eq:S51.2 optimal coeff a_1^(3)-1}\\
		a_{\textrm{Adap},2}^{(3)} & =3\frac{81+47\sqrt{g}+76g+20g^{\frac{3}{2}}}{4(1+\sqrt{g})^{7}},\label{eq:S51.3 optimal coeff a_2^(3)-1}\\
		a_{\textrm{Adap},3}^{(3)} & =-5\frac{25+7\sqrt{g}}{4(1+\sqrt{g})^{7}},\label{eq:S51.4 optimal coeff a_3^(3)-1}
	\end{align}
	for $M=3$. 
	
	Let us now check that the obtained coefficients minimize $\varepsilon_{\textrm{L2}}^{(M)}$
	in the subspace determined by the variables $\left\{ a_{m}^{(M)}\right\} _{m=0}^{M-1}$.
	For $M=1$, the second derivative $\frac{\partial^{2}\varepsilon_{\textrm{L2}}^{(1)}}{\partial a_{0}^{(1)2}}=2\left(1-g\right)$
	is positive if $g\leq1$, and therefore $a_{\textrm{Adap},0}^{(1)}$ minimizes
	$\varepsilon_{\textrm{L2}}^{(1)}$ with respect to $a_{0}^{(1)}$.
	To see if $a_{\textrm{Adap},0}^{(2)}$ and $a_{\textrm{Adap},1}^{(2)}$ in Eqs. (\ref{eq:S44 optimal coeff a_0^(1)})
	and (\ref{eq:S45 optimal coeff a_1^(1)}) minimize $\varepsilon_{\textrm{L2}}^{(2)}$,
	we evaluate the determinant of the Hessian matrix of $\varepsilon_{\textrm{L2}}^{(2)}$
	and the second partial derivative $\frac{\partial^{2}\varepsilon_{\textrm{L2}}^{(2)}}{\partial a_{0}^{(2)2}}$,
	and check their positivity. Since $\frac{\partial^{2}\varepsilon_{\textrm{L2}}^{(2)}}{\partial a_{0}^{(2)2}}=2(1-g)\geq0$,
	we only need to check the determinant of the Hessian matrix 
	\begin{equation}
		H=\left(\begin{array}{cc}
			\frac{\partial^{2}\varepsilon_{\textrm{L2}}^{(2)}}{\partial a_{0}^{(2)2}} & \frac{\partial^{2}\varepsilon_{\textrm{L2}}^{(2)}}{\partial a_{0}^{(2)}\partial a_{1}^{(2)}}\\
			\frac{\partial^{2}\varepsilon_{\textrm{L2}}^{(2)}}{\partial a_{1}^{(2)}\partial a_{0}^{(2)}} & \frac{\partial^{2}\varepsilon_{\textrm{L2}}^{(2)}}{\partial a_{1}^{(2)2}}
		\end{array}\right)=\left(\begin{array}{cc}
			2\left(1-g\right) & 1-g^{2}\\
			1-g^{2} & \frac{2}{3}(1-g^{3})
		\end{array}\right).\label{eq:S51.5 Hessian matrix for M=00003D2}
	\end{equation}
	Such a determinant is given by $\textrm{det}(H)=\frac{4}{3}\left[1-g\right]\left[1-g^{3}\right]-\left[1-g^{2}\right]^{2}$,
	which is also positive in the interval $0\leq g\leq1$. Finally, for
	$M=3$ we obtain the Hessian matrix 
	\begin{equation}
		H=\left(\begin{array}{ccc}
			\frac{\partial^{2}\varepsilon_{\textrm{L2}}^{(3)}}{\partial a_{0}^{(3)2}} & \frac{\partial^{2}\varepsilon_{\textrm{L2}}^{(3)}}{\partial a_{0}^{(3)}\partial a_{1}^{(3)}} & \frac{\partial^{2}\varepsilon_{\textrm{L2}}^{(3)}}{\partial a_{0}^{(3)}\partial a_{2}^{(3)}}\\
			\frac{\partial^{2}\varepsilon_{\textrm{L2}}^{(3)}}{\partial a_{1}^{(3)}\partial a_{0}^{(3)}} & \frac{\partial^{2}\varepsilon_{\textrm{L2}}^{(3)}}{\partial a_{1}^{(3)2}} & \frac{\partial^{2}\varepsilon_{\textrm{L2}}^{(3)}}{\partial a_{1}^{(3)}\partial a_{2}^{(3)}}\\
			\frac{\partial^{2}\varepsilon_{\textrm{L2}}^{(3)}}{\partial a_{2}^{(3)}\partial a_{0}^{(3)}} & \frac{\partial^{2}\varepsilon_{\textrm{L2}}^{(3)}}{\partial a_{2}^{(3)}\partial a_{1}^{(3)}} & \frac{\partial^{2}\varepsilon_{\textrm{L2}}^{(3)}}{\partial a_{2}^{(3)2}}
		\end{array}\right)=\left(\begin{array}{ccc}
			2(1-g) & 1-g^{2} & \frac{2}{3}(1-g^{3})\\
			1-g^{2} & \frac{2}{3}(1-g^{3}) & \frac{1}{2}(1-g^{4})\\
			\frac{2}{3}(1-g^{3}) & \frac{1}{2}(1-g^{4}) & \frac{2}{5}(1-g^{5})
		\end{array}\right).\label{eq:S51.6 Hessian matrix for M=00003D3}
	\end{equation}
	The eigenvalues of this matrix are plotted in Supplementary Figure 4. Since all of them are positive in the interval $0\leq g\leq1$,
	we conclude that Eqs. (\ref{eq:S51.1 optimal coeff a_0^(3)-1})-(\ref{eq:S51.4 optimal coeff a_3^(3)-1})
	provide the coefficients $a_{\textrm{Adap},m}^{(3)}$ that minimize $\varepsilon_{\textrm{L2}}^{(3)}$
	with respect to $\left\{ a_{m}^{(M)}\right\} _{m=0}^{2}$. 
	\begin{figure}
		\centering{}\includegraphics[scale=0.5]{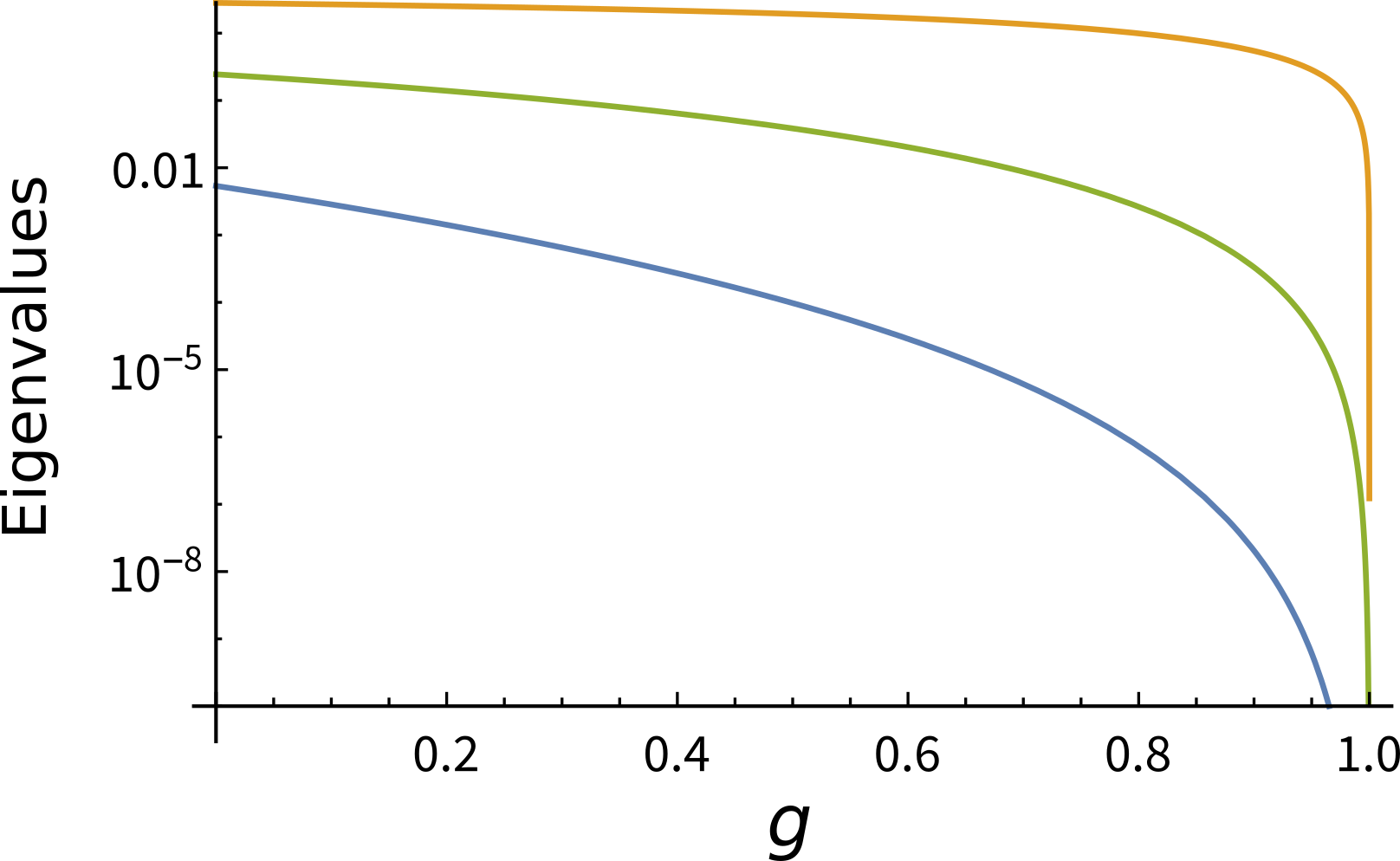}\caption{Eigenvalues of the Hessian matrix (\ref{eq:S51.6 Hessian matrix for M=00003D3}),
			as a function of $g=g(\mu)$. }
	\end{figure}

	\section*{Supplementary Note 5: KIK QEM applied to the transverse Ising model }
	
	In this supplementary note, we simulate the application of the KIK method to
	a noisy implementation of the transverse Ising model. This model is
	characterized by the Hamiltonian 
	\begin{align}
		H & =g\sum_{i=1}^{n}X_{i}+J\sum_{i=1}^{n-1}Z_{i}\otimes Z_{i+1}\nonumber \\
		& :=gH_{X}+JH_{ZZ}\label{eq:1}
	\end{align}
	where $H_{ZZ}$ accounts for the magnetic interactions between nearest-neighbor
	spins, and the transverse magnetic fields are represented by the local
	term $H_{X}$. We consider $n=5$ spins and set $g=0.2$ and $J=0.1$.
	There is no particular reason for choosing these parameter values.
	Similar performance was obtained with simulations using alternative
	values (not shown here). 
	
	For the simulation, we assume that the target evolution $U=e^{-iHT}$
	is implemented in a quantum computer via Trotterization \cite{trotter1959product}.
	Taking into account that the $X$ and $Z$ Pauli matrices do not commute,
	an approximation to $U$ involving $n$ Trotter steps has the form
	\begin{equation}
		U\approx U_{n}:=\left(e^{-igH_{X}\frac{T}{n}}e^{-iJH_{ZZ}\frac{T}{n}}\right)^{n},\label{eq:2}
	\end{equation}
	where $T$ is the total evolution time. Assuming $T=10$ and 10 Trotter
	steps, we have that 
	\begin{align}
		U & \approx U_{10}\nonumber \\
		& :=\left(e^{-igH_{X}}e^{-iJH_{ZZ}}\right)^{10}.\label{eq:3}
	\end{align}
	
	In what follows, $I_{j}$ will denote the identity matrix of dimension
	$2^{j}\times2^{j}$, with $I=I_{n}$. We remark that our goal is not
	to test the accuracy of the Trotter approximation (\ref{eq:3}), but
	to study the performance of the KIK method to mitigate errors in a
	noisy implementation of $U_{10}$. To this end, we model the noisy
	evolution $\mathcal{K}$ associated with $U_{10}$ using
	\begin{align}
		\mathcal{K} & =\left(e^{-ig\mathcal{H}_{X}+\xi\mathcal{L}}e^{-iJ\mathcal{H}_{ZZ}+\xi\mathcal{L}}\right)^{10},\label{eq:4}\\
		\mathcal{H}_{X} & =H_{X}\otimes I-I\otimes\left(H_{X}\right)^{\textrm{T}},\label{eq:5}\\
		\mathcal{H}_{ZZ} & =H_{ZZ}\otimes I-I\otimes\left(H_{ZZ}\right)^{\textrm{T}},\label{eq:6}
	\end{align}
	where 
	\begin{align}
		\mathcal{{\normalcolor L}} & {\normalcolor =S\otimes S^{\ast}-\frac{1}{2}S^{\dagger}S\otimes I-\frac{1}{2}I\otimes\left(S^{\dagger}S\right)^{\textrm{T}},}\label{eq:7}\\
		S & =0.5S_{1}+1.7S_{2}+0.3S_{3}+2S_{4}+S_{5},\label{eq:8}\\
		S_{j} & ={\color{red}{\normalcolor I_{j-1}}}\otimes A\otimes{\color{red}{\normalcolor I_{n-j}}},\textrm{ for }1\leq j\leq5,\label{eq:9}\\
		A & {\color{blue}{\normalcolor =\left(\begin{array}{cc}
					0 & 1\\
					0 & 0
				\end{array}\right)}.}\label{eq:10}
	\end{align}
	The initial state in our simulation is the ground state $\rho=|0\rangle\langle0|^{\otimes5}$.\\
	\\
	\fakesubsection{Error mitigation}
	\textbf{Error mitigation.} To assess the performance of the KIK QEM, we use the fidelity between
	\begin{figure}
		\centering{}\includegraphics[scale=0.55]{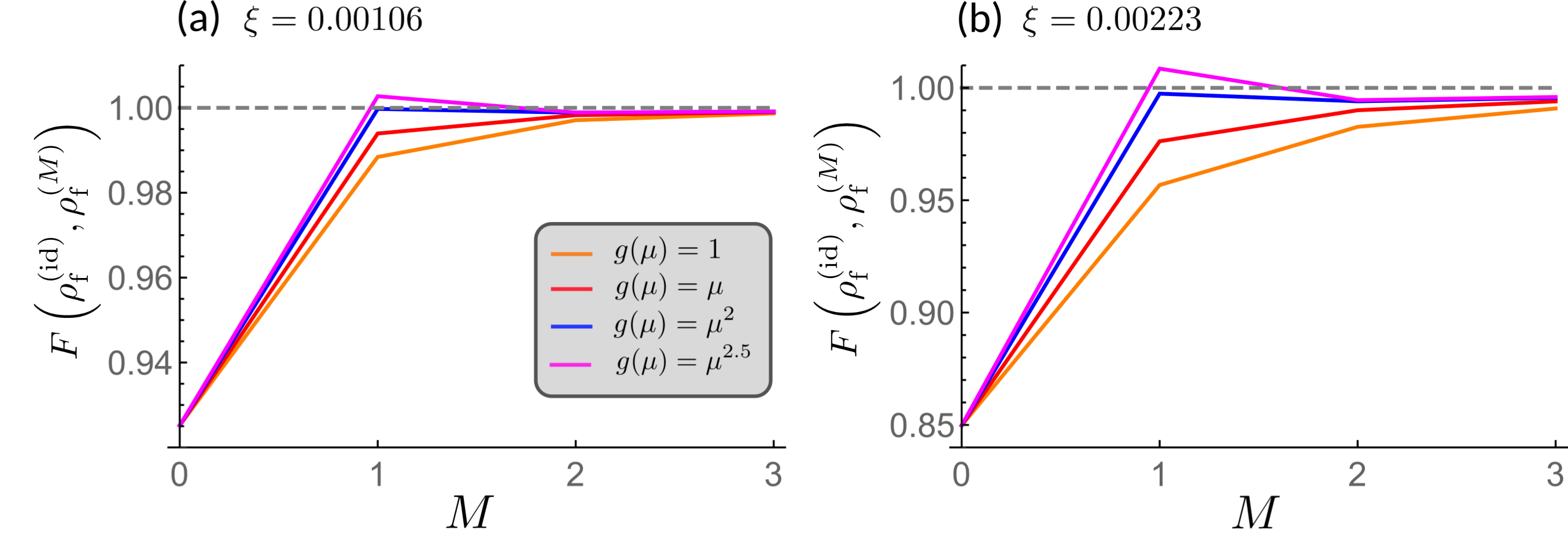}\caption{Error-mitigated fidelity $F\left(\rho_{\textrm{f}}^{(\textrm{id})},\rho_{\textrm{f}}^{(M)}\right)$,
			between the ideal final state $|\rho_{\textrm{f}}^{(\textrm{id})}\rangle=\mathcal{U}_{10}|\rho\rangle$
			and the error-mitigated final state $|\rho_{\textrm{f}}^{(M)}\rangle:=\mathcal{U}_{\textrm{KIK}}^{(M)}|\rho\rangle$.
			Each curve is obtained by evaluating the coefficients $a_{\textrm{Adap},m}^{(M)}[g(\mu)]$
			in the $M$th-order approximation ${\normalcolor {\color{red}{\normalcolor \mathcal{U}_{\textrm{KIK}}^{(M)}=}}}\sum_{m=0}^{M}a_{\textrm{Adap},m}^{(M)}[g(\mu)]\mathcal{K}\left(\mathcal{K}_{\textrm{I}}\mathcal{K}\right)^{m}$
			at the functions $g(\mu)$ indicated in Supplementary Figure 5(a).
			Supplementary Figures 5(a) and 5(b) stand respectively
			for $\xi=0.00106$ and $\xi=0.00223$, and both are characterized
			by the color code of Supplementary Figure 5(a). }
	\end{figure} 
	the ideal final state $|\rho_{\textrm{f}}^{(\textrm{id})}\rangle:=\mathcal{U}_{10}|\rho\rangle$
	(being $\mathcal{U}_{10}$ the Liouville-space representation of $U_{10}$)
	and the error-mitigated final state $|\rho_{\textrm{f}}^{(M)}\rangle:=\mathcal{U}_{\textrm{KIK}}^{(M)}|\rho\rangle$,
	which reads \cite{nielsen2002quantum}
	\begin{equation}
		F\left(\rho_{\textrm{f}}^{(\textrm{id})},\rho_{\textrm{f}}^{(M)}\right)=\left[\textrm{Tr}\sqrt{\sqrt{\rho_{\textrm{f}}^{(\textrm{id})}}\rho_{\textrm{f}}^{(M)}\sqrt{\rho_{\textrm{f}}^{(\textrm{id})}}}\right]^{2}.\label{eq:11}
	\end{equation}
	Notice that the fidelity must be computed using the density matrices
	$\rho_{\textrm{f}}^{(\textrm{id})}$ and $\rho_{\textrm{f}}^{(M)}$, and not their vector
	representations $|\rho_{\textrm{f}}^{(\textrm{id})}\rangle$ and $|\rho_{\textrm{f}}^{(M)}\rangle$.
	We consider values of $\xi$ given by $\xi=0.00223$ and $\xi=0.00106$,
	which lead to fidelities of unmitigated final states ($M=0$) 
	\begin{align}
		F\left(\rho_{\textrm{f}}^{(\textrm{id})},\rho_{\textrm{f}}^{(0)}\right) & =0.85,\textrm{ for }\xi=0.00223,\label{eq:12}\\
		F\left(\rho_{\textrm{f}}^{(\textrm{id})},\rho_{\textrm{f}}^{(0)}\right) & =0.925,\textrm{ for }\xi=0.00106.\label{eq:13}
	\end{align}
	
	QEM is performed by choosing functions $\{g(\mu)\}=\{1,\mu,\mu^{2},\mu^{2.5}\}$,
	and $1\leq M\leq3$. We recall that the function $g(\mu)$ determines
	the lower limit of integration in $\varepsilon_{\textrm{L2}}^{(M)}$
	(see the main text for the definition), and that for each $g(\mu)$
	QEM is applied with ${\normalcolor {\color{red}{\normalcolor \mathcal{U}_{\textrm{KIK}}^{(M)}=}}}\sum_{m=0}^{M}a_{\textrm{Adap},m}^{(M)}[g(\mu)]\mathcal{K}\left(\mathcal{K}_{\textrm{I}}\mathcal{K}\right)^{m}$,
	using coefficients $a_{\textrm{Adap},m}^{(M)}[g(\mu)]$ evaluated at $g=g(\mu)$
	(cf. Supplementary Note 4). In particular, $g(\mu)=1$ corresponds
	to Taylor mitigation. Keeping in mind Eq. (\ref{eq:4}), the inverse
	$\mathcal{K}_{\textrm{I}}$ is given by 
	\begin{equation}
		\mathcal{K}_{\textrm{I}}=\left(e^{iJ\mathcal{H}_{ZZ}+\xi\mathcal{L}}e^{ig\mathcal{H}_{X}+\xi\mathcal{L}}\right)^{10}.\label{eq:14}
	\end{equation}
	
	Supplementary Figure 5 shows the fidelity $F\left(\rho_{\textrm{f}}^{(\textrm{id})},\rho_{\textrm{f}}^{(M)}\right)$
	as a function of $M$, for $\xi=0.00106$ and $\xi=0.00223$. Overall,
	we observe that the best performance among the tested functions $g(\mu)$
	is achieved by $g(\mu)=\mu^{2}$ (blue curve). In particular, $F\left(\rho_{\textrm{f}}^{(\textrm{id})},\rho_{\textrm{f}}^{(M)}\right)$
	reaches a value extremely close to one already for $M=1$. A clearer
	comparison between the different functions $g(\mu)$ is possible by
	looking at the ratio between the infidelity before QEM and after QEM,
	\begin{equation}
		r_{F}(M)=\frac{1-F\left(\rho_{\textrm{f}}^{(\textrm{id})},\rho_{\textrm{f}}^{(0)}\right)}{1-F\left(\rho_{\textrm{f}}^{(\textrm{id})},\rho_{\textrm{f}}^{(M)}\right)},\label{eq:15}
	\end{equation}
	which quantifies the infidelity suppression provided by the KIK method.
	This quantity is plotted in Supplementary Figure 6. In this figure,
	we see that $g(\mu)=\mu^{2}$ outperforms $g(\mu)=1$ and $g(\mu)=\mu$
	for all $1\leq M\leq3$. Although the ratio $r_{F}(M)$ corresponding
	to $g(\mu)=\mu^{2}$ is slightly below that associated with $g(\mu)=\mu^{2.5}$,
	in the case of $M=2,3$, $g(\mu)=\mu^{2}$ yields a substantially
	larger $r_{F}(M)$ if $M=1$.
	
	\begin{figure}
		\centering{}\includegraphics[scale=0.55]{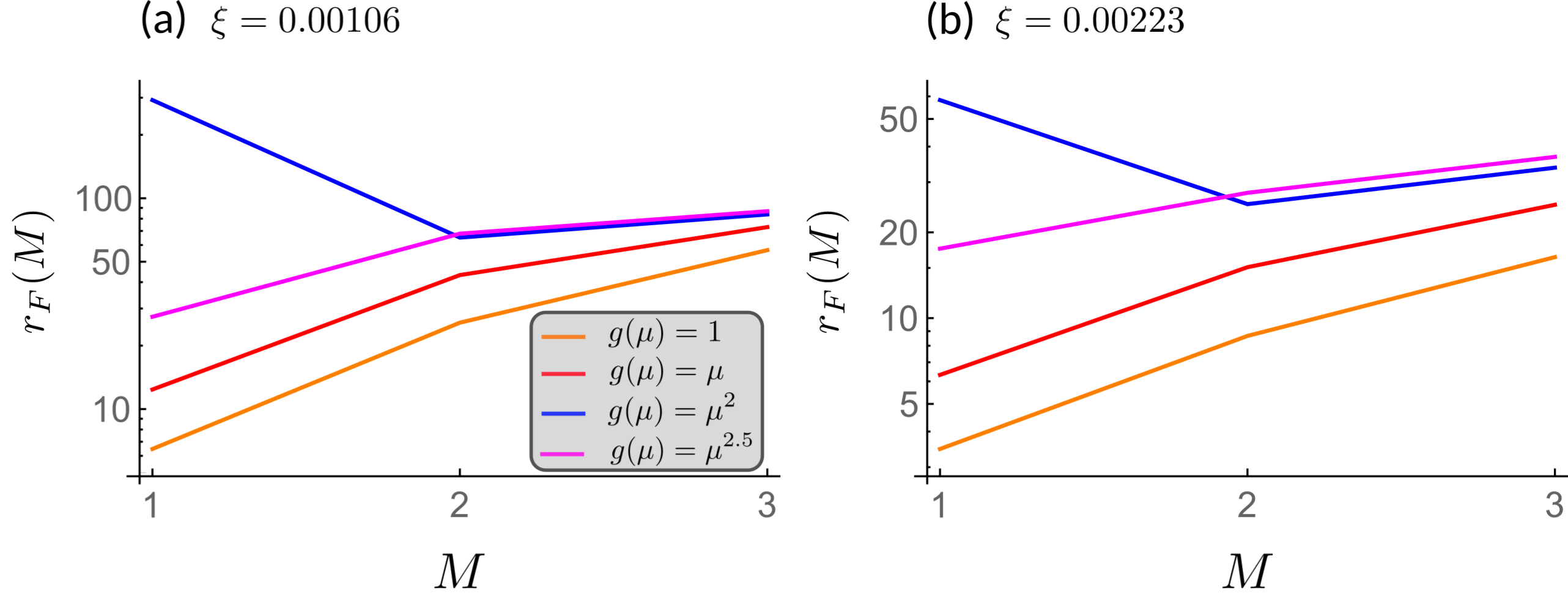}\caption{Enhancement ratio (\ref{eq:15}), which quantifies the infidelity
			suppression obtained by the KIK method. The color code in Supplementary Figure 6(b)
			is the same as in 6(a). }
	\end{figure}
	
	Finally, it is interesting to observe how the different functions
	$g(\mu)$ produce physically consistent fidelities $F\left(\rho_{\textrm{f}}^{(\textrm{id})},\rho_{\textrm{f}}^{(M)}\right)\leq1$
	as $M$ increases. In particular, we can see that the unphysical fidelity
	$F\left(\rho_{\textrm{f}}^{(\textrm{id})},\rho_{\textrm{f}}^{(1)}\right)>1$ corresponding
	to $g(\mu)=\mu^{2.5}$ is quickly fixed by going to the next mitigation
	order $M=2$. The explanation for this behavior is as follows. The
	quality of the approximation $\sum_{m=0}^{M}a_{\textrm{Adap},m}^{(M)}[g(\mu)]\left(\mathcal{K}_{\textrm{I}}\mathcal{K}\right)^{m}$
	to $\left(\mathcal{K}_{\textrm{I}}\mathcal{K}\right)^{-1/2}$ is determined
	by how well the polynomial $\sum_{m=0}^{M}a_{\textrm{Adap},m}^{(M)}[g(\mu)]\lambda^{m}$
	approximates the function $\lambda^{-1/2}$, where $\lambda$ is an
	eigenvalue of $\mathcal{K}_{\textrm{I}}\mathcal{K}$. If $g(\mu)$ is too small,
	as compared to the smallest eigenvalue of $\left(\mathcal{K}_{\textrm{I}}\mathcal{K}\right)^{-1/2}$,
	a polynomial $\sum_{m=0}^{M}a_{\textrm{Adap},m}^{(M)}[g(\mu)]\left(\mathcal{K}_{\textrm{I}}\mathcal{K}\right)^{m}$
	with low $M$ may be a rough approximation to $\left(\mathcal{K}_{\textrm{I}}\mathcal{K}\right)^{-1/2}$.
	This leads to undesired effects such as the aforementioned fidelity
	$F\left(\rho_{\textrm{f}}^{(\textrm{id})},\rho_{\textrm{f}}^{(1)}\right)>1$ (note that
	in our example the function $g(\mu)=\mu^{2.5}$ yields the smallest
	$g(\mu)$ for any value of $\mu$). However, by increasing $M$ we
	can always improve the approximation in the whole interval $(g(\mu),1)$,
	which by assumption contains all the eigenvalues of $\left(\mathcal{K}_{\textrm{I}}\mathcal{K}\right)^{-1/2}$.
	This would explain not only the recovery of a physical fidelity but
	also that for $M=2$ and $M=3$ the maximum values of this quantity
	correspond to $g(\mu)=\mu^{2.5}$. 
	
	On the contrary, when $g(\mu)$ is larger than the smallest eigenvalue
	of $\left(\mathcal{K}_{\textrm{I}}\mathcal{K}\right)^{-1/2}$, the interval
	$(g(\mu),1)$ does not contain all the eigenvalues of $\left(\mathcal{K}_{\textrm{I}}\mathcal{K}\right)^{-1/2}$
	and increasing $M$ does not necessarily improves the QEM. This is
	likely the reason that $g(\mu)=\mu^{2}$ yields fidelities smaller
	for $M=2$ and $M=3$, as compared to $M=1$. 
	
	\section*{Supplementary Note 6: Description of the experimental procedures }
	
	Here, we provide additional details and complementary information
	concerning the experiments described in the main text. In the first subsection we describe techniques to tackle different classes
	of errors, which complement the KIK error mitigation. Next, we explain in detail the steps to estimate
	error-mitigated expectation values, as well as the corresponding uncertainties.
	Then, information specific to the CNOT calibration experiment and
	the ten-swap experiment is given in the following two subsections. We conclude in the last subsection 
	with a simulated calibration of the CNOT gate. This simulation shows
	results very similar to those of the CNOT calibration experiment,
	and provides additional evidence of the usefulness of KIK-based calibration.\\
	\\
	\fakesubsection{Integration with complementary error mitigation methods}
	\textbf{Integration with complementary error mitigation methods.} We consider three complementary techniques for addressing errors that
	can affect an experiment in the three stages of its execution. Namely,
	the preparation of the initial state, the evolution, and the measurement. 
	
	In the IBM quantum processors the preparation of the single-qubit
	(default) computational state $|0\rangle$ may possess a small deviation angle $\delta\theta$ \cite{landa2022experimental}.
	As a result, the prepared state is given by $|\psi(\delta\theta)\rangle=\textrm{cos}\left(\frac{\delta\theta}{2}\right)|0\rangle+e^{i\varphi}\textrm{sin}\left(\frac{\delta\theta}{2}\right)|1\rangle$.
	To cope with this error, we apply rotations $R_{Z}(\pm\pi/2)=e^{\mp i\frac{\pi}{4}Z}$
	on each qubit, in such a way that $R_{Z}(\pi/2)$ and $R_{Z}(-\pi/2)$
	are equally distributed among the number of shots. This produces a
	mixed state 
	\begin{equation}
		\varrho=\textrm{cos}^{2}\left(\frac{\delta\theta}{2}\right)|0\rangle\langle0|+\textrm{sin}^{2}\left(\frac{\delta\theta}{2}\right)|1\rangle\langle1|,\label{eq:16 rho single-qubit dephased}
	\end{equation}
	whose population $\textrm{Tr}\left(|0\rangle\langle0|\varrho\right)=\textrm{cos}^{2}\left(\frac{\delta\theta}{2}\right)$
	coincides with $|\langle0|\psi(\delta\theta)\rangle|^{2}$. However,
	the coherent error associated with the rotation angles $\delta\theta$
	and $\varphi$ is eliminated through this procedure. Since both experiments
	involve circuits acting on two qubits, there are four possible rotation
	operations characterized by the pairs of angles $\{(\pm\pi/2,\pm\pi/2)\}$.
	This is illustrated by the green squares in Supplementary Figure 7(a).
	\begin{figure}
		\centering{}\includegraphics[scale=0.6]{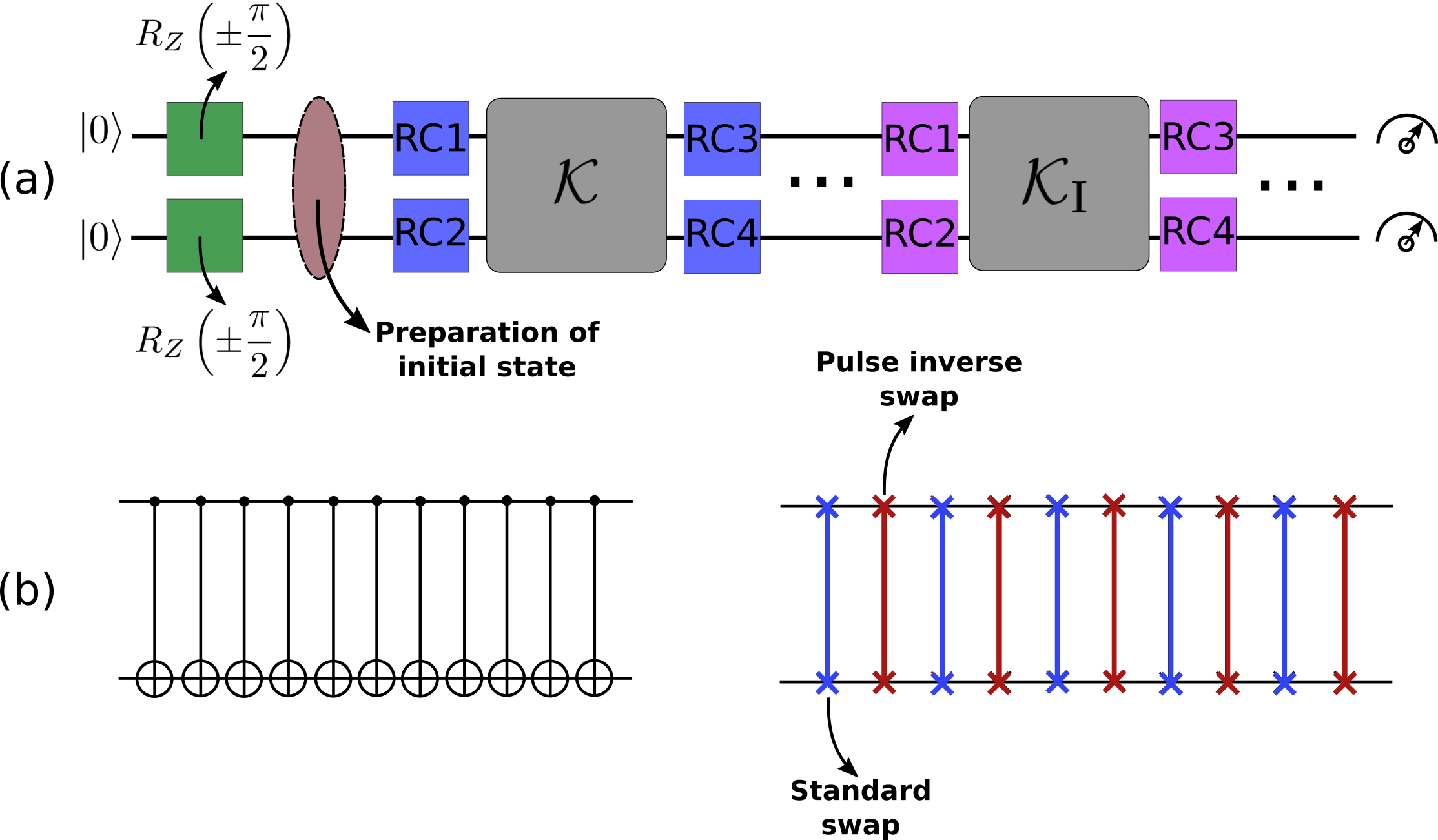}\caption{Schematic of the QEM circuits used in the experiments. (a) General
			form of a KIK circuit $\mathcal{K}\left(\mathcal{K}_{\textrm{I}}\mathcal{K}\right)^{m}$.
			For both experiments, initial rotations $R_{Z}(\pm\pi/2)$ (green
			squares) are applied on each qubit to mitigate a potential coherent
			error affecting the state $|00\rangle$. After that, the initial state
			for the CNOT calibration experiment is prepared (this preparation
			is not part of the ten-swap experiment), and the evolutions $\mathcal{K}$
			and $\mathcal{K}_{\textrm{I}}$ are interleaved by RC operations (only in the
			case of the ten-swap experiment). Blue and magenta squares distinguish
			potentially different RC realizations, chosen from Supplementary Table 1.
			(b) The evolution $\mathcal{K}$ used for the CNOT calibration experiment
			(left), and for the ten-swap experiment (right).}
	\end{figure}
	
	Coherent errors present during the evolution stage of a quantum circuit
	can be specially harmful for quantum computing \cite{wallman2015estimating}.
	Randomized compiling (RC) \cite{wallman2016noise} is a standard
	method that allows to transform these errors into stochastic noise,
	which is amenable to the application of QEM using the KIK method.
	The essential idea of RC is to replace the target circuit by an average
	over compiled realizations that are logically equivalent to the original
	evolution. Each RC realization is obtained by randomly replacing and/or
	adding some gates that leave the noise-free target circuit invariant.
	However, in the presence of coherent errors, different RC realizations
	produce different evolutions, and coherent errors characteristic of
	single realizations can be averaged out into stochastic noise. 
	
	In our method, the implementation $\mathcal{U}_{\textrm{KIK}}^{(M)}$ of the
	KIK formula contains both the target evolution $\mathcal{K}$ and
	its inverse $\mathcal{K}_{\textrm{I}}$. Therefore, for each KIK power $\mathcal{K}\left(\mathcal{K}_{\textrm{I}}\mathcal{K}\right)^{m}$
	we independently applied RC on $\mathcal{K}$ and $\mathcal{K}_{\textrm{I}}$,
	as indicated in Supplementary Figure 7(a). This is done by using
	the Pauli gates shown in Supplementary Table 1, which are performed
	before and after the execution of $\mathcal{K}$ and $\mathcal{K}_{\textrm{I}}$
	in the ten-swap experiment. A sequence of ten swap gates is logically
	equivalent to the identity operation, and one can straightforwardly
	check that each RC realization in Supplementary Table 1 reproduces
	this operation. Furthermore, we remark that an independent application
	of RC on $\mathcal{K}$ and $\mathcal{K}_{\textrm{I}}$ is important to perform
	an effective randomization of the coherent errors affecting both evolutions.
	This ensures that the first Magnus terms appearing in the evolutions
	$\mathcal{K}$ and $\mathcal{K}_{\textrm{I}}\mathcal{K}$ (cf. Eqs. (\ref{eq:S23 K with Magnus expansion})
	and (\ref{eq:S24 KIK with Magnus expansion})) are still related through
	Eq. (\ref{eq:S25 Omega1(T)=00003D(1/2)Omega1(2T)}), and thus validates
	the KIK formula with the randomized versions of $\mathcal{K}$ and
	$\mathcal{K}_{\textrm{I}}$. Along the same line, we avoid the simplification
	of RC layers in any sequence $\mathcal{K}\left(\mathcal{K}_{\textrm{I}}\mathcal{K}\right)^{m}$.
	This means that consecutive single-qubit gates used for RC are not
	merged into a single gate. Otherwise, the independence of the randomization
	would be affected by preventing that the gates of the original pair
	act separately on $\mathcal{K}$ and $\mathcal{K}_{\textrm{I}}$. 
	
	On the other hand, for the CNOT calibration experiment we omitted
	the use of RC. The reason is that, in this experiment, we explored
	an alternative mitigation of coherent errors affecting the calibrated
	gate, which operates differently from RC. More specifically, a KIK-based
	mitigation of stochastic noise in the calibration process can improve
	the quality of the calibration, which translates into a calibrated
	gate with reduced coherent errors. 
	
	\begin{table}
		\centering{}%
		\begin{tabular}{|c|c|}
			\hline 
			RC1 & RC2\tabularnewline
			\hline 
			\hline 
			$I$ & $I$\tabularnewline
			\hline 
			$I$ & $X$\tabularnewline
			\hline 
			$I$ & $Y$\tabularnewline
			\hline 
			$I$ & $Z$\tabularnewline
			\hline 
		\end{tabular}%
		\begin{tabular}{c}
			\tabularnewline
			\tabularnewline
			\tabularnewline
			\tabularnewline
			\tabularnewline
		\end{tabular}%
		\begin{tabular}{|c|c|}
			\hline 
			RC1 & RC2\tabularnewline
			\hline 
			\hline 
			$X$ & $I$\tabularnewline
			\hline 
			$X$ & $X$\tabularnewline
			\hline 
			$X$ & $Y$\tabularnewline
			\hline 
			$X$ & $Z$\tabularnewline
			\hline 
		\end{tabular}%
		\begin{tabular}{c}
			\tabularnewline
			\tabularnewline
			\tabularnewline
			\tabularnewline
			\tabularnewline
		\end{tabular}%
		\begin{tabular}{|c|c|}
			\hline 
			RC1 & RC2\tabularnewline
			\hline 
			\hline 
			$Y$ & $I$\tabularnewline
			\hline 
			$Y$ & $X$\tabularnewline
			\hline 
			$Y$ & $Y$\tabularnewline
			\hline 
			$Y$ & $Z$\tabularnewline
			\hline 
		\end{tabular}%
		\begin{tabular}{c}
			\tabularnewline
			\tabularnewline
			\tabularnewline
			\tabularnewline
			\tabularnewline
		\end{tabular}%
		\begin{tabular}{|c|c|}
			\hline 
			RC1 & RC2\tabularnewline
			\hline 
			\hline 
			$Z$ & $I$\tabularnewline
			\hline 
			$Z$ & $X$\tabularnewline
			\hline 
			$Z$ & $Y$\tabularnewline
			\hline 
			$Z$ & $Z$\tabularnewline
			\hline 
		\end{tabular}\caption{Possible combinations of RC operations used for the evolutions $\mathcal{K}$
			and $\mathcal{K}_{\textrm{I}}$ in the ten-swap experiment. Here, RC1 and RC2
			stand for Pauli gates that are performed as indicated in Supplementary Figure 7(a). Since in this circuit the ideal unitary associated with $\mathcal{K}$
			and $\mathcal{K}_{\textrm{I}}$ is the identity, the gate RC3 must coincide
			with RC1, and the gate RC4 must coincide with RC2.}
	\end{table}
	
	A final ingredient for our QEM experiments is the mitigation of readout
	errors. For a circuit executed on $N$ qubits, the measurement process
	outputs counts of the states $|i_{1}i_{2}...i_{N}\rangle$ ($i_{j}\in\{0,1\}$
	for $1\leq j\leq N$) in the computational basis. A readout error
	occurs when the state registered upon the measurement is incorrect.
	For example, for a single qubit whose final state is $|1\rangle$,
	some counts may erroneously register the state as being $|0\rangle$. 
	
	Let us denote a general $N$-bit string by $\boldsymbol{k}=(k_{1}k_{2}...k_{N})$,
	and the corresponding computational state by $|\boldsymbol{k}\rangle$.
	When a quantum circuit produces a final state $\sigma$, readout errors
	cause wrong estimates of the probabilities $p_{\boldsymbol{k}}=\textrm{Tr}\left(|\boldsymbol{k}\rangle\langle\boldsymbol{k}|\sigma\right)$.
	A simple way to relate the ideal probabilities $p_{\boldsymbol{k}}$
	and the erroneous ones is by considering the probability distributions
	associated with each computational state. If $p\left(\boldsymbol{l}|\boldsymbol{k}\right)$
	denotes the conditional probability to register the state $|\boldsymbol{l}\rangle$,
	given that the actual state is $|\boldsymbol{k}\rangle$, for the
	final state $\sigma$ the probability to measure $|\boldsymbol{l}\rangle$
	in the presence of readout errors is given by
	
	\begin{equation}
		q_{\boldsymbol{l}}=\sum_{\boldsymbol{k}}p\left(\boldsymbol{l}|\boldsymbol{k}\right)p_{\boldsymbol{k}}.\label{eq:17 probabilities before readout mitigation}
	\end{equation}
	Thus, the information about the readout errors is contained in a $2^{N}\times2^{N}$
	matrix $\mathbf{M}$ with entries $p\left(\boldsymbol{l}|\boldsymbol{k}\right)$,
	whose columns and rows are associated with $\boldsymbol{k}$ and $\boldsymbol{l}$,
	respectively. To counteract the effect of the readout errors, we can
	invert the ``measurement matrix'' $\mathbf{M}$. In this way, the
	vector of error-free probabilities can be obtained by applying the
	inverse $\mathbf{M}^{-1}$ to the vector of measured probabilities.
	
	Although the procedure described above is not scalable in $N$, because
	the size of $\mathbf{M}$ is exponential in the number of qubits,
	our experiments involve two qubits and admit an efficient estimation
	of the $p\left(\boldsymbol{l}|\boldsymbol{k}\right)$. To this end,
	we prepared the four computational states $\{|ij\rangle\}_{i,j=0}^{1}$
	using circuits with the appropriate $X$ gates. For example, $|01\rangle$
	is prepared by applying $X$ on the second qubit. The measurement
	matrices were experimentally determined and then inverted for readout
	error mitigation in both the CNOT calibration experiment and the ten-swap
	experiment. The number of shots invested in the estimation of the
	distributions $\{p\left(\boldsymbol{l}|\boldsymbol{k}\right)\}_{\boldsymbol{l}}$
	is given in Supplementary Table 2. 
	
	As a final comment, it is important to mention that the study of methods
	to efficiently cope with readout errors in circuits containing a large
	number of qubits is an active area of research. For example, in many
	cases the noisy probability distribution $q_{\boldsymbol{l}}$ is
	mostly concentrated around the ideal distribution $p_{\boldsymbol{k}}$.
	This allows to construct a measurement matrix of reduced dimension
	that is defined over the subspace of bit strings with $q_{\boldsymbol{l}}\neq0$
	\cite{nation2021scalable}. In particular, the maximum size of
	such a matrix is determined by the maximum number of shots used to
	sample $q_{\boldsymbol{l}}$, and is independent of the number of
	qubits. \\
	\\
	\fakesubsection{Statistical analysis of experimental data}
	\textbf{Statistical analysis of experimental data.} The experimental estimates of different probability distributions
	are computed as frequencies over the number of shots. Ultimately,
	the goal is to estimate probability distributions $\{p_{\boldsymbol{k}}^{(m,i)}\}_{\boldsymbol{k}}$
	for each $m$ and $i$, where $p_{\boldsymbol{k}}^{(m,i)}$ denotes
	the probability to measure $|\boldsymbol{k}\rangle$ when the circuit
	$\mathcal{K}\left(\mathcal{K}_{\textrm{I}}\mathcal{K}\right)^{m}$ is implemented
	in combination with the $i$th RC realization. These
	probabilities are the key element for the computation of expectation
	values and the application of QEM. 
	
	In the case of the ten-swap experiment, we applied 16 RC realizations
	of each circuit $\mathcal{K}\left(\mathcal{K}_{\textrm{I}}\mathcal{K}\right)^{m}$.
	Hence, for any $m$ the expectation value $\textrm{Tr}\left(O\sigma\right)$
	of an observable $O$ is estimated as 
	\begin{align}
		\left\langle O\right\rangle _{m} & =\frac{1}{N_{\textrm{RC}}}\sum_{i=1}^{N_{\textrm{RC}}}\left\langle O\right\rangle _{m,i},\label{eq:18}\\
		\left\langle O\right\rangle _{m,i} & =\sum_{\boldsymbol{k}}p_{\boldsymbol{k}}^{(m,i)}O_{\boldsymbol{k}},\label{eq:19}
	\end{align}
	where $N_{\textrm{RC}}=16$. Importantly, for observables that are not diagonal
	in the computational basis, the final state $\sigma$ in $\textrm{Tr}\left(O\sigma\right)$
	must contain a rotation that performs the corresponding change of
	basis. For example, to measure the observable $O=Y_{1}$ on qubit
	1, for the CNOT calibration experiment, before the measurement we
	applied a rotation $R_{X_{1}}(\pi/2)=e^{-i\frac{\pi}{4}X_{1}}$ on
	this qubit.
	
	Since different RC realizations correspond to independent experiments,
	the variance for $\left\langle O\right\rangle _{m}$ reads 
	
	\begin{align}
		\textrm{Var}\left(\left\langle O\right\rangle _{m}\right) & =\frac{1}{N_{\textrm{RC}}^{2}}\sum_{i=1}^{N_{\textrm{RC}}}\textrm{Var}\left(\left\langle O\right\rangle _{m,i}\right)\nonumber \\
		& =\frac{1}{N_{\textrm{RC}}^{2}}\sum_{i=1}^{N_{\textrm{RC}}}\left[\frac{\sum_{\boldsymbol{k}}p_{\boldsymbol{k}}^{(m,i)}O_{\boldsymbol{k}}^{2}-\left\langle O\right\rangle _{m,i}^{2}}{N_{i}}\right],\label{eq:20}
	\end{align}
	where $N_{i}$ is the number of shots used in the $i$th RC realization.
	In the CNOT calibration experiment RC was not implemented, as mentioned
	before. However, Eqs. (\ref{eq:18})-(\ref{eq:20}) are still applicable
	in this case, with $N_{\textrm{RC}}=8$ being the number of circuits associated
	with different initial rotations,
	and the index $i$ labeling any of these circuits. 
	
	The KIK method provides the error-mitigated expectation value 
	\begin{equation}
		\left\langle O\right\rangle _{M}=\sum_{m=0}^{M}a_{m}^{(M)}\left\langle O\right\rangle _{m},\label{eq:22 error-mitigated expectation value}
	\end{equation}
	with the coefficients $a_{m}^{(M)}$ chosen depending on the QEM strategy.
	Namely, Taylor mitigation, or adaptive mitigation. The independence
	of the experiments associated with different circuits $\mathcal{K}\left(\mathcal{K}_{I}\mathcal{K}\right)^{m}$
	leads to the variance 
	
	\begin{equation}
		\textrm{Var}\left(\left\langle O\right\rangle _{M}\right)=\sum_{m=0}^{M}\left(a_{m}^{(M)}\right)^{2}\textrm{Var}\left(\left\langle O\right\rangle _{m}\right).\label{eq:23 variance of error-mitigated expect value}
	\end{equation}
	The error bars in the plots of the main text and the plots of Supplementary Figures 8 and 9 correspond to one standard deviation.
	That is, half of each error bar is given by $\sqrt{\textrm{Var}\left(\left\langle O\right\rangle _{M}\right)}$.\\
	\\
	\fakesubsection{CNOT calibration experiment}
	\textbf{CNOT calibration experiment.} The calibration experiment was performed on the IBM processor Jakarta,
	using the qubits labeled by 0 and 1. The qubit 0 was employed as control
	for the CNOT gate and the qubit 1 as target. In the IBM processors,
	the two-qubit interaction used to generate the CNOT gate is effectively
	implemented via the so called cross-resonance interaction \cite{alexander2020qiskit}
	$H_{\textrm{CR}}=Z\otimes X$, where $Z$ and $X$ are the Pauli matrices $Z=\left(\begin{array}{cc}
		1 & 0\\
		0 & -1
	\end{array}\right)$ and $X=\left(\begin{array}{cc}
		0 & 1\\
		1 & 0
	\end{array}\right)$ acting on the control qubit and target qubit, respectively. The CNOT
	thus involves a $\pi/2$ rotation with the Hamiltonian $H_{\textrm{CR}}$.
	This operation is performed in the quantum processor by applying a
	microwave pulse characterized by various calibrated parameters such
	as amplitude and duration. However, the values obtained for these
	parameters may be affected by systematic errors, due to noise present
	in the measurements used for calibration. 
	
	We focus on the calibration of the pulse amplitude, using the KIK
	method to mitigate the effect of noise. As explained in the main text,
	we prepare the initial state $\frac{1}{\sqrt{2}}\left(|0\rangle+|1\rangle\right)\otimes|0\rangle$
	and measure the observable $Y=\left(\begin{array}{cc}
		0 & -i\\
		i & 0
	\end{array}\right)$ on the target qubit for different pulse amplitudes. This choice of
	initial state and observable is convenient because it produces a calibration
	curve that is very well described by a straight line. We measured
	the expectation value $\left\langle Y_{1}\right\rangle $ for the
	amplitude factors $F\in\{0.98,0.9866,0.9933,1,1.0066,1.0133,1.02\}$,
	where $F=1$ corresponds to the default amplitude used by IBM, and
	fitted a straight line to the resulting data points. The calibrated
	amplitude factor is extracted from the intersection of the fitted
	line with the $x$ axis, which corresponds to $\left\langle Y_{1}\right\rangle =0$.
	The range of amplitudes $F$ was chosen to be sufficiently narrow,
	so that one could observe the linear behavior previously described. 
	
	To increase the precision of the calibration, the experiment was performed
	using a sequence of 11 CNOT gates. Ideally, this circuit is equivalent
	to a single CNOT, and the repetition of CNOTs has the effect of amplifying
	the variations of $\left\langle Y_{1}\right\rangle $ associated with
	different values of $F$. Accordingly, the inverse $\mathcal{K}_{\textrm{I}}$
	for the left circuit in Supplementary Figure 7(b) consists of a
	sequence of gates such that each of them is the pulse inverse of a
	single CNOT. 
	
	In the processor Jakarta, the maximum number of shots per circuit
	is 32000. For the mitigation of readout errors, we repeated four times
	the circuits employed in the preparation of each computational state,
	which yields a total of $4\times32000=128000$ shots used to estimate
	each probability distribution $\{p\left(\boldsymbol{l}|\boldsymbol{k}\right)\}_{\boldsymbol{l}}$.
	The circuits $\mathcal{K}\left(\mathcal{K}_{\textrm{I}}\mathcal{K}\right)^{m}$,
	used for QEM, were preceded by any of the four rotation operations
	used to mitigate the preparation coherent error. These rotations were
	executed before the preparation of the initial state, as seen in Supplementary Figure 7(a). We repeated two times the circuit corresponding to any rotation.
	Hence, $8\times32000=256000$ shots were employed to measure the expectation
	value of $Y_{1}$ on each $\mathcal{K}\left(\mathcal{K}_{\textrm{I}}\mathcal{K}\right)^{m}$,
	for each amplitude $F$. Since we applied KIK mitigation up to order
	3, the value of $m$ varies between 0 and 3. These experimental details
	are summarized in Supplementary Table 2.\\
	\\ 
	\fakesubsection{Ten-swap experiment}
	\textbf{Ten-swap experiment.} In this experiment, 
	\begin{table}
		\centering{}%
		\begin{tabular}{|c|c|c|c|}
			\cline{2-4} \cline{3-4} \cline{4-4} 
			\multicolumn{1}{c|}{} & shots per KIK circuit & shots for readout mitigation & RC operations\tabularnewline
			\hline 
			CNOT calibration experiment & 256000 & 512000 & n/a\tabularnewline
			\hline 
			ten-swap experiment & 320000 & 240000 & 16\tabularnewline
			\hline 
		\end{tabular}\caption{Summary of experimental resources. The `shots per
			KIK circuit' are the number of shots associated
			with each circuit \textcolor{blue}{${\normalcolor \mathcal{K}\left(\mathcal{K}_{\textrm{I}}\mathcal{K}\right)^{m}}$},
			for $0\protect\leq m\protect\leq3$. For both experiments, these shots
			include the initial rotations, and also RC operations in the case
			of the ten-swap experiment. The `shots for readout
			mitigation' are the total number of shots used to
			compute $\mathbf{M}^{-1}$.}
	\end{table}
	the error-mitigated quantity was the probability
	for the system to remain in the initial state $|00\rangle$, after
	applying a sequence of ten swap gates. The IBM processor employed
	was Quito. In practice, three CNOT gates were used to implement each
	swap, as shown in Fig. 4(b) of the main text. Apart from the application
	of RC, detailed below, we reduced the impact of coherent errors by
	alternating a normal swap with a pulse-based inverse swap (see Supplementary Figure 7(b)). This strategy has been previously applied to gates that are
	their own inverse, with the purpose of mitigating coherent errors
	such as overrotations \cite{zhang2022hidden}. Since a swap
	gate is its own inverse, our ideal target circuit is not modified
	by interleaving standard swaps with their pulse-inverse counterparts.
	
	Let us now describe in more detail the distribution of the different
	circuits for this experiment. As in the previous case, any of the
	circuits ${\normalcolor \mathcal{K}\left(\mathcal{K}_{\textrm{I}}\mathcal{K}\right)^{m}}$
	was preceded by one of the four rotation operations applied on the
	state $|00\rangle$. For $0\leq m\leq3$, a circuit ${\normalcolor \mathcal{K}\left(\mathcal{K}_{\textrm{I}}\mathcal{K}\right)^{m}}$
	is accompanied by a rotation operation and a RC operation chosen at
	random from Supplementary Table 1. We repeated four times each
	possible rotation, which results in a total of 16 circuits per each
	value of $m$, involving 16 (not necessarily different) RC realizations.
	Keeping in mind that the maximum number of shots for the IBM Quito
	device is 20000, the total number of shots used for each circuit ${\normalcolor \mathcal{K}\left(\mathcal{K}_{\textrm{I}}\mathcal{K}\right)^{m}}$ was
	$16\times20000=320000$. For readout mitigation, the circuits that
	prepare the computational states were repeated three times. Therefore,
	$3\times20000=60000$ shots were employed to estimate each probability
	distribution $\{p\left(\boldsymbol{l}|\boldsymbol{k}\right)\}_{\boldsymbol{l}}$. 
	
	We also include here the error mitigation curves for the initial states
	$|01\rangle$, $|10\rangle$, and $|11\rangle$, which complement
	the curves shown in Fig. 4(a) of the main text. These plots are given
	in Supplementary Figure 8. As in the case of the
	state $|00\rangle$, we can see that adaptive mitigation with the
	function $g(\mu)=\mu^{2}$ leads to the best results. We also remark
	that the slightly worse results corresponding to the state $|11\rangle$
	may be due to several experimental factors. Namely, imperfect averaging
	of the initial state (see Eq. (\ref{eq:16 rho single-qubit dephased})),
	drifts in gate parameters that characterize the single-qubit gates
	used for RC, and drifts in the measurement matrix $\mathbf{M}$. However,
	for the present experiment we do not have sufficient information for
	determining the most dominant factor. 
	
	\begin{figure}
		
		\centering{}\includegraphics[scale=0.7]{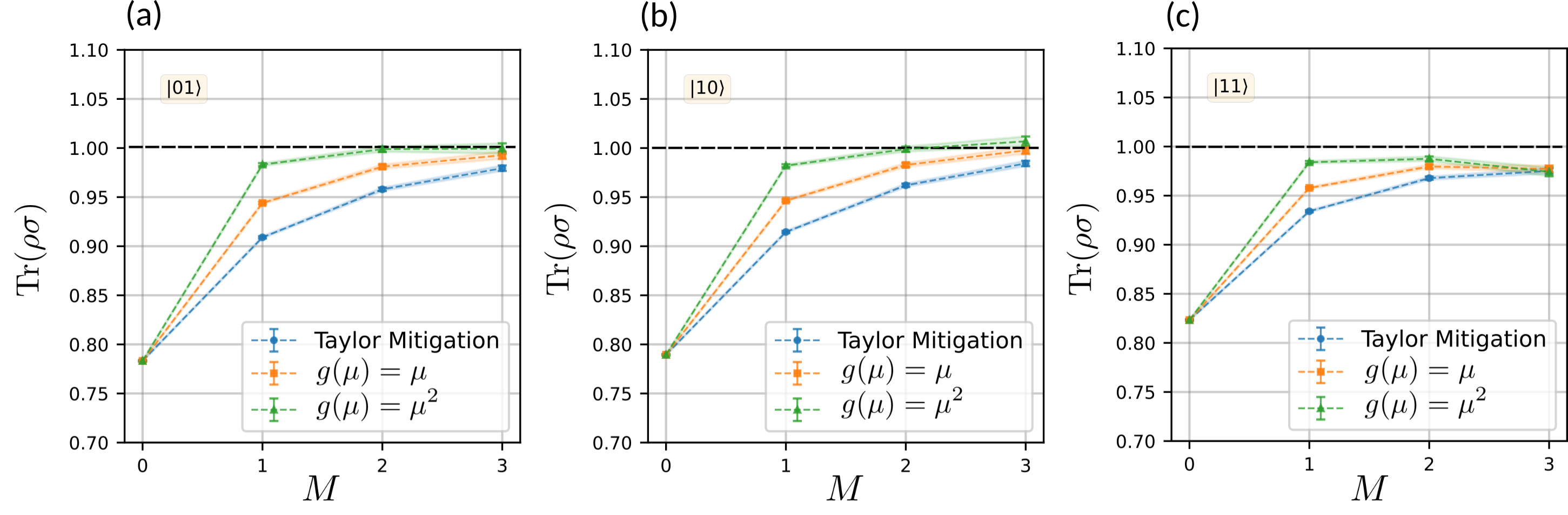}\caption{Experimental QEM in the IBM processor Quito. Error-mitigated survival
			probability for the ten-swap experiment, for initial states $|01\rangle$(a),
			$|10\rangle$(b), and $|11\rangle$(c). The ideal survival probability
			is 1 (dashed black line). Green and orange curves show adaptive QEM,
			and the blue curve stands for mitigation assuming weak noise (Taylor
			mitigation). The thin colored areas connect small experimental error
			bars. Here, we see again that Taylor mitigation is outperformed by
			adaptive mitigation.}
	\end{figure}
	
	In order to study the effect of coherent errors in the ten-swap circuit,
	we executed an additional experiment on the IBM processor Quito. As
	in the case of Supplementary Figure 8(c), the quantity measured
	was the survival probability for the system to remain in the initial
	state $|11\rangle$. The results presented in Supplementary Figure 9 were obtained by applying adaptive mitigation with $g(\mu)=\mu^{2}$. Coherent errors are manifested
	in Supplementary Figure 9(a) by the significant
	separation between different QEM curves, which correspond to different
	initial rotations. Furthermore, Supplementary Figure 9(b)
	shows how the application of RC (blue curve) produces substantially
	more accurate QEM, for all $1\leq M\leq3$. It is also worth stressing
	that in the absence of KIK QEM the enhancement provided by RC disappears,
	as evidenced by the matching of the blue and orange curves at $M=0$. 
	
	Each curve in Supplementary Figure 9(a) shows
	the survival probability for a given rotation operation, obtained
	by averaging over the associated RC realizations. Thus, if use the
	subscript $r$ to label any of the four possible initial rotations,
	and $i_{r}$ to label RC realizations that accompany the rotation
	$r$, for $M$th order mitigation the corresponding survival probability
	is given by 
	\begin{equation}
		\left\langle O\right\rangle _{M,r}=\sum_{m=0}^{M}a_{\textrm{Adap},m}^{(M)}\left(\mu^{2}\right)\left\langle O\right\rangle _{m,r},\label{eq:24}
	\end{equation}
	where $O=\rho=|00\rangle\langle00|$ and 
	\begin{align}
		\left\langle O\right\rangle _{m,r} & =\frac{1}{4}\sum_{i_{r}=1}^{4}\left\langle O\right\rangle _{m,i_{r}}\nonumber \\
		& =\frac{1}{4}\sum_{i_{r}=1}^{4}\sum_{\boldsymbol{k}}p_{\boldsymbol{k}}^{(m,i_{r})}O_{\boldsymbol{k}}.\label{eq:25}
	\end{align}
	Once again, the calculation of the corresponding variances takes into
	account the independence of different RC realizations for a given
	rotation. 
	
	As compared to Supplementary Figure 8(c), the
	blue curve in Supplementary Figure 9(b) features
	a performance more consistent with that observed in Supplementary Figures 8(a)
	and 8(b). This provides strong evidence that the
	modest performance observed in Supplementary Figure 8(c)
	is not intrinsic to the state $|11\rangle$, but possibly related
	to experimental factors already mentioned. We also remark that the
	orange curve is obtained by omitting the application of RC in any
	of the four repetitions of each initial rotation operation. The construction
	of both curves in Supplementary Figure 9(b) involves
	the same number of shots per KIK circuit, as per Supplementary Table 2.\\
	\\ 
	\fakesubsection{Simulation of a noisy calibration of a CNOT gate, using the KIK method}
	\textbf{Simulation of a noisy calibration of a CNOT gate, using the KIK method.}
	To complement the CNOT calibration  
	\begin{figure}
		\centering{}\includegraphics[scale=0.75]{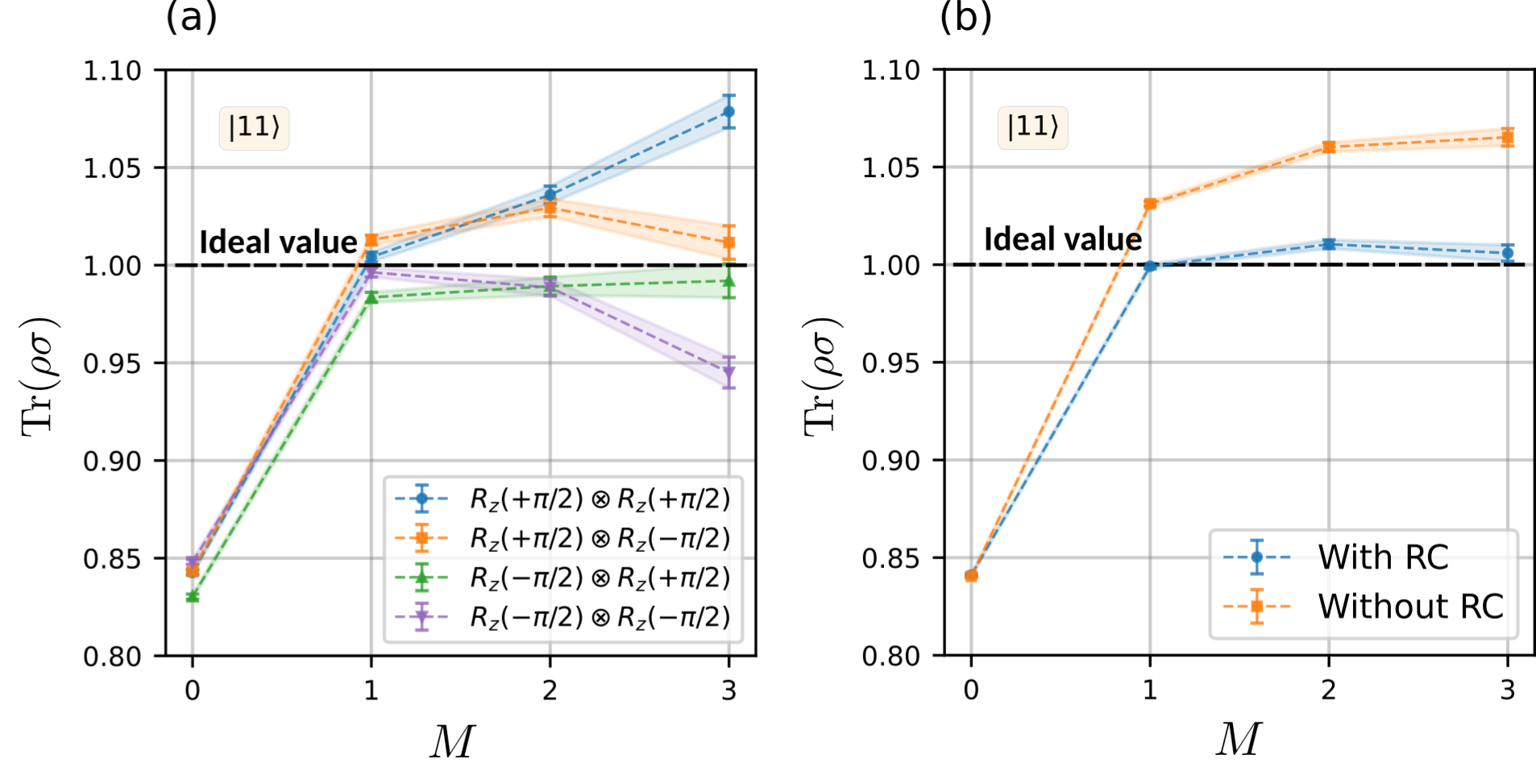}\caption{Effect of coherent errors on the error-mitigated survival probability
			for a ten-swap experiment. The initial state is $|11\rangle$ and
			the ideal survival probability is 1 (dashed black lines). The thin
			colored areas connect small experimental error bars. All the plots
			show adaptive mitigation using $g(\mu)=\mu^{2}$. (a) Error-mitigated
			survival probability for different initial rotations. By comparing
			these results with the blue curve in Supplementary Figure 9(b),
			it becomes clear that averaging over the initial rotations substantially
			mitigates the preparation coherent error and enhances the QEM. (b)
			Error-mitigated survival probabilities obtained with RC (blue curve)
			and without RC (orange curve). The QEM accuracy is significantly increased
			by applying RC. }
	\end{figure}
	experiment, we present now a simulation
	of the same calibration process. The ideal CNOT gate is simulated
	through the unitary evolution 
	\begin{equation}
		U=R_{Z}^{(0)}(-\pi/2)e^{-i\frac{\theta}{2}H_{\textrm{CR}}}R_{X}^{(1)}(-\pi/2),\label{eq:26}
	\end{equation}
	with the cross-resonance interaction $H_{\textrm{CR}}=Z\otimes X$, and the
	rotations $R_{Z}^{(0)}(-\pi/2)=e^{i\frac{\pi}{4}Z}$ and $R_{X}^{(1)}(-\pi/2)=e^{i\frac{\pi}{4}X}$
	acting on the target qubit and the control qubit, respectively. The
	angle $\theta$ depends on the strength and the duration of the pulse
	used to generate the cross-resonance interaction. An ideal (noise-free)
	CNOT gate corresponds to the angle $\theta=\pi/2$. 
	
	For the effect of noise, we consider the dissipator
	\begin{align}
		\mathcal{L} & =\xi\sum_{i=1}^{2}\gamma_{i}\left[A_{i}\otimes A_{i}^{\ast}-\frac{1}{2}A_{i}^{\dagger}A_{i}\otimes I-\frac{1}{2}I\otimes\left(A_{i}^{\dagger}A_{i}\right)^{\textrm{T}}\right],\label{eq:27 dissipator}\\
		A_{1} & =Z,\label{eq:21 A1 in dissipator}\\
		A_{2} & =\left(\begin{array}{cc}
			0 & 1\\
			0 & 0
		\end{array}\right),\label{eq:29 A1 in dissipator}
	\end{align}
	where $\gamma_{1}=1$ and $\gamma_{2}=1/10$. In Supplementary Figures 10
	and 11 we show the calibration curves for $\xi=2/100$
	(Supplementary Figure 10) and $\xi=1/100$ (Supplementary Figure 11).
	Since the action of noise produces an imperfect CNOT gate, the angle
	$\theta$ obtained under a noisy calibration is in general different
	from $\pi/2$. We express this angle as $\theta=A_{\theta}\frac{\pi}{2}$,
	and take the ``angle amplitude'' $A_{\theta}$ as the parameter
	for calibration. 
	
	\begin{figure}
		
		\begin{centering}
			\includegraphics[scale=0.7]{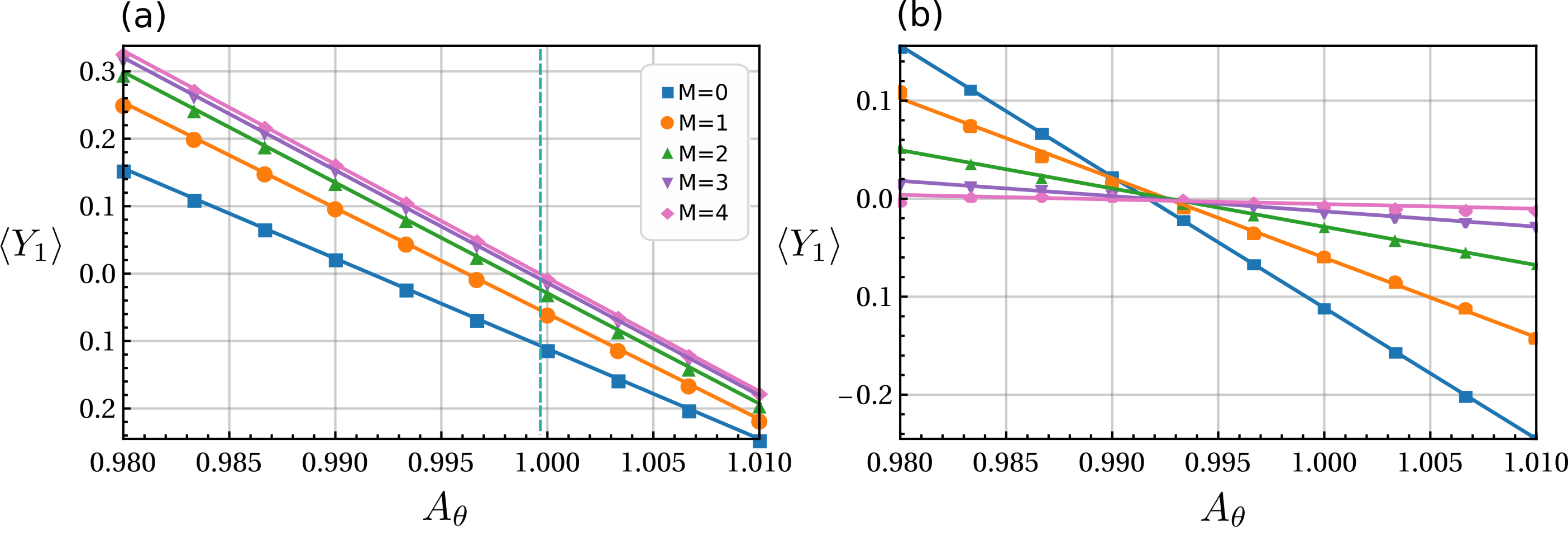}\caption{Simulation of the calibration of a noisy CNOT gate, for noise strength
				$\xi=2/100$. The curves show the expectation value of $Y_{1}=I\otimes Y$
				as a function of the angle amplitude $A_{\theta}$, which determines
				the angle $\theta=A_{\theta}\frac{\pi}{2}$ in the cross-resonance
				interaction $H_{CR}$ (cf. Eq. (\ref{eq:26})). The calibrated amplitude
				corresponds to $\left\langle Y_{1}\right\rangle =0$ (cyan dashed
				line). Different curves stand for different mitigation orders $0\protect\leq M\protect\leq4$.
				Figures (a) and (b) correspond to the application of the pulse inverse
				and the circuit inverse $\mathcal{K}_{\textrm{I}}=\mathcal{K}$, respectively. }
			
			\par\end{centering}
	\end{figure}
	
	As in the experimental case, the initial state and observable used
	for calibration are $\frac{1}{\sqrt{2}}\left(|0\rangle+|1\rangle\right)\otimes|0\rangle$
	and $Y_{1}=I\otimes Y$, respectively. Furthermore, the target circuit
	consists of a sequence of 11 CNOTs, in order to increase the sensitivity
	of $\left\langle Y_{1}\right\rangle $ to variations of $A_{\theta}$.
	Therefore, the noisy target circuit and its inverse are given by 
	\begin{align}
		\mathcal{K} & =\left(e^{-i\frac{\pi}{2}A_{\theta}\mathcal{H}_{\textrm{CR}}+\mathcal{L}}\right)^{11},\label{eq:30}\\
		\mathcal{K}_{\textrm{I}} & =\left(e^{i\frac{\pi}{2}A_{\theta}\mathcal{H}_{\textrm{CR}}+\mathcal{L}}\right)^{11},\label{eq:31}
	\end{align}
	where $\mathcal{H}_{\textrm{CR}}$ is the Liouville space representation of
	$H_{\textrm{CR}}$. 
	
	\begin{figure}
		\centering{}\includegraphics[scale=0.7]{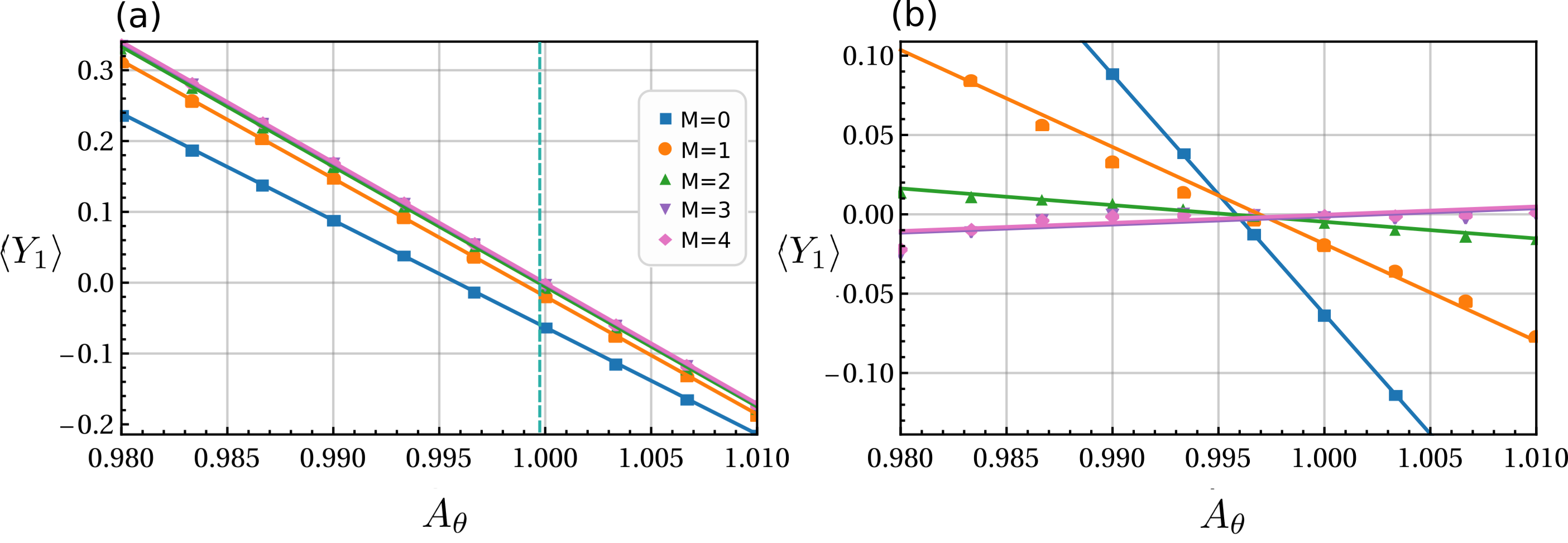}\caption{Simulation of the calibration of a noisy CNOT gate, for noise strength
			$\xi=1/100$. The curves show the expectation value of $Y_{1}=I\otimes Y$
			as a function of the angle amplitude $A_{\theta}$, which determines
			the angle $\theta=A_{\theta}\frac{\pi}{2}$ in the cross resonance
			interaction $H_{\textrm{CR}}$ (cf. Eq. (\ref{eq:26})). The calibrated amplitude
			corresponds to $\left\langle Y_{1}\right\rangle =0$ (cyan dashed
			line). Different curves stand for different mitigation orders $0\protect\leq M\protect\leq4$.
			Figures (a) and (b) correspond to the application of the pulse inverse
			and the circuit inverse $\mathcal{K}_{\textrm{I}}=\mathcal{K}$, respectively.}
	\end{figure}
	
	The amplitudes $A_{\theta}$ obtained for error mitigation orders
	$0\leq M\leq4$ are given in Supplementary Table 3. Due to the
	small values of the noise strength $\xi$, we apply Taylor mitigation
	with the coefficients $a_{\textrm{Tay},m}^{(M)}$. In Supplementary Table 3,
	we can observe that $A_{\theta}$ tends to one as $M$ increases.
	This reflects the fact that higher mitigation orders produce an evolution
	$\mathcal{U}_{\textrm{KIK}}^{(M)}$ closer to the ideal CNOT gate, and therefore
	the resulting calibrated angle $\theta$ also approaches its noise-free
	value $\theta=\pi/2$. On the other hand, for the noisy calibration
	corresponding to $M=0$, the associated angle presents the largest
	deviation from $\theta=\pi/2$. As a consequence, such a biased angle
	gives rise to a coherent error. 
	
	\begin{table}
		
		\begin{centering}
			\begin{tabular}{|c|c|c|c|c|c|}
				\hline 
				$M$ (mitigation order) & 0 & 1 & 2 & 3 & 4\tabularnewline
				\hline 
				\hline 
				$A_{\theta}$ for $\xi=2/100$ & 0.991671 & 0.996210 & 0.998235 & 0.999181 & 0.999631\tabularnewline
				\hline 
				$A_{\theta}$ for $\xi=1/100$ & 0.995830 & 0.998836 & 0.999673 & 0.999909 & 0.999977\tabularnewline
				\hline 
			\end{tabular}\caption{Angle amplitudes $A_{\theta}$ obtained with KIK-based calibration
				of a CNOT gate. $A_{\theta}$ determines the angle $\theta=A_{\theta}\frac{\pi}{2}$
				in Eq. (\ref{eq:26}), and $A_{\theta}=1$ for a noise-free CNOT.
				$\xi$ gives the noise strength in the dissipator (\ref{eq:27 dissipator}). }
			\par\end{centering}
	\end{table}
	
	To test the calibrated gate, we simulate the average gate fidelity
	after applying $M$th order Taylor mitigation to the noisy evolution
	\begin{equation}
		\mathcal{K}_{\textrm{CNOT}}=e^{-i\frac{\theta_{M}}{2}\mathcal{H}_{\textrm{CR}}+\mathcal{L}},\label{eq:32}
	\end{equation}
	where $\theta_{M}$ is the angle calibrated via $M$th order mitigation.
	In other words, this strategy consists of using the same order of
	mitigation for the calibration of the CNOT and for the evaluation
	of its average fidelity. 
	
	The average gate fidelity between a quantum channel $\Lambda$ and
	a unitary evolution $U$ is defined by 
	\begin{equation}
		F(\Lambda,U)=\int d\psi\langle\psi|U^{\dagger}\Lambda(|\psi\rangle\langle\psi|)U|\psi\rangle,\label{eq:32.1}
	\end{equation}
	where the integral is taken over all the pure states $|\psi\rangle$
	with respect to the Haar measure. Here, $\Lambda(|\psi\rangle\langle\psi|)$
	denotes the state obtained by applying $\Lambda$ on $|\psi\rangle\langle\psi|$.
	For quantum channels acting on $N$ qubits, the fidelity (\ref{eq:32.1})
	can be computed by using the Pauli transfer matrices of $\Lambda$
	and $U$. The Pauli transfer matrix $R_{\Lambda}$ for a general quantum
	channel $\Lambda$ has elements \cite{greenbaum2015introduction}
	\begin{equation}
		\left(R_{\Lambda}\right)_{ij}=\frac{1}{d}\textrm{Tr}\left[P_{i}\Lambda\left(P_{j}\right)\right],\label{eq:32.2}
	\end{equation}
	where $P_{i},P_{j}\in\{I,X,Y,Z\}^{\otimes N}$ are Pauli operators
	and $d=2^{N}$ is the dimension of the Hilbert space. 
	
	In this way, the fidelities $F_{M}$ in Supplementary Tables 4
	and 5 are calculated as \cite{greenbaum2015introduction}
	\begin{equation}
		F(\Lambda,U)=\frac{\textrm{Tr}\left(R_{\Lambda}^{-1}R_{U}\right)+d}{d(d+1)},\label{eq:33}
	\end{equation}
	with $U=e^{-i\frac{\theta}{2}H_{\textrm{CR}}}$ and the channel $\Lambda$
	corresponding to $M$th order mitigation, 
	\begin{align}
		\mathcal{U}_{\textrm{KIK}}^{(M)} & =\sum_{m=0}^{M}a_{\textrm{Tay},m}^{(M)}\mathcal{K}_{\textrm{CNOT}}\left(\mathcal{K}_{\textrm{I},\textrm{CNOT}}\mathcal{K}_{\textrm{CNOT}}\right)^{m},\label{eq:33.0}\\
		\mathcal{K}_{\textrm{I},\textrm{CNOT}} & =e^{i\frac{\theta_{M}}{2}\mathcal{H}_{\textrm{CR}}+\mathcal{L}}.\label{eq:33.01}
	\end{align}
	Using the Liouville space formalism, the matrix elements of $R_{\Lambda}$
	and $R_{U}$ are given by
	\begin{align}
		\left(R_{\Lambda}\right)_{ij} & =\frac{1}{d}\langle P_{i}|\mathcal{U}_{\textrm{KIK}}^{(M)}|P_{j}\rangle,\label{eq:33.1}\\
		\left(R_{U}\right)_{ij} & =\frac{1}{d}\langle P_{i}|\mathcal{U}|P_{j}\rangle,\label{eq:33.2}
	\end{align}
	where $\mathcal{U}=U\otimes U^{\ast}$ and $|P_{j}\rangle$ is the
	vector representation of $P_{j}$. On the other hand, $F_{0}$ is
	computed by replacing $\mathcal{K}_{\textrm{CNOT}}$ and $\mathcal{K}_{\textrm{I},\textrm{CNOT}}$
	in Eq. (\ref{eq:33.0}), by $e^{-i\frac{\theta_{0}}{2}\mathcal{H}_{\textrm{CR}}+\mathcal{L}}$
	and $e^{i\frac{\theta_{0}}{2}\mathcal{H}_{\textrm{CR}}+\mathcal{L}}$, respectively.
	
	We choose Taylor mitigation because the corresponding coefficients
	do not depend on the initial state. In contrast, in the case of adaptive
	mitigation the quantity $\mu=\langle\rho|\mathcal{K}_{\textrm{I},\textrm{CNOT}}\mathcal{K}_{\textrm{CNOT}}|\rho\rangle$
	depends on the initial state $\rho$, which makes it difficult to
	take into account when evaluating the gate fidelity. 
	
	\begin{table}
		\centering{}%
		\begin{tabular}{|c|c|c|c|}
			\hline 
			$M$ (mitigation order) & $F_{0}$ (noisy calibration) & $F_{M}$ (KIK-based calibration) & $F_{0}$ with RC\tabularnewline
			\hline 
			\hline 
			0 & 0.967325 & 0.967325 & 0.967325\tabularnewline
			\hline 
			1 & 0.997270 & 0.997298 & 0.997388\tabularnewline
			\hline 
			2 & 0.999692 & 0.999725 & 0.999743\tabularnewline
			\hline 
			3 & 0.999934 & 0.999968 & 0.999972\tabularnewline
			\hline 
			4 & 0.999961 & 0.999996 & 0.999997\tabularnewline
			\hline 
		\end{tabular}\caption{KIK-mitigated gate fidelity (\ref{eq:33}) (simulation) for a CNOT
			gate affected by noise (\ref{eq:27 dissipator}), with noise strength
			$\xi=2/100$. $F_{0}$ is the fidelity obtained when the gate is calibrated
			without KIK error mitigation. $F_{M}$ is the fidelity when $M$th
			order mitigation is performed alongside $M$th order KIK calibration.
			\textquotedblleft$F_{0}$ with RC\textquotedblright{} refers to the
			fidelity obtained without KIK-based calibration, and by combining
			KIK mitigation with RC. In this case the fidelity is the average over
			fidelities associated with all the 16 RC operations described in Supplementary Table
			1. }
	\end{table}
	
	Supplementary Tables 4 and 5 show the simulated
	fidelities for $\xi=2/100$ and $\xi=1/100$, respectively. The second
	column in these tables gives the error-mitigated fidelities without
	a KIK-based calibration, i.e. when $\theta_{M}=\theta_{0}$. The third
	column contains the values simulated by following the $M$-order mitigation
	strategy previously described. Finally, fidelities obtained without
	KIK-based calibration and RC integrated into the KIK mitigation are
	presented in the fourth column. Both tables show that KIK-based calibration
	and RC yield similar fidelities, which surpass the values obtained
	when none of these methods is employed. The effect of these two approaches
	is different though. On the one hand, RC mitigates the coherent error
	that characterizes the biased angle $\theta_{0}$ by postprocessing
	measurement data. On the other hand, in the KIK-based calibration
	this coherent error is reduced at a physical level. 
	
	\begin{table}
		\centering{}%
		\begin{tabular}{|c|c|c|c|}
			\hline 
			$M$ (mitigation order) & $F_{0}$ (noisy calibration) & $F_{M}$ (KIK-based calibration) & $F_{0}$ with RC\tabularnewline
			\hline 
			\hline 
			0 & 0.983434 & 0.983434 & 0.983434\tabularnewline
			\hline 
			1 & 0.999288 & 0.999296 & 0.999320\tabularnewline
			\hline 
			2 & 0.999954 & 0.999963 & 0.999965\tabularnewline
			\hline 
			3 & 0.999989 & 0.999998 & 0.999998\tabularnewline
			\hline 
			4 & 0.999991 & 1.000000 & 1.000000\tabularnewline
			\hline 
		\end{tabular}\caption{KIK-mitigated gate fidelity (\ref{eq:33}) (simulation) for a CNOT
			gate affected by noise (\ref{eq:27 dissipator}), with noise strength
			$\xi=1/100$. The description of each column is as in Supplementary Table 4. }
	\end{table}

	\section*{Supplementary Note 7: Numerical example to illustrate saturation of the KIK formula}
	
	In this section, we consider a numerical example where the accuracy
	of QEM using the KIK method saturates at approximately the mitigation
	order $M=4$. We quantify the accuracy by the relative error 
	
	\begin{equation}
		\frac{\varepsilon_{\textrm{KIK}}^{(M)}}{|\langle A|\mathcal{U}|\rho\rangle|}=\frac{\left|\langle A|\mathcal{U}|\rho\rangle-\langle A|\mathcal{U}_{\textrm{KIK}}^{(M)}|\rho\rangle\right|}{|\langle A|\mathcal{U}|\rho\rangle|},\label{eq:1 relative error}
	\end{equation}
	where $\varepsilon_{\textrm{KIK}}^{(M)}=\left|\langle A|\mathcal{U}|\rho\rangle-\langle A|\mathcal{U}_{\textrm{KIK}}^{(M)}|\rho\rangle\right|$
	is the difference between the ideal expectation value $\langle A|\mathcal{U}|\rho\rangle$,
	and the error-mitigated expectation value $\langle A|\mathcal{U}_{\textrm{KIK}}^{(M)}|\rho\rangle$,
	for the observable $A$ and the initial state $\rho$. In our example,
	the error (\ref{eq:1 relative error}) first decreases as a function
	of $M$, and then attains a value approximately constant for $M\geq4$.
	This is shown in Supplementary Figure 12. We apply QEM with Taylor
	mitigation, characterized by $\mathcal{U}_{\textrm{KIK}}^{(M)}=\sum_{m=0}^{M}a_{\textrm{Tay},m}^{(M)}\mathcal{K}\left(\mathcal{K}_{\textrm{I}}\mathcal{K}\right)^{m}$,
	and the coefficients $a_{\textrm{Tay},m}^{(M)}$ in Eq. (\ref{eq:S39 coefficients of the truncated Taylor expansion}).
	
	Our example refers to a four-qubit system characterized by the time-independent
	Hamiltonian 
	\begin{align}
		H & =X\otimes X\otimes I\otimes I+I\otimes X\otimes X\otimes I+I\otimes I\otimes X\otimes X,\label{eq:2 Hamiltonian}\\
		\mathcal{H} & =H\otimes I-I\otimes H^{\textrm{T}},\label{eq:3 Liouville space Hamiltonian}
	\end{align}
	where $X=\left(\begin{array}{cc}
		0 & 1\\
		1 & 0
	\end{array}\right)$ and $I=\left(\begin{array}{cc}
		1 & 0\\
		0 & 1
	\end{array}\right)$. Here, $\mathcal{H}$ is the Liouville space representation of $H$,
	taking into account Eq. (\ref{eq:S5 triple-prod identity}). Each
	qubit is subjected to spontaneous emission, which can be described
	via a GKLS (Gorini-Kossakowski-Sudarshan-Lindblad) master equation
	with dissipator $\hat{L}$ such that 
	\begin{align}
		\hat{L}[\rho] & =\sum_{k=1}^{4}\xi_{k}\left(A_{k}\rho A_{k}^{\dagger}-\frac{1}{2}A_{k}^{\dagger}A_{k}\rho-\frac{1}{2}\rho A_{k}^{\dagger}A_{k}\right),\label{eq:4 Dissipator}\\
		A_{1} & =\left(\begin{array}{cc}
			0 & 1\\
			0 & 0
		\end{array}\right)\otimes I\otimes I\otimes I,\label{eq:4.1 A1}\\
		A_{2} & =I\otimes\left(\begin{array}{cc}
			0 & 1\\
			0 & 0
		\end{array}\right)\otimes I\otimes I,\label{eq:4.2 A2}\\
		A_{3} & =I\otimes I\otimes\left(\begin{array}{cc}
			0 & 1\\
			0 & 0
		\end{array}\right)\otimes I,\label{eq:4.3 A3}\\
		A_{4} & =I\otimes I\otimes I\otimes\left(\begin{array}{cc}
			0 & 1\\
			0 & 0
		\end{array}\right).\label{eq:4.4 A4}
	\end{align}
	We assume that all the qubits are affected by the same relaxation
	rate $\xi=\xi_{k}=0.02$, for $1\leq k\leq4$. Using the vectorization
	rule (cf. Eq. (\ref{eq:S5 triple-prod identity})), in Liouville space
	the dissipator (\ref{eq:4 Dissipator}) takes the form 
	\begin{equation}
		\mathcal{L}=\xi\sum_{k=1}^{4}\left(A_{k}\otimes A_{k}^{\ast}-\frac{1}{2}A_{k}^{\dagger}A_{k}\otimes I-\frac{1}{2}I\otimes A_{k}^{\dagger}A_{k}\right),\label{eq:5 Liouville space dissipator}
	\end{equation}
	where $A_{k}^{\ast}$ denotes the complex conjugate of $A_{k}$. 
	
	Assuming a time unit $T=1$ for the evolution time, we have that 
	\begin{align}
		\mathcal{U} & =e^{-i\mathcal{H}},\label{eq:6.1 U}\\
		\mathcal{K} & =e^{-i\mathcal{H}+\mathcal{L}},\label{eq:6.2 K}\\
		\mathcal{K}_{\textrm{I}} & =e^{i\mathcal{H}+\mathcal{L}}.\label{eq:6.3 KI}
	\end{align}
	The total system starts in the ground state $\rho=\left(\begin{array}{cc}
		1 & 0\\
		0 & 0
	\end{array}\right)^{\otimes4}$, whose vector form is given by $|\rho\rangle=(1,0,0,...,0)^{\textrm{T}}$.
	The observable $A$ is the projector onto the ground state, and thus
	$\langle A|=(1,0,0,...,0)$. 
	\begin{figure}
		\centering{}\includegraphics[scale=0.6]{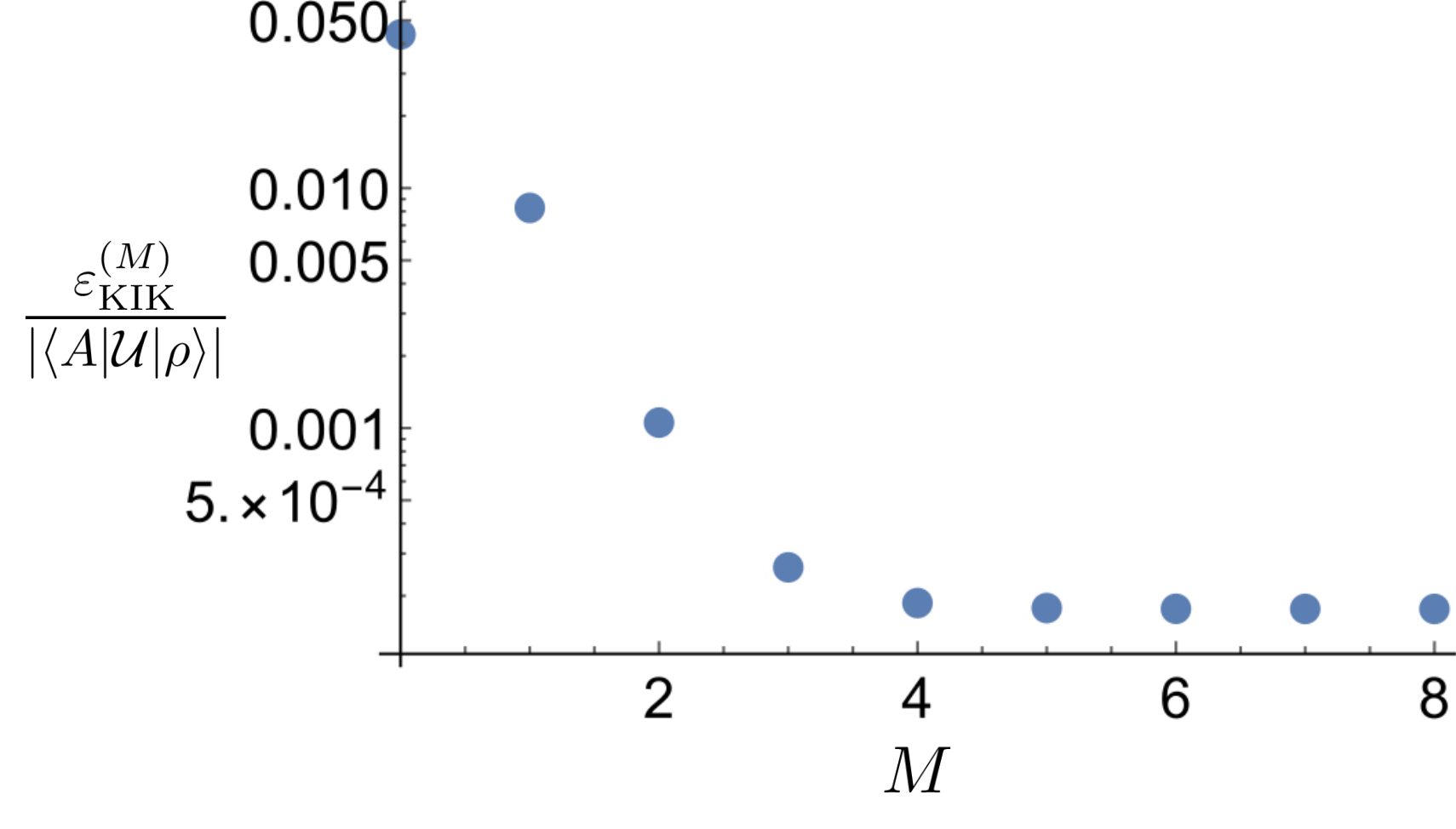}\caption{Log scale plot of the relative error (\ref{eq:1 relative error}),
			for the QEM example   described in the present supplementary note. }
	\end{figure}
	
	The saturation of the relative error can be understood as being a
	consequence of the approximations involved in the derivation of the
	KIK formula. More specifically, it is related to the fact that in
	the derivation of Eq. (\ref{eq:S26 KIK formula for noise chanel})
	we have discarded higher-order Magnus terms $\Omega_{n\geq2}$. Nevertheless,
	it is worth noting that, according to Supplementary Figure 12, a
	relative error as small as $\sim10^{-4}$ can already be achieved
	with a mitigation order $M=3$. When the initial error is smaller
	(i.e. when $\xi$ is smaller), the improvement is even more substantial. 
	
	\section*{Supplementary Note 8: Derivation of upper bounds for the performance of the KIK method}
	
	A complete assessment of a QEM technique requires establishing bounds
	for the remaining error in the estimation of expectation values. Furthermore,
	these bounds should be applicable to general observables $A$ and
	circuits $\mathcal{K}$ of arbitrary size. The goal of the present
	section is to derive such bounds, in the case of QEM using the KIK
	method. In the fisrt two subsections we will state some technical results and physically
	realistic assumptions, which will be used for the derivation of our
	bounds in the last two subsections.\\ 
	\\
	\fakesubsection{Lower bound on the smallest eigenvalue of $\mathcal{K}_{\textrm{I}}\mathcal{K}$}
	\textbf{Lower bound on the smallest eigenvalue of $\mathcal{K}_{\textrm{I}}\mathcal{K}$.} Assuming that $\mathcal{K}_{\textrm{I}}\mathcal{K}$ is diagonalizable, i.e.
	$\mathcal{K}_{\textrm{I}}\mathcal{K}=\sum_{k}\lambda_{k}|k)(k|$, where $|k)$
	and $(k|$ denote respectively right and left eigenvectors (note that
	in general we may have $(k|\neq|k)^{\dagger}$), we will obtain the
	bound 
	\begin{equation}
		\textrm{min}_{k}\lambda_{k}\geq e^{-2\intop_{0}^{T}\left\Vert \mathcal{L}(t)\right\Vert dt}.\label{eq:S60 bound on minimum eigenv of KIK}
	\end{equation}
	We recall that $\mathcal{L}(t)$ is the dissipator introduced in Eq.
	(\ref{eq:S10 mast eq in L space 2}), and $T$ is the evolution time
	for $\mathcal{K}$. The norm $\left\Vert \ast\right\Vert $ in Eq
	(\ref{eq:S60 bound on minimum eigenv of KIK}) is the spectral norm. 
	
	First, we consider the equivalent expression of Eq. (\ref{eq:S20 Master Eq in int picture}),
	\begin{equation}
		\frac{d}{dt}\tilde{\mathcal{K}}_{\textrm{int}}(t)=\tilde{\mathcal{L}}_{\textrm{int}}(t)\tilde{\mathcal{K}}_{\textrm{int}}(t),\label{eq:S61 mast equation for Kint}
	\end{equation}
	with the interaction-picture evolution $\tilde{\mathcal{K}}_{\textrm{int}}(t)=\tilde{\mathcal{U}}^{\dagger}(t)\tilde{\mathcal{K}}(t)$,
	and dissipator $\tilde{\mathcal{L}}_{\textrm{int}}(t)=\tilde{\mathcal{U}}^{\dagger}(t)\tilde{\mathcal{L}}(t)\tilde{\mathcal{U}}(t)$.
	We point out that, in this case, we consider only the time interval
	$0\leq t\leq T$. The inverse evolution $\tilde{\mathcal{K}}_{\textrm{int}}^{-1}(t)=\tilde{\mathcal{K}}^{-1}(t)\tilde{\mathcal{U}}(t)$
	satisfies the equation 
	\begin{equation}
		\frac{d}{dt}\tilde{\mathcal{K}}_{\textrm{int}}^{-1}(t)=-\tilde{\mathcal{L}}_{\textrm{int}}(T-t)\tilde{\mathcal{K}}_{\textrm{int}}^{-1}(t),\label{eq:S62 mast eq for inverse Kint^-1}
	\end{equation}
	which will be used in the derivation of (\ref{eq:S60 bound on minimum eigenv of KIK}).
	Here, it is important to remark that $\tilde{\mathcal{K}}_{\textrm{int}}^{-1}(t)$
	is the mathematical inverse of $\tilde{\mathcal{K}}_{\textrm{int}}(t)$
	and must not be confused with $\mathcal{K}_{\textrm{I}}$.
	
	By taking the spectral norm at both sides of Eq. (\ref{eq:S62 mast eq for inverse Kint^-1}),
	we get 
	\begin{align}
		\left\Vert \frac{d}{dt}\tilde{\mathcal{K}}_{\textrm{int}}^{-1}(t)\right\Vert  & =\left\Vert \tilde{\mathcal{L}}_{\textrm{int}}(T-t)\tilde{\mathcal{K}}_{\textrm{int}}^{-1}(t)\right\Vert \nonumber \\
		& \leq\left\Vert \tilde{\mathcal{L}}(T-t)\right\Vert \left\Vert \tilde{\mathcal{K}}^{-1}(t)\right\Vert ,\label{eq:S63 espec norm inequality for equat of Kint^-1}
	\end{align}
	where the inequality follows from the submultiplicativity and unitary
	invariance of the spectral norm. On the other hand, we can apply the
	reverse triangle inequality $\left\Vert x-y\right\Vert \geq\left|\left\Vert x\right\Vert -\left\Vert y\right\Vert \right|$
	to obtain 
	\begin{align}
		\left\Vert \frac{d}{dt}\tilde{\mathcal{K}}_{\textrm{int}}^{-1}(t)\right\Vert  & =\textrm{lim}_{\Delta t\rightarrow0}\left\Vert \frac{\tilde{\mathcal{K}}_{\textrm{int}}^{-1}(t+\Delta t)-\tilde{\mathcal{K}}_{\textrm{int}}^{-1}(t)}{\Delta t}\right\Vert \nonumber \\
		& \geq\textrm{lim}_{\Delta t\rightarrow0}\left|\frac{\left\Vert \tilde{\mathcal{K}}_{\textrm{int}}^{-1}(t+\Delta t)\right\Vert -\left\Vert \tilde{\mathcal{K}}_{\textrm{int}}^{-1}(t)\right\Vert }{\Delta t}\right|\nonumber \\
		& =\frac{d}{dt}\left\Vert \tilde{\mathcal{K}}_{\textrm{int}}^{-1}(t)\right\Vert =\frac{d}{dt}\left\Vert \tilde{\mathcal{K}}^{-1}(t)\right\Vert .\label{eq:S64}
	\end{align}
	By combining this result with Eq. (\ref{eq:S63 espec norm inequality for equat of Kint^-1}),
	we have the inequality
	\begin{equation}
		\frac{d}{dt}\left\Vert \tilde{\mathcal{K}}^{-1}(t)\right\Vert \leq\left\Vert \tilde{\mathcal{L}}(T-t)\right\Vert \left\Vert \tilde{\mathcal{K}}^{-1}(t)\right\Vert ,\label{eq:S65}
	\end{equation}
	which upon integration yields 
	\begin{equation}
		\left\Vert \tilde{\mathcal{K}}^{-1}(T)\right\Vert \leq e^{\intop_{0}^{T}\left\Vert \mathcal{\tilde{\mathcal{L}}}(T-t)\right\Vert dt}.\label{eq:S66}
	\end{equation}
	
	Taking into account that $\tilde{\mathcal{K}}(T)=\mathcal{K}$ and
	$\mathcal{\tilde{\mathcal{L}}}(T-t)=\mathcal{L}(T-t)$ for $0\leq t\leq T$
	(cf. Eq. (\ref{eq:S17 H(t) and L(t) for t in (0,2T)})), the change
	of variable $t'=T-t$ straightforwardly leads to 
	\begin{equation}
		\left\Vert \mathcal{K}^{-1}\right\Vert \leq e^{\intop_{0}^{T}\left\Vert \mathcal{L}(t)\right\Vert dt}.\label{eq:S67}
	\end{equation}
	Consider now the singular value decomposition (SVD) $\mathcal{K}=\mathcal{V}\mathcal{S}\mathcal{W}$,
	where $\mathcal{S}$ is a diagonal matrix with the singular values
	of $\mathcal{K}$ and $\mathcal{V}$ and $\mathcal{W}$ are unitary
	matrices. Then, the SVD of $\mathcal{K}^{-1}$ reads $\mathcal{W}^{\dagger}\mathcal{S}^{-1}\mathcal{V}^{\dagger}$,
	and the spectral norm (maximum singular value) of $\mathcal{K}^{-1}$
	is given by $\left\Vert \mathcal{K}^{-1}\right\Vert =1/s_{\textrm{min}}$,
	where $s_{\textrm{min}}$ is the minimum singular value of $\mathcal{K}$.
	Accordingly, we can rewrite (\ref{eq:S67}) as 
	\begin{equation}
		s_{\textrm{min}}\geq e^{-\intop_{0}^{T}\left\Vert \mathcal{L}(t)\right\Vert dt}.\label{eq:S68}
	\end{equation}
	
	The singular values of $\mathcal{K}$ are the eigenvalues of $\left(\mathcal{K}^{\dagger}\mathcal{K}\right)^{\frac{1}{2}}$,
	where $\mathcal{K}^{\dagger}$ is the Hermitian conjugate of $\mathcal{K}$.
	If the evolution $\mathcal{K}_{\textrm{I}}$ coincides with $\mathcal{K}^{\dagger}$,
	we have that $s_{\textrm{min}}$ is also the minimum eigenvalue of  $\left(\mathcal{K}_{\textrm{I}}\mathcal{K}\right)^{\frac{1}{2}}$,
	or, equivalently, the minimum eigenvalue of $\mathcal{K}_{\textrm{I}}\mathcal{K}$
	satisfies the inequality (\ref{eq:S60 bound on minimum eigenv of KIK}).
	Hence, our last step in proving this inequality is to show that $\mathcal{K}_{\textrm{I}}=\mathcal{K}^{\dagger}$,
	under sound physical conditions. This is done in the next subsection.\\ 
	\\
	\fakesubsection{Sufficient condition for $\mathcal{K}_{\textrm{I}}=\mathcal{K}^{\dagger}$}
	\textbf{Sufficient condition for $\mathcal{K}_{\textrm{I}}=\mathcal{K}^{\dagger}$.}
	Suppose that the dissipator $\mathcal{L}(t)$ is Hermitian, i.e. $\mathcal{L}^{\dagger}(t)=\mathcal{L}(t)$
	for $0\leq t\leq T$. Then, from the definition of the first-order
	Magnus term (\ref{eq:S22 Omega1}) it also follows that 
	\begin{equation}
		\Omega_{1}^{\dagger}(T)=\Omega_{1}(T).\label{eq:S69 Hermiticity of first Magnus term}
	\end{equation}
	If we can prove that, up to first order in the Magnus expansion (where
	$\Omega(T)$ is approximated by $\Omega_{1}(T)$), 
	\begin{equation}
		\mathcal{K}_{\textrm{I}}\approx e^{\Omega_{1}(T)}\mathcal{U}^{\dagger},\label{eq:S70 KI=00003De^Omega1*U^+}
	\end{equation}
	then, Eqs. (\ref{eq:S23 K with Magnus expansion}) and (\ref{eq:S70 KI=00003De^Omega1*U^+})
	imply that $\mathcal{K}_{\textrm{I}}=\mathcal{K}^{\dagger}$ within this approximation. 
	
	Before discussing under which circumstances the property $\mathcal{L}^{\dagger}(t)=\mathcal{L}(t)$
	is satisfied, let us first derive Eq. (\ref{eq:S70 KI=00003De^Omega1*U^+}).
	We start by writing the formal solution to the inverse evolution $\mathcal{K}_{\textrm{I}}$.
	Denoting the corresponding Magnus expansion by $\Omega_{I}$, and
	assuming for simplicity that $\mathcal{L}_{\textrm{I}}(t)$ acts on the time
	interval $(0,T)$ (as opposed to the interval $(T,2T)$, used in the
	derivation of the KIK formula), Eq. (\ref{eq:S23 K with Magnus expansion})
	leads to 
	\begin{align}
		\mathcal{K}_{\textrm{I}} & =\mathcal{U}^{\dagger}e^{\Omega_{\textrm{I}}(T)}\nonumber \\
		& \approx\mathcal{U}^{\dagger}e^{\Omega_{\textrm{I},1}(T)},\label{eq:S71}
	\end{align}
	with the first-order Magnus term 
	\begin{align}
		\Omega_{\textrm{I},1}(T) & =\int_{0}^{T}\mathcal{L}_{\textrm{I},\textrm{int}}(t)dt\nonumber \\
		& =\int_{0}^{T}\mathcal{U}(t)\mathcal{L}(T-t)\mathcal{U}^{\dagger}(t)dt.\label{eq:S72}
	\end{align}
	Writing $\mathcal{U}(t)$ as $\mathcal{U}(t)=\mathcal{U}(T)\mathcal{U}^{\dagger}(T-t)$,
	and performing the change of variable $t'=T-t$ in the integral, we
	obtain 
	\begin{align}
		\Omega_{\textrm{I},1}(T) & =-\mathcal{U}(T)\left(\int_{T}^{0}\mathcal{U}^{\dagger}(t')\mathcal{L}(t')\mathcal{U}(t')dt'\right)\mathcal{U}^{\dagger}(T)\nonumber \\
		& =\mathcal{U}\left(\int_{0}^{T}\mathcal{U}^{\dagger}(t')\mathcal{L}(t')\mathcal{U}(t')dt'\right)\mathcal{U}^{\dagger}\nonumber \\
		& =\mathcal{U}\Omega_{1}(T)\mathcal{U}^{\dagger}.\label{eq:S73 Omega_I,1=00003DUOmega_1U^+}
	\end{align}
	Note that we also used the simplified notation $\mathcal{U}(T)=\mathcal{U}$.
	In this way, we conclude that 
	\begin{align}
		\mathcal{K}_{\textrm{I}} & \approx\mathcal{U}^{\dagger}e^{\mathcal{U}\Omega_{1}(T)\mathcal{U}^{\dagger}}\nonumber \\
		& =e^{\Omega_{1}(T)}\mathcal{U}^{\dagger}.\label{eq:S74}
	\end{align}
	Thus, the hermiticity of $\mathcal{L}(t)$ leads to $\mathcal{K}_{\textrm{I}}=\mathcal{K}^{\dagger}$,
	within our approximation. 
	
	Now, let us come back to the subject of the Hermiticity of $\mathcal{L}(t)$.
	A sufficient condition to have this property is that 
	\begin{equation}
		\mathcal{L}(t)=\sum_{\vec{\mathbf{k}}}\alpha_{\vec{\mathbf{k}}}(t)\left(P_{\vec{\mathbf{k}}}\otimes P_{\vec{\mathbf{k}}}^{\textrm{T}}-\mathcal{I}\right),\label{eq:S75 Pauli dissipator in L space}
	\end{equation}
	where $P_{\vec{\mathbf{k}}}=\sigma_{\mathbf{k}_{1}}^{(1)}\otimes\sigma_{\mathbf{k}_{2}}^{(2)}\otimes...\otimes\sigma_{\mathbf{k}_{n}}^{(n)}\otimes...\otimes\sigma_{\mathbf{k}_{N}}^{(N)}$
	is a Pauli operator acting on $N$ qubits, $\sigma_{\mathbf{k}_{n}}^{(n)}\in\left\{ I,X,Y,Z\right\} $
	is a Pauli matrix acting on the $n$th qubit, and the time-dependent
	coefficients $\alpha_{\vec{\mathbf{k}}}(t)$ are real. We note that a time-independent version of this dissipator was recently considered
	in Ref. \cite{van2023probabilistic}. For a given state $\rho$,
	we can check by direct application of the identity (\ref{eq:S5 triple-prod identity})
	that Eq. (\ref{eq:S75 Pauli dissipator in L space}) is the Liouville
	space representation of the superoperator $\hat{L}(t)$ such that
	\begin{equation}
		\hat{L}(t)[\rho]=\sum_{\vec{\mathbf{k}}}\alpha_{\vec{\mathbf{k}}}(t)(P_{\vec{\mathbf{k}}}\rho P_{\vec{\mathbf{k}}}-\rho).\label{eq:S76 Pauli dissipator in H space}
	\end{equation}

	The noise model of Ref. \cite{van2023probabilistic} relies on
	a time-independent $\hat{L}(t)=\hat{L}$, characterized by time-independent
	coefficients $\alpha_{\vec{\mathbf{k}}}(t)=\alpha_{\vec{\mathbf{k}}}$. In addition,
	$\alpha_{\vec{\mathbf{k}}}\neq0$ only if the index $\vec{\mathbf{k}}=\{\mathbf{k}_{1},\mathbf{k}_{2},...,\mathbf{k}_{N}\}$
	contains at most two components $\mathbf{k}_{n}$ different from 0, which limits
	correlated errors to occur at most between pairs of qubits. On the
	other hand, the more general dissipator (\ref{eq:S75 Pauli dissipator in L space})
	guarantees the condition $\mathcal{K}_{\textrm{I}}=\mathcal{K}^{\dagger}$.
	We remark that the time-dependent coefficients $\alpha_{\vec{\mathbf{k}}}(t)$
	in Eq. (\ref{eq:S75 Pauli dissipator in L space}) can also be different
	from zero for arbitrary indices $\vec{\mathbf{k}}$, thereby allowing for correlated
	errors between any number of qubits in the system. 
	
	The Pauli channel (\ref{eq:S75 Pauli dissipator in L space}) arises
	naturally when executing RC for the purpose of mitigating coherent
	errors. In others words, by applying RC on a general dissipator, which
	can contain non-diagonal components in the Pauli operator representation,
	a dissipator of the form (\ref{eq:S75 Pauli dissipator in L space})
	is obtained. Since we exploit RC in our experiments, the effective
	noise is described by a Pauli channel, which according to the above
	results implies $\mathcal{K}_{\textrm{I}}=\mathcal{K}^{\dagger}$.\\\\
	\fakesubsection{First error bounds for Adaptive mitigation and Taylor mitigation}
	\textbf{First error bounds for Adaptive mitigation and Taylor mitigation.} To begin this subsection, we recall that the error-mitigated evolution
	$\mathcal{U}_{\textrm{KIK}}=\mathcal{K}\left(\mathcal{K}_{\textrm{I}}\mathcal{K}\right)^{-\frac{1}{2}}$
	is implemented through $M$-order approximations of the form $\mathcal{U}_{\textrm{KIK}}^{(M)}=\sum_{m=0}^{M}a_{m}^{(M)}\mathcal{K}\left(\mathcal{K}_{\textrm{I}}\mathcal{K}\right)^{m}$.
	Our goal is to obtain an upper bound on the error 
	\begin{equation}
		\varepsilon_{\textrm{KIK}}^{(M)}=\left|\langle A|\mathcal{U}|\rho\rangle-\langle A|\mathcal{U}_{\textrm{KIK}}^{(M)}|\rho\rangle\right|,\label{eq:S77 order-M mitigation error}
	\end{equation}
	which quantifies how much the ideal expectation value $\langle A|\mathcal{U}|\rho\rangle$
	deviates from the error-mitigated expectation value $\langle A|\mathcal{U}_{\textrm{KIK}}^{(M)}|\rho\rangle$.
	Here, $A$ is an arbitrary observable and $\rho$ is an arbitrary
	initial state. 
	
	We will start by deriving a bound for the error associated with adaptive
	mitigation, with the coefficients ${\color{red}{\normalcolor a_{\textrm{Adap},m}^{(M)}}}[g(\mu)]$
	evaluated at $g(\mu)=\mu$, and $1\leq M\leq3$. The possibility of
	obtaining tighter bounds with different choices of $g(\mu)$ is left
	as an open problem. Later on, we will derive another bound that, despite
	being looser, has the advantage of being independent of $\mu$, and
	is also valid for both adaptive mitigation with $g(\mu)=\mu$ and
	Taylor mitigation. 
	
	Using our conventional approximations $\mathcal{K}\approx\mathcal{U}e^{\Omega_{1}}$
	and $\mathcal{K}_{\textrm{I}}\mathcal{K}\approx e^{2\Omega_{1}}$, we can write
	$\mathcal{U}_{\textrm{KIK}}^{(M)}$ as 
	\begin{align}
		\mathcal{U}_{\textrm{KIK}}^{(M)} & \approx\mathcal{U}\sum_{m=0}^{M}{\color{red}{\normalcolor a_{\textrm{Adap},m}^{(M)}}}(\mu)e^{(2m+1)\Omega_{1}}\nonumber \\
		& \approx\mathcal{U}\sum_{m=0}^{M}{\color{red}{\normalcolor a_{\textrm{Adap},m}^{(M)}}}(\mu)\left(\mathcal{K}_{\textrm{I}}\mathcal{K}\right)^{m+1/2}.\label{eq:S78 U_KIK^(M) in terms of U}
	\end{align}
	Therefore, 
	\begin{align}
		\varepsilon_{\textrm{KIK}}^{(M)} & =\left|\langle A|\left(\mathcal{U}-\mathcal{U}_{\textrm{KIK}}^{(M)}\right)|\rho\rangle\right|\nonumber \\
		& \leq\sqrt{\langle A|A\rangle}\sqrt{\langle\rho|\rho\rangle}\left\Vert \mathcal{U}\left(\mathcal{I}-\sum_{m=0}^{M}{\color{red}{\normalcolor a_{\textrm{Adap},m}^{(M)}}}(\mu)\left(\mathcal{K}_{\textrm{I}}\mathcal{K}\right)^{m+1/2}\right)\right\Vert \nonumber \\
		& =\sqrt{\langle A|A\rangle}\left\Vert \left(\mathcal{I}-\sum_{m=0}^{M}{\color{red}{\normalcolor a_{\textrm{Adap},m}^{(M)}}}(\mu)\left(\mathcal{K}_{\textrm{I}}\mathcal{K}\right)^{m+1/2}\right)\right\Vert ,\label{eq:S79 upper bound on order-M mitigation error}
	\end{align}
	where the first inequality follows from the definition (\ref{eq:S78 U_KIK^(M) in terms of U})
	and the definition of the spectral norm $\left\Vert *\right\Vert $.
	In the last line, we use the unitary invariance of this norm. Moreover,
	we assume an initial pure state $\rho$, which implies $\langle\rho|\rho\rangle=1$.
	Assuming as before that $\mathcal{L}(t)$ satisfies Eq. (\ref{eq:S75 Pauli dissipator in L space}),
	and hence $\mathcal{K}_{\textrm{I}}=\mathcal{K}^{\dagger}$, we have that the
	operator $\mathcal{K}_{\textrm{I}}\mathcal{K}$ is Hermitian. Since this implies
	that $\mathcal{I}-\sum_{m=0}^{M}{\color{red}{\normalcolor a_{\textrm{Adap},m}^{(M)}}}(\mu)\left(\mathcal{K}_{\textrm{I}}\mathcal{K}\right)^{m+1/2}$
	is also Hermitian, the corresponding spectral norm is simply the absolute
	value of its maximum eigenvalue. In this way, Eq. (\ref{eq:S79 upper bound on order-M mitigation error})
	leads to 
	
	\begin{equation}
		\varepsilon_{\textrm{KIK}}^{(M)}\leq\sqrt{\langle A|A\rangle}\textrm{max}_{k}\left|1-\sum_{m=0}^{M}{\color{red}{\normalcolor a_{\textrm{Adap},m}^{(M)}}}(\mu)\left(\lambda_{k}\right)^{m+1/2}\right|,\label{eq:S79.1 uppor bound on E_KIK^(M) 1}
	\end{equation}
	where we have again denoted the eigenvalues of $\mathcal{K}_{\textrm{I}}\mathcal{K}=\sum_{k}\lambda_{k}|k\rangle\langle k|$
	by $\{\lambda_{k}\}$. Since these eigenvalues may be difficult to
	evaluate both experimentally and theoretically, we establish another
	upper bound to $\varepsilon_{\textrm{KIK}}^{(M)}$ that depends on the single
	quantity $\intop_{0}^{T}\left\Vert \mathcal{L}(t)\right\Vert dt$,
	instead of all the eigenvalues $\lambda_{k}$. 
	
	We start by looking at the behavior of the function
	$f_{M}(\mu,\lambda):=\left|1-\sum_{m=0}^{M}{\color{red}{\normalcolor a_{\textrm{Adap},m}^{(M)}}}(\mu)\left(\lambda\right)^{m+1/2}\right|$
	in the plots of Supplementary Figure 13. This will allow us to derive the bound 
	\begin{equation}
		\varepsilon_{\textrm{KIK}}^{(M)}\leq\sqrt{\langle A|A\rangle}\left|1-\sum_{m=0}^{M}{\color{red}{\normalcolor a_{\textrm{Adap},m}^{(M)}}}(\mu)\left(e^{-2\intop_{0}^{T}\left\Vert \mathcal{L}(t)\right\Vert dt}\right)^{m+1/2}\right|,\label{eq:S79.4}
	\end{equation}
	for $M=1,2,3$. To this end, we note that the right hand side of Eq. \eqref{eq:S79.1 uppor bound on E_KIK^(M) 1} can be written as $\sqrt{\langle A|A\rangle}$$\textrm{max}_{k}f_{M}(\mu,\lambda_{k})$. Thus, we will obtain the bound \eqref{eq:S79.4} by proving the inequality  $\textrm{max}_{k}f_{M}(\mu,\lambda_{k})\leq f_{M}\left(\mu,e^{-2\intop_{0}^{T}\left\Vert \mathcal{L}(t)\right\Vert dt}\right)$, for $M=1,2,3$. First, we will show that this inequality holds when the eigenvalue that maximizes $f_{M}(\mu,\lambda_{k})$ is smaller
	than $\mu$, and then the proof will be extended to the complementary case (i.e. when the aforementioned eigenvalue is larger than $\mu$).
	
	For the first case we resort to the plots (a), (b) and (c) in Supplementary Figure 13. These plots show that if $\lambda\leq\mu$, the
	function $f_{M}(\mu,\lambda)$ is monotonically decreasing with respect
	to $\lambda$. Therefore, the inequality $\textrm{min}_{k}\lambda_{k}\geq e^{-2\intop_{0}^{T}\left\Vert \mathcal{L}(t)\right\Vert dt}$ (cf. Eq. (\ref{eq:S60 bound on minimum eigenv of KIK})) implies that \begin{equation}
		f_{M}(\mu,\lambda_{k})\leq f_{M}(\mu,\textrm{min}_{k}\lambda_{k})\leq f_{M}\left(\mu,e^{-2\intop_{0}^{T}\left\Vert \mathcal{L}(t)\right\Vert dt}\right),\label{eq:S79.3}
	\end{equation}
	for any eigenvalue $\lambda_{k}$ such that $\lambda_{k}\leq\mu$. 
	
	If $\lambda\geq\mu$, the plots (d), (e) and (f) in Supplementary Figure 13
	show that $f_{M}(\mu,\lambda)\leq f_{M}(\mu,\mu)$. Since
	
	\begin{equation}
		\mu=\langle\rho|\mathcal{K}_{\textrm{I}}\mathcal{K}|\rho\rangle=\sum_{k}\lambda_{k}|\langle k|\rho\rangle|^{2}\geq\textrm{min}_{k}\lambda_{k},\label{eq:S79.2 mu>=00003Dmin_k(lambda_k)}
	\end{equation}
	the monotonicity observed in Supplementary Figures 13(a)-13(c) leads to $f_{M}(\mu,\lambda)\leq f_{M}(\mu,\mu)\leq f_{M}(\mu,\textrm{min}_{k}\lambda_{k})$, where the rightmost inequality is a consequence of \eqref{eq:S79.2 mu>=00003Dmin_k(lambda_k)} and the leftmost inequality follows by hypothesis ($\lambda\geq\mu$).
	Therefore, the inequalities (\ref{eq:S79.3})
	also hold when $\lambda_{k}\geq\mu$. This implies that $\textrm{max}_{k}f_{M}(\mu,\lambda_{k})\leq f_{M}\left(\mu,e^{-2\intop_{0}^{T}\left\Vert \mathcal{L}(t)\right\Vert dt}\right)$,
	because (\ref{eq:S79.3}) is valid for any eigenvalue $\lambda_{k}$.
	
	Next, we derive another bound that is independent of both $\lambda_{k}$
	and $\mu$. In this case, we will use the monotonicity of $f_{M}(\mu,\lambda)$ with respect to $\mu$, for $\mu\geq\lambda$ (see Supplementary Figures 13(a)-13(c)). By combining  $\textrm{min}_{k}\lambda_{k}\geq e^{-2\intop_{0}^{T}\left\Vert \mathcal{L}(t)\right\Vert dt}$ (Eq. \eqref{eq:S60 bound on minimum eigenv of KIK}) with Eq. \eqref{eq:S79.2 mu>=00003Dmin_k(lambda_k)}, we have that $\mu,1\geq$$e^{-2\intop_{0}^{T}\left\Vert \mathcal{L}(t)\right\Vert dt}$ and therefore    $f_{M}\left(\mu,e^{-2\intop_{0}^{T}\left\Vert \mathcal{L}(t)\right\Vert dt}\right)\leq f_{M}\left(1,e^{-2\intop_{0}^{T}\left\Vert \mathcal{L}(t)\right\Vert dt}\right)$. By combining this result with Eq. (\ref{eq:S79.4}), we obtain the bound
	\begin{align}
		\varepsilon_{\textrm{KIK}}^{(M)} & \leq\sqrt{\langle A|A\rangle}f_{M}\left(\mu,e^{-2\intop_{0}^{T}\left\Vert \mathcal{L}(t)\right\Vert dt}\right)\nonumber \\
		& \leq\sqrt{\langle A|A\rangle}f_{M}\left(1,e^{-2\intop_{0}^{T}\left\Vert \mathcal{L}(t)\right\Vert dt}\right)\nonumber \\
		& =\sqrt{\langle A|A\rangle}\left|1-\sum_{m=0}^{M}{\color{red}{\normalcolor a_{\textrm{Adap},m}^{(M)}}}(1)\left(e^{-2\intop_{0}^{T}\left\Vert \mathcal{L}(t)\right\Vert dt}\right)^{m+1/2}\right|,\label{eq:S79.5 upper bound on E_KIK^(M) 2}
	\end{align}
	for $M=1,2,3$. Although this bound is looser than (\ref{eq:S79.4}),
	it only depends on $\intop_{0}^{T}\left\Vert \mathcal{L}(t)\right\Vert dt$.
	Since this quantity is the integral of the spectral norm of the dissipator
	$\mathcal{L}(t)$, over the time $T$ consumed by the evolution $\mathcal{K}$,
	it can be seen as a quantifier of the total error rate in our approach.
	Hence, the bound (\ref{eq:S79.5 upper bound on E_KIK^(M) 2}) has
	the advantage of being given only in terms of this error rate. 
	
	Let us see now that this bound is also applicable to Taylor mitigation.
	For $M$ arbitrary, we have that 
	\begin{align}
		\mathcal{U}_{\textrm{KIK}}^{(M)} & \approx\mathcal{U}\sum_{m=0}^{M}{\color{red}{\normalcolor a_{\textrm{Tay},m}^{(M)}}}\left(\mathcal{K}_{\textrm{I}}\mathcal{K}\right)^{m+1/2},\nonumber \\
		\varepsilon_{\textrm{KIK}}^{(M)} & =\left|\langle A|\left(\mathcal{U}-\mathcal{U}_{\textrm{KIK}}^{(M)}\right)|\rho\rangle\right|\nonumber \\
		& \leq\sqrt{\langle A|A\rangle}\left\Vert \mathcal{U}\left(\mathcal{I}-\sum_{m=0}^{M}{\color{red}{\normalcolor a_{\textrm{Tay},m}^{(M)}}}\left(\mathcal{K}_{\textrm{I}}\mathcal{K}\right)^{m+1/2}\right)\right\Vert .\label{eq:S79.6}
	\end{align}
	If $1\leq M\leq3$, a direct calculation allows us to corroborate
	that (cf. Eqs. (\ref{eq:S39 coefficients of the truncated Taylor expansion})
	and (\ref{eq:S44 optimal coeff a_0^(1)})-(\ref{eq:S51.4 optimal coeff a_3^(3)-1}))\textcolor{blue}{{}
		${\normalcolor a_{\textrm{Tay},m}^{(M)}={\color{red}{\normalcolor a_{\textrm{Adap},m}^{(M)}}}(1)}.$
	}Therefore, in this case we can rewrite (\ref{eq:S79.6}) as 
	\begin{equation}
		\varepsilon_{\textrm{KIK}}^{(M)}\leq\sqrt{\langle A|A\rangle}\left\Vert \mathcal{U}\left(\mathcal{I}-\sum_{m=0}^{M}{\color{red}{\normalcolor a_{\textrm{Adap},m}^{(M)}}}(1)\left(\mathcal{K}_{\textrm{I}}\mathcal{K}\right)^{m+1/2}\right)\right\Vert .\label{eq:S79.7}
	\end{equation}
	Finally, it is not difficult to check that $f_{M}(1,\lambda)=\left|1-\sum_{m=0}^{M}{\color{red}{\normalcolor a_{\textrm{Adap},m}^{(M)}}}(1)\left(\lambda\right)^{m+1/2}\right|$
	is monotonically decreasing in $\lambda$, which leads to the inequalities 
	
	\begin{equation}
		f_{M}\left(1,e^{-2\intop_{0}^{T}\left\Vert \mathcal{L}(t)\right\Vert dt}\right)\geq f_{M}(1,\textrm{min}_{k}\lambda_{k})\geq f_{M}(1,\lambda_{k}).\label{eq:S79.8}
	\end{equation}
	From this result and Eq. (\ref{eq:S79.7}), it follows that the bound
	(\ref{eq:S79.5 upper bound on E_KIK^(M) 2}) is also valid for Taylor
	mitigation.\\ 
	\\
	\fakesubsection{Alternative bound for Taylor mitigation}
	\textbf{Alternative bound for Taylor mitigation.} Now, we derive another bound 
	\begin{figure}
		\centering{}\includegraphics[scale=0.35]{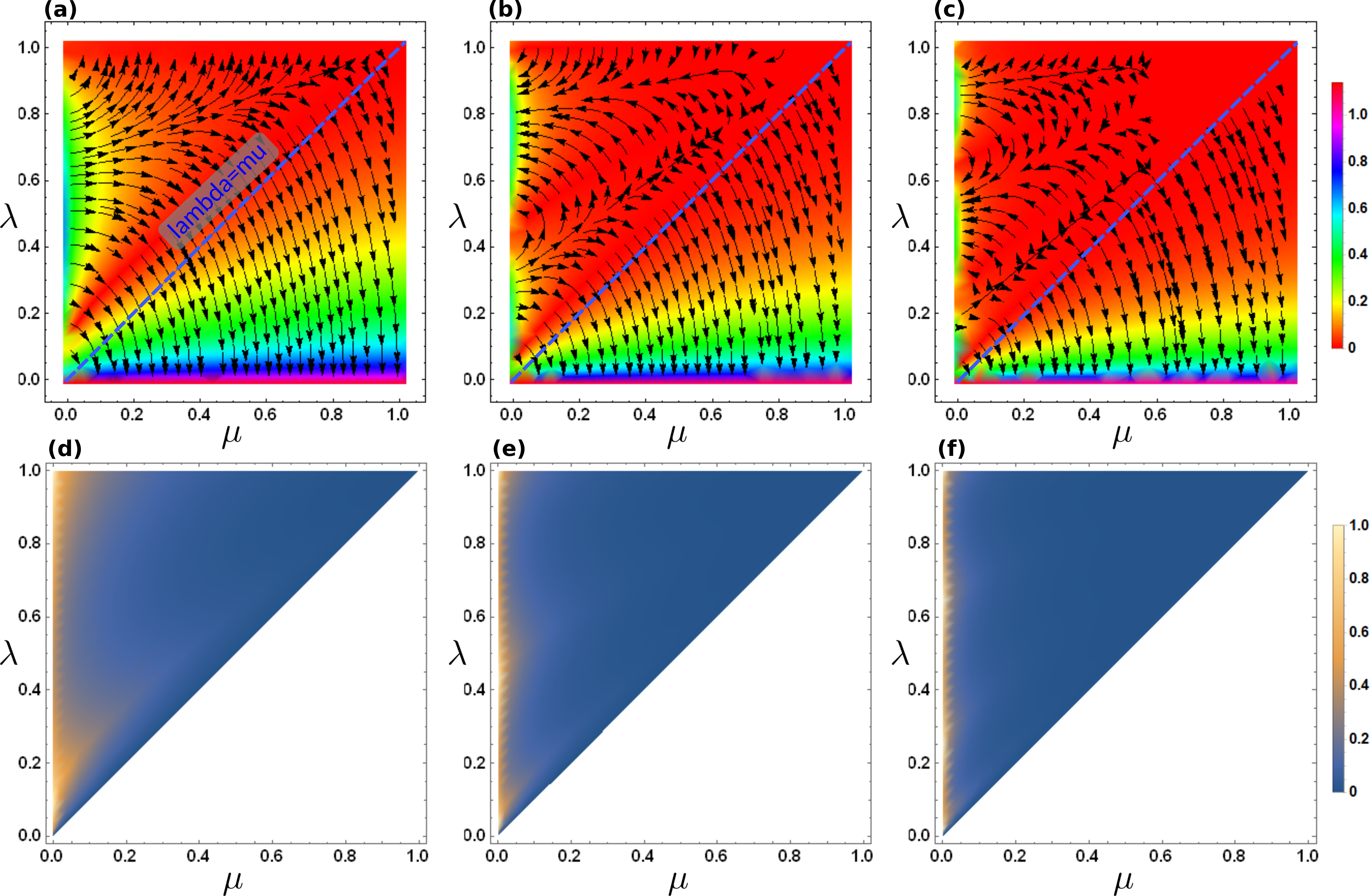}\caption{Plots used in the derivation of the bounds (\ref{eq:S79.4}) and (\ref{eq:S79.5 upper bound on E_KIK^(M) 2}).
			(a), (b), and (c) are color density plots of $f_{M}(\mu,\lambda):=\left|1-\sum_{m=0}^{M}{\color{red}{\normalcolor a_{\textrm{Adap},m}^{(M)}}}(\mu)\left(\lambda\right)^{m+1/2}\right|$,
			for $M=1$, $M=2$, and $M=3$, respectively. The streamlines depict
			the gradient of $f_{M}(\mu,\lambda)$. In particular, we can see that
			for $\lambda\protect\leq\mu$ the functions $f_{M}(\mu,\lambda)$
			are monotonically decreasing with respect to $\lambda$, and monotonically
			increasing with respect to $\mu$. The plots (d), (e) and (f) are
			color density plots of $f_{M}(\mu,\mu)-f_{M}(\mu,\lambda)$ (with
			$1\protect\leq M\protect\leq3$ increasing from left to right), for
			$\lambda\protect\geq\mu$, and show that in this interval $f_{M}(\mu,\mu)\protect\geq f_{M}(\mu,\lambda)$.}
	\end{figure}
	on $\varepsilon_{\textrm{KIK}}^{(M)}$, applicable
	to Taylor error mitigation. While this bound is looser than Eq. (\ref{eq:S79.5 upper bound on E_KIK^(M) 2}),
	it holds for any $M\geq1$ and not only for $1\leq M\leq3$. To this
	end,\textcolor{blue}{{} }we apply the Taylor remainder 
	\begin{equation}
		R_{M}(x)=\intop_{a}^{x}\frac{1}{M!}\left(\frac{d^{M+1}f}{dx^{M+1}}\right)_{x=t}(x-t)^{M}dt,\label{eq:S80 Taylor remainder 1}
	\end{equation}
	which gives the error $R_{M}(x)=f(x)-P_{M}(x)$ when approximating
	a function $f(x)$ with the $M$-degree Taylor polynomial 
	\begin{equation}
		P_{M}(x)=\sum_{m=0}^{M}\frac{1}{m!}\left(\frac{d^{m}f}{dx^{m}}\right)_{x=a}(x-a)^{m}.\label{eq:S81 Taylor polynomial}
	\end{equation}
	
	In Taylor mitigation, we approximate the eigenvalues $f(\lambda)=\lambda^{-\frac{1}{2}}$
	of $\left(\mathcal{K}_{\textrm{I}}\mathcal{K}\right)^{-\frac{1}{2}}$, using
	the Taylor polynomial $P_{M}(\lambda)=\sum_{m=0}^{M}\frac{1}{m!}\left(\frac{(2m-1)!!}{2^{m}}\lambda^{-m-\frac{1}{2}}\right)_{\lambda=1}(\lambda-1)^{m}$.
	The corresponding Taylor remainder is 
	\begin{equation}
		R_{M}(\lambda)=\intop_{1}^{\lambda}\frac{1}{M!}\frac{(2M+1)!!}{2^{M+1}}t^{-M-\frac{3}{2}}(\lambda-t)^{M}dt.\label{eq:S82 Taylor remainder 2}
	\end{equation}
	The eigenvalues of the Hermitian operator $\mathcal{K}_{\textrm{I}}\mathcal{K}=\mathcal{K}^{\dagger}\mathcal{K}$
	must satisfy $\lambda\leq1$. Otherwise, many applications of the
	evolution $\mathcal{K}_{\textrm{I}}\mathcal{K}$ would lead to a non-physical
	operation, characterized by divergent eigenvalues. Taking this into
	account, the absolute value of $R_{M}(\lambda)$ is upper bounded
	by 
	\begin{align}
		\left|R_{M}(\lambda)\right| & \leq\frac{1}{M!}\frac{(2M+1)!!}{2^{M+1}}\lambda^{-M-\frac{3}{2}}\intop_{\lambda}^{1}\left|(\lambda-t)^{M}\right|dt\nonumber \\
		& =\frac{1}{M!}\frac{(2M+1)!!}{2^{M+1}}\lambda^{-M-\frac{3}{2}}\intop_{\lambda}^{1}(t-\lambda)^{M}dt\nonumber \\
		& =\frac{(2M+1)!!}{2^{M+1}(M+1)!}\lambda^{-M-\frac{3}{2}}(1-\lambda)^{M+1}.\label{eq:S83 bound on Taylor remainder}
	\end{align}
	
	Using the diagonal form $\mathcal{K}_{\textrm{I}}\mathcal{K}=\sum_{k}\lambda_{k}|k\rangle\langle k|$,
	we can express the approximation $\sum_{m=0}^{M}a_{\textrm{Tay},m}^{(M)}\left(\mathcal{K}_{\textrm{I}}\mathcal{K}\right)^{m}$
	to the inverse $\left(\mathcal{K}_{\textrm{I}}\mathcal{K}\right)^{-\frac{1}{2}}$
	as 
	\begin{equation}
		\sum_{m=0}^{M}a_{\textrm{Tay},m}^{(M)}\left(\mathcal{K}_{\textrm{I}}\mathcal{K}\right)^{m}=\left(\mathcal{K}_{\textrm{I}}\mathcal{K}\right)^{-\frac{1}{2}}+\sum_{k}\left(\pm\left|R_{M}(\lambda_{k})\right|\right)|k\rangle\langle k|.\label{eq:S84 Taylor polynomial in terms of Taylor remainder}
	\end{equation}
	Therefore, 
	\begin{align}
		\mathcal{U}_{\textrm{KIK}}^{(M)} & =\mathcal{K}\left[\sum_{m=0}^{M}a_{\textrm{Tay},m}^{(M)}\left(\mathcal{K}_{\textrm{I}}\mathcal{K}\right)^{m}\right]\nonumber \\
		& =\mathcal{U}\left(\mathcal{K}_{\textrm{I}}\mathcal{K}\right)^{\frac{1}{2}}\left[\left(\mathcal{K}_{\textrm{I}}\mathcal{K}\right)^{-\frac{1}{2}}+\sum_{k}\left(\pm\left|R_{M}(\lambda_{k})\right|\right)|k\rangle\langle k|\right]\nonumber \\
		& =\mathcal{U}+\sum_{k}\left(\pm\lambda_{k}^{\frac{1}{2}}\left|R_{M}(\lambda_{k})\right|\right)|k\rangle\langle k|,\label{eq:S85 U_KIK^(M) in terms of taylor remainder}
	\end{align}
	where we write $\mathcal{K}$ as $\mathcal{K}=\mathcal{U}\left(\mathcal{K}_{\textrm{I}}\mathcal{K}\right)^{\frac{1}{2}}$in
	the second line, and $\left(\mathcal{K}_{\textrm{I}}\mathcal{K}\right)^{\frac{1}{2}}$
	as $\left(\mathcal{K}_{\textrm{I}}\mathcal{K}\right)^{\frac{1}{2}}=\sum_{k}\lambda_{k}^{\frac{1}{2}}|k\rangle\langle k|$
	in the third line. Accordingly, for Taylor mitigation we have that
	\begin{align}
		\varepsilon_{\textrm{KIK}}^{(M)} & =\left|\langle A|\left(\mathcal{U}-\mathcal{U}_{\textrm{KIK}}^{(M)}\right)|\rho\rangle\right|\nonumber \\
		& \leq\frac{(2M+1)!!}{2^{M+1}(M+1)!}\sqrt{\langle A|A\rangle}\left\Vert \sum_{k}\left(\frac{1}{\lambda_{k}}-1\right)^{M+1}|k\rangle\langle k|\right\Vert \nonumber \\
		& =\frac{(2M+1)!!}{2^{M+1}(M+1)!}\sqrt{\langle A|A\rangle}\left(\frac{1}{\textrm{min}_{k}\lambda_{k}}-1\right)^{M+1}\nonumber \\
		& \leq\frac{(2M+1)!!}{2^{M+1}(M+1)!}\sqrt{\langle A|A\rangle}\left(e^{2\intop_{0}^{T}\left\Vert \mathcal{L}(t)\right\Vert dt}-1\right)^{M+1},\label{eq:S86 upper bound on Taylor mitigation error}
	\end{align}
	where we assume again $\langle\rho|\rho\rangle=1$, and Eqs. (\ref{eq:S83 bound on Taylor remainder})
	and (\ref{eq:S60 bound on minimum eigenv of KIK}) are respectively
	applied in the second line and the last line. 
	
	In the following section, we will take advantage of an important property
	of the coefficients $a_{\textrm{Adap},m}^{(M)}(\mu)$ and $a_{\textrm{Tay},m}^{(M)}$,
	in order to further tighten the bounds (\ref{eq:S79.4}), (\ref{eq:S79.5 upper bound on E_KIK^(M) 2})
	and (\ref{eq:S86 upper bound on Taylor mitigation error}).\\ 
	\\
	\fakesubsection{Traceless observables and second (tighter) error bounds for Adaptive
		error mitigation and Taylor error mitigation}
	\textbf{Traceless observables and second (tighter) error bounds for Adaptive
		error mitigation and Taylor error mitigation.} Given an arbitrary observable $A$, we already know that the error-mitigated
	expectation value in the case of $M$th order mitigation
	is given by 
	\begin{equation}
		\left\langle A\right\rangle _{\textrm{mit}}=\left\langle A\left|\sum_{m=0}^{M}a_{m}^{(M)}\mathcal{K}\left(\mathcal{K}_{\textrm{I}}\mathcal{K}\right)^{m}\right|\rho\right\rangle ,\label{eq:S87 Taylor error-mitigated expectation value-1}
	\end{equation}
	where $a_{m}^{(M)}=a_{\textrm{Adap},m}^{(M)}(\mu)$ for adaptive mitigation
	and $a_{m}^{(M)}=a_{\textrm{Tay},m}^{(M)}$ for Taylor mitigation.
	
	Now, let us see that the error $\varepsilon_{\textrm{KIK}}^{(M)}$ is invariant
	under a transformation $A\rightarrow A+bI$, where $I$ is the identity
	operator and $b$ is a real number. This shifts the trace of $A$
	by the value $b\textrm{Tr}(I)$. Letting $\varepsilon_{\textrm{KIK}}^{(M)}(b)$
	denote the error corresponding to the observable $A+bI$, we have
	that 
	\begin{align}
		\varepsilon_{\textrm{KIK}}^{(M)}(b) & =\left|\left\langle A+bI\left|\sum_{m=0}^{M}a_{m}^{(M)}\mathcal{K}\left(\mathcal{K}_{\textrm{I}}\mathcal{K}\right)^{m}\right|\rho\right\rangle -\left\langle A+bI\left|\mathcal{U}\right|\rho\right\rangle \right|\nonumber \\
		& =\left|\sum_{m=0}^{M}a_{m}^{(M)}\langle A|\mathcal{K}\left(\mathcal{K}_{\textrm{I}}\mathcal{K}\right)^{m}|\rho\rangle-\left\langle A\left|\mathcal{U}\right|\rho\right\rangle +b\sum_{m=0}^{M}a_{m}^{(M)}-b\right|,\label{eq:S88 error in Taylor mitigation for A+bI-1}
	\end{align}
	where in the second line we use the fact that $\left\langle I\left|\mathcal{K}\left(\mathcal{K}_{\textrm{I}}\mathcal{K}\right)^{m}\right|\rho\right\rangle =\textrm{Tr}\left(\rho^{(m)}\right)=1$,
	for a trace-preserving evolution $\mathcal{K}\left(\mathcal{K}_{\textrm{I}}\mathcal{K}\right)^{m}$.
	Here, $\rho^{(m)}$ is the state resulting from applying the circuit
	$\mathcal{K}\left(\mathcal{K}_{\textrm{I}}\mathcal{K}\right)^{m}$ on $\rho$. 
	
	In the case of adaptive mitigation, we have that, for all $M$, $\sum_{m=0}^{M}a_{m}^{(M)}=\sum_{m=0}^{M}a_{\textrm{Adap},m}^{(M)}(\mu)=1$
	by construction. Since in the limit of zero noise the Taylor approximation
	converges to $\left(\mathcal{K}_{\textrm{I}}\mathcal{K}\right)^{-1}=\mathcal{I}$,
	it also follows from Eq.\textcolor{blue}{{} (\ref{eq:S38 truncated expansion of N^-1})}
	that 
	\begin{equation}
		\sum_{m=0}^{M}a_{\textrm{Tay},m}^{(M)}=1,\textrm{ for all }M.\label{eq:S89 Sum_m(a_Tay_m)=00003D1 in the limit of zero noise-1}
	\end{equation}
	Therefore, Eq. (\ref{eq:S88 error in Taylor mitigation for A+bI-1})
	is equivalent to (both for adaptive mitigation and Taylor mitigation)
	\begin{equation}
		\varepsilon_{\textrm{KIK}}^{(M)}(b)=\varepsilon_{\textrm{KIK}}^{(M)}(0),\label{eq:S90 invariance of error for Taylor mitigation-1}
	\end{equation}
	for all $b$ real. This result implies that the error $\varepsilon_{\textrm{KIK}}^{(M)}(0)$,
	associated with the actual observable $A$, can be evaluated using
	instead the shifted observable $A+bI$. By combining this property
	with Eqs. (\ref{eq:S79.4}), (\ref{eq:S79.5 upper bound on E_KIK^(M) 2}),
	and\textcolor{blue}{{} (\ref{eq:S86 upper bound on Taylor mitigation error})},
	we can obtain the families of $b$-dependent bounds
	\begin{align}
		\varepsilon_{\textrm{KIK}}^{(M)}(0) & \leq\sqrt{\textrm{Tr}\left[\left(A+bI\right)^{2}\right]}\left|1-\sum_{m=0}^{M}a_{\textrm{Adap},m}^{(M)}(\mu)\left(e^{-2\intop_{0}^{T}\left\Vert \mathcal{L}(t)\right\Vert dt}\right)^{m+1/2}\right|,\label{eq:S90.1 family of b-bounds for the adapted mitigation error}\\
		\varepsilon_{\textrm{KIK}}^{(M)}(0) & \leq\frac{(2M+1)!!}{2^{M+1}(M+1)!}\sqrt{\textrm{Tr}\left[\left(A+bI\right)^{2}\right]}\left(e^{2\intop_{0}^{T}\left\Vert \mathcal{L}(t)\right\Vert dt}-1\right)^{M+1},\label{eq:S91 family of b-bounds for the Taylor error-1}
	\end{align}
	where we have written $\langle A+bI|A+bI\rangle$ as $\textrm{Tr}\left[\left(A+bI\right)^{2}\right]$.\textcolor{blue}{{}
	}In the case of Eq. (\ref{eq:S90.1 family of b-bounds for the adapted mitigation error}),
	values of $0\leq\mu<1$ yield bounds for adapative mitigation, as
	per Eq. (\ref{eq:S79.4}), and $\mu=1$ yields bounds applicable to
	adaptive mitigation and Taylor mitigation, cf. Eq. (\ref{eq:S79.5 upper bound on E_KIK^(M) 2}).
	On the other hand, Eq. (\ref{eq:S91 family of b-bounds for the Taylor error-1})
	is a consequence of the Taylor-mitigation bound (\ref{eq:S86 upper bound on Taylor mitigation error}).
	
	The tightest bounds in (\ref{eq:S90.1 family of b-bounds for the adapted mitigation error})
	and (\ref{eq:S91 family of b-bounds for the Taylor error-1}) are
	obtained by minimizing $\textrm{Tr}\left[\left(A+bI\right)^{2}\right]=\textrm{Tr}\left(A^{2}\right)+2b\textrm{Tr}\left(A\right)+b^{2}\textrm{Tr}\left(I\right)$,
	with respect to $b$. A simple calculation shows that the minimum
	is attained by $b$ such that the observable $A+bI$ is traceless,
	i.e. $\textrm{Tr}\left(A+bI\right)=0$. The resulting expression is
	\begin{equation}
		b=-\frac{\textrm{Tr}\left(A\right)}{\textrm{Tr}\left(I\right)}.\label{eq:S92 optimal b-1}
	\end{equation}
	By substituting this result into Eqs. (\ref{eq:S90.1 family of b-bounds for the adapted mitigation error})
	and (\ref{eq:S91 family of b-bounds for the Taylor error-1}), we
	obtain the optimal bounds
	\begin{align}
		\varepsilon_{\textrm{KIK}}^{(M)}(0) & \leq\sqrt{\textrm{Tr}\left(A^{2}\right)-\frac{\left[\textrm{Tr}\left(A\right)\right]^{2}}{\textrm{Tr}\left(I\right)}}\left|1-\sum_{m=0}^{M}a_{\textrm{Adap},m}^{(M)}(\mu)\left(e^{-2\intop_{0}^{T}\left\Vert \mathcal{L}(t)\right\Vert dt}\right)^{m+1/2}\right|,\label{eq:S93}\\
		\varepsilon_{\textrm{KIK}}^{(M)}(0) & \leq\frac{(2M+1)!!}{2^{M+1}(M+1)!}\sqrt{\textrm{Tr}\left(A^{2}\right)-\frac{\left[\textrm{Tr}\left(A\right)\right]^{2}}{\textrm{Tr}\left(I\right)}}\left(e^{2\intop_{0}^{T}\left\Vert \mathcal{L}(t)\right\Vert dt}-1\right)^{M+1}.\label{eq:S94}
	\end{align}
	\fakesubsection{Summary}
	\textbf{Summary.} To conclude this supplementary note, we merge the bounds previously derived
	into a single expression. That is, 
	\begin{align}
		\varepsilon_{\textrm{KIK}}^{(M)} & \leq\sqrt{\textrm{Tr}\left(A^{2}\right)-\frac{\left[\textrm{Tr}\left(A\right)\right]^{2}}{\textrm{Tr}\left(I\right)}}\left|1-\sum_{m=0}^{M}{\color{red}{\normalcolor a_{\textrm{Adap},m}^{(M)}}}(\mu)e^{-2(m+1/2)\intop_{0}^{T}\left\Vert \mathcal{L}(t)\right\Vert dt}\right|,\textrm{for }M=1,2,3,\label{eq:S94.1}\\
		& \leq\sqrt{\textrm{Tr}\left(A^{2}\right)-\frac{\left[\textrm{Tr}\left(A\right)\right]^{2}}{\textrm{Tr}\left(I\right)}}\left|1-\sum_{m=0}^{M}{\color{red}{\normalcolor a_{\textrm{Adap},m}^{(M)}}}(1)e^{-2(m+1/2)\intop_{0}^{T}\left\Vert \mathcal{L}(t)\right\Vert dt}\right|,\textrm{\textrm{for }}M=1,2,3,\label{eq:S94.2}\\
		& \leq\frac{(2M+1)!!}{2^{M+1}(M+1)!}\sqrt{\textrm{Tr}\left(A^{2}\right)-\frac{\left[\textrm{Tr}\left(A\right)\right]^{2}}{\textrm{Tr}\left(I\right)}}\left(e^{2\intop_{0}^{T}\left\Vert \mathcal{L}(t)\right\Vert dt}-1\right)^{M+1},\label{eq:S95}
	\end{align}
	where Eq. (\ref{eq:S94.1}) is equivalent to Eq. (\ref{eq:S93}),
	and (\ref{eq:S94.2}) follows by applying the optimization strategy
	of the previous section to Eq. (\ref{eq:S79.5 upper bound on E_KIK^(M) 2}).
	Equation (\ref{eq:S95}) corresponds to Eq. (\ref{eq:S94}), and is
	the looser bound according to Supplementary Figure 14. We also recall
	that the bound (\ref{eq:S94.2}) is valid for both adaptive mitigation
	and Taylor mitigation, in addition to being independent of $\mu$.
	On the other hand, the bound (\ref{eq:S95}) is valid for Taylor mitigation
	and for all $M\geq1$. From this bound we can also see that the error
	$\varepsilon_{\textrm{KIK}}^{(M)}$ is exponentially decreasing in $M$ if
	$\intop_{0}^{T}\left\Vert \mathcal{L}(t)\right\Vert dt<\frac{1}{2}\textrm{ln}(2)$,
	since the prefactor $\frac{(2M+1)!!}{2^{M+1}(M+1)!}$ satisfies $\frac{(2M+1)!!}{2^{M+1}(M+1)!}\leq\frac{3}{8}$.
	
	\begin{figure}
		
		\begin{centering}
			\includegraphics[scale=0.6]{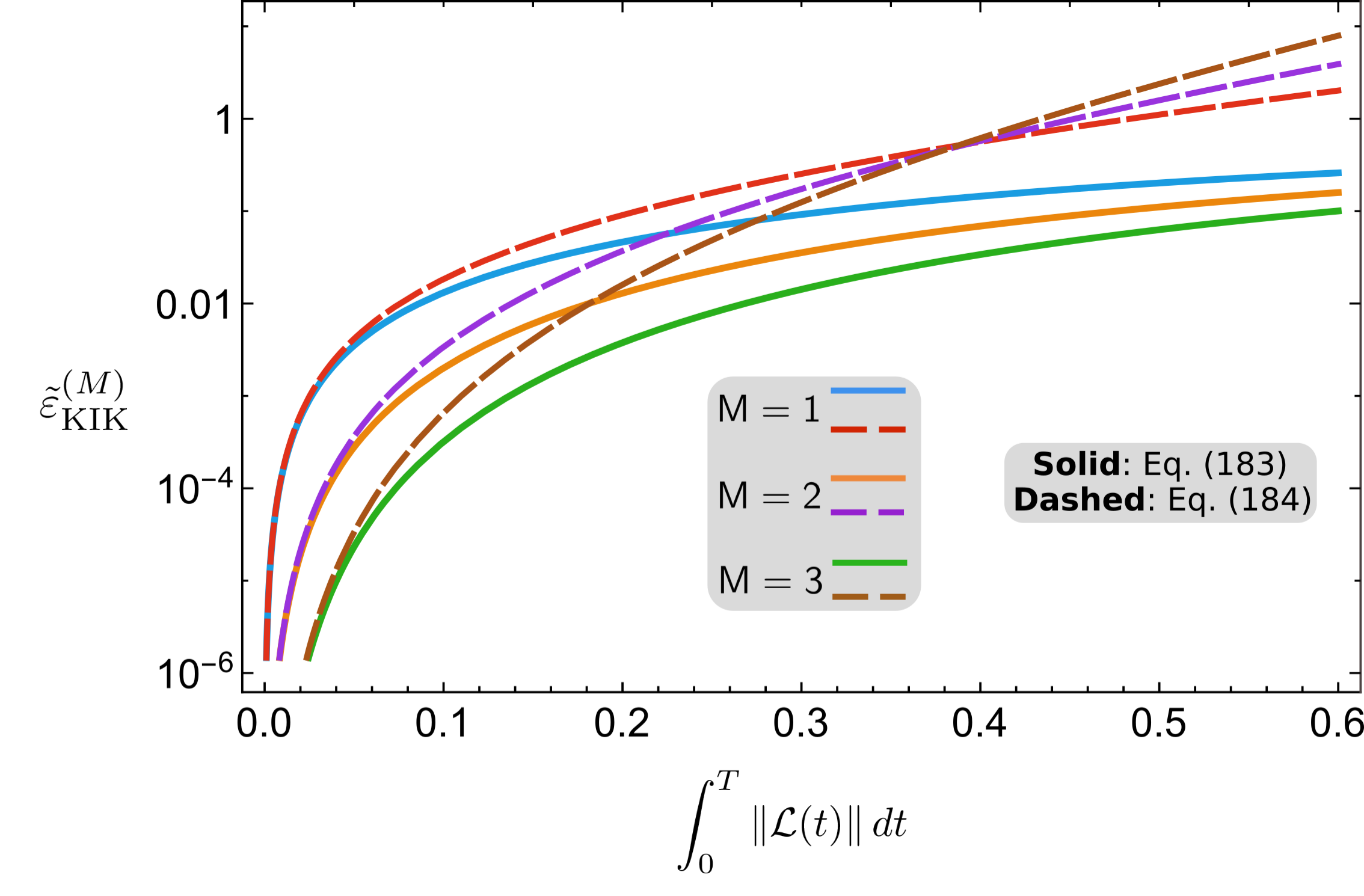}\caption{Upper bounds (\ref{eq:S94.2}) and (\ref{eq:S95}) on $\tilde{\varepsilon}_{\textrm{KIK}}^{(M)}:=\frac{\varepsilon_{\textrm{KIK}}^{(M)}}{\sqrt{\textrm{Tr}\left(A^{2}\right)-\frac{\left[\textrm{Tr}\left(A\right)\right]^{2}}{\textrm{Tr}\left(I\right)}}}$,
				for mitigation orders $M=1,2,3$. }
			\par\end{centering}
	\end{figure}
	
	We also stress that, once coherent noise is converted into incoherent
	noise, via RC, the quantity $\intop_{0}^{T}\left\Vert \mathcal{L}(t)\right\Vert dt$
	accounts for the total error rate affecting the evolution $\mathcal{K}$,
	irrespective of the depth or the width of the corresponding circuit.
	Therefore, the bounds (\ref{eq:S94.1})-(\ref{eq:S95}) are meaningful
	to assess the performance of the KIK method applied to circuits $\mathcal{K}$
	of arbitrary size. It is also important to note that generic QEM methods
	can be useful so long as the total error rate is not excessively high
	\cite{endo2021hybrid}. In the context of the KIK method,
	we can have scalable QEM if $\intop_{0}^{T}\left\Vert \mathcal{L}(t)\right\Vert dt$
	is kept below a fixed value for
	increasingly large circuits, and if for this value the bound (\ref{eq:S94.1})
	is sufficiently small.

	


\end{document}